\documentclass{aa}
\bibpunct{(}{)}{;}{a}{}{,}
\usepackage[varg]{txfonts}
\usepackage{graphicx,gensymb}
\usepackage{subcaption}
\usepackage[normalem]{ulem}
\usepackage{amsmath, amssymb, amsfonts}
\usepackage{rotating} 
\usepackage{adjustbox}
\usepackage{ulem}
\usepackage{upgreek}
\usepackage{float}
\begin{document}
\pagestyle{empty}
\title{Probing the Mpc-scale environment of hyperluminous infrared galaxies at $2<z<4$}

\author{F.~Gao\inst{\ref{inst1},\ref{inst2}}
\and L.~Wang\inst{\ref{inst1},\ref{inst2}}
\and A.~F.~Ramos Padilla\inst{\ref{inst1},\ref{inst2}}
\and D.~Clements\inst{\ref{inst3}}
\and D.~Farrah\inst{\ref{inst4},\ref{inst5}}
\and T.~Huang\inst{\ref{inst6},\ref{inst7}}
}

\institute{Kapteyn Astronomical Institute, University of Groningen, Postbus 800, 9700 AV Groningen, The Netherlands\label{inst1}
\and 
SRON Netherlands Institute for Space Research, Landleven 12, 9747 AD, Groningen, The Netherlands\label{inst2}
\and
Imperial College London, Blackett Laboratory, Prince Consort Road, London, SW7 2AZ, UK\label{inst3}
\and
Department of Physics and Astronomy, University of Hawai’i, 2505 Correa Road, Honolulu, HI 96822, USA\label{inst4}
\and
Institute for Astronomy, 2680 Woodlawn Drive, University of Hawai’i, Honolulu, HI 96822, USA\label{inst5}
\and
Department of Space and Astronautical Science, Graduate University for Advanced Studies, SOKENDAI, Shonankokusaimura, Hayama, Miura District, Kanagawa 240-0193, Japan\label{inst6}
\and
Institute of Space and Astronautical Science, Japan Aerospace Exploration Agency, 3-1-1 Yoshinodai, Chuo-ku, Sagamihara, Kanagawa 252-5210, Japan\label{inst7}
}

\abstract{Protoclusters are progenitors of galaxy clusters and are important for studying how halo mass and stellar mass assemble in the early universe. Finding signposts of such over-dense regions, for example bright dusty star-forming galaxies (DSFG), is a popular method to identify protocluster candidates.}{Hyperluminous infrared galaxies (HLIRGs), which are ultra-massive and show extreme levels of dusty star formation and/or black hole accretion, are expected to reside in overdense regions with massive halos. We study the Mpc-scale environment of the largest HLIRG sample to date (526 HLIRGs over 26 deg$^2$) and investigate whether they predominantly live in overdense regions.}{We first explore the surface density of \textit{Herschel} 250 $\mu$m sources around HLIRGs and compare with that around random positions. Then, we compare the spatial distribution of neighbours around HLIRGs with that around randomly selected galaxies using a deep IRAC-selected catalogue with good-quality photometric redshifts. We also use a redshift-matched quasar sample and submillimeter galaxy (SMG) sample to validate our method, as previous clustering studies have measured the host halo masses of these populations. Finally, we adopt a Friends of Friends (FOF) algorithm to seek (proto)clusters that host HLIRGs.}{We find that HLIRGs tend to have more bright star-forming neighbours (with 250 $\mu$m flux density $>$10 mJy) within 100$\arcsec$ projected radius ($\sim 0.8$ Mpc at $2<z<4$) than a random galaxy at a $3.7\sigma$ significance. In our 3D analysis, we find relatively weak excess of IRAC-selected sources within 3 Mpc around HLIRGs compared with random galaxy neighbours, mainly influenced by photometric redshift uncertainty and survey depth. We find a more significant difference (at a 4.7$\sigma$ significance) in the number of Low Frequency Array (LOFAR)-detected neighbours in the deepest ELAIS-N1 (EN1) field. HLIRGs at $3<z<4$ show stronger excess compared to HLIRGs at $2<z<3$ ($0.13\pm{0.04}$ and $0.14\pm{0.01}$ neighbours around HLIRGs and random positions at $2<z<3$ respectively, and $0.08\pm{0.04}$ and $0.05\pm{0.01}$ neighbours around HLIRGs and random positions at $3<z<4$ respectively), consistent with cosmic downsizing. Finally, we select and present a list of 30 most promising protocluster candidates for future follow-up observations. }{}

\maketitle
\thispagestyle{empty}

\section{Introduction}\label{intro}

In the local universe, the morphology-density relation is well established: galaxies that reside in dense regions such as clusters are more likely to be massive ellipticals and have lower star-formation rates (SFR), while galaxies in less crowded regions are more likely to be less massive spirals with ongoing star formation \citep{1980ApJ...236..351D, 2002MNRAS.334..673L, 2003ApJ...584..210G,2004MNRAS.353..713K,2005ApJ...621..673T, 2010ApJ...721..193P,2011MNRAS.416.1680C,2013ApJS..206....3S}. Numerous studies have also revealed that massive early-type galaxies formed most of their stellar mass in the early universe \citep[galaxy down-sizing, e.g.,][]{1996AJ....112..839C, 2005ApJ...621..673T,2006MNRAS.366..499D,2008A&A...482...21C,2008ApJ...675..234P,2010ApJ...709..644I}. Therefore, overdense regions at the present day such as galaxy clusters that host massive early-type galaxies must have had rapid star-formation activity at higher redshifts, when clusters are less virialized \citep[often referred as protoclusters;][]{2016A&ARv..24...14O}. For example, \citet{ 2007A&A...468...33E} used data from the Great Observatories Origins Deep Survey(GOODS) and demonstrated that SFR increases as galaxy density increases at z$\sim 1$, opposite to what have been found for local galaxies. By studying galaxy members in a Spitzer-selected cluster at z = 1.62, \citet{2010ApJ...719L.126T} found that the fraction of star-forming galaxies (SFGs) triples from the lowest to the highest density regions. A recent work by \citet{2020arXiv200903324L} which studied 6730 spectroscopically confirmed SFGs at $2<z<5$ found a positive trend between average SFR and local environment.

Protoclusters are progenitors of today's most massive structures. They play an important role in answering many key questions, such as how halo and stellar mass assemble across cosmic time, how environment affects the evolution of massive galaxies and so on. To study the nature of protoclusters, we first need an efficient method to find them. However, identifying protoclusters is more challenging than identifying their descendant clusters, which can be found by searching for overdensity of red massive galaxies or  hot intracluster medium (ICM) through the Sunyaev–Zel'Dovich effect. In contrast, the lack of red massive galaxies and mature ICM make it difficult to trace protoclusters \citep[see review][and references therein]{2016A&ARv..24...14O}.

By definition protoclusters can be selected by tracing overdensity of galaxies in a small region (on the scale corresponding to the expected size of massive halos scale of several Mpc), ideally spectroscopically confirmed. However, good-quality photometric redshifts have also been used \citep[e.g.,][]{2011Natur.470..233C, 2014ApJ...782L...3C, 2016ApJ...833...15F,2018A&A...615A..77L}. An alternative method is to trace surface overdensity of galaxies that occupy a narrow redshift slice. For example, \citet{2014ApJ...796..126L} reported a large-scale structure containing three protoclusters at $z=3.78$ in the Bo$\rm \ddot{o}$tes field by searching for overdensity of Lyman $\alpha$ emitter (LAE) candidates. In their work, deep imaging survey revealed 65 LAEs within a comoving volume of $72 \times 72 \times 25$ Mpc$^3$. These protoclusters may evolve into clusters with a total stellar mass of a few times of 10$^{14} \, \mathrm{M_{\sun}}$ to 10$^{15}\, \mathrm{M_{\sun}}$ at the present day. \citet{2018PASJ...70S..12T} selected 179 protocluster candidates at $z\sim 4$ by tracing surface overdensity of Lyman break galaxies (LBGs), based on data from the Hyper SuprimeCam Subaru strategic program (HSC-SSP) covering a wide area of 121 deg$^2$. These methods provide a simple and direct way to identify potential protoclusters, and thus contribute to a population census of protoclusters. However, they usually require a large-area sky survey to conduct a systematic search for overdense regions due to their rarity. Moreover, as redshift increases, protoclusters occupying a given volume will cover a larger sky area, making it more challenging to find them in a survey with limited sky coverage. In addition, selections based on overdensities of LAEs or LBGs may not be able to uncover overdensities of dusty star-forming galaxies which dominate the star-forming population at high redshifts \citep[e.g., see][for a review]{2014PhR...541...45C} due to severe obscuration in the optical bands. Owing to these difficulties, identification of protoclusters at high redshifts is so far largely dependent on serendipitous findings, by targeting signposts which are expected to reside in dense regions. Popular signposts include radio-loud galaxies \citep[e.g.,][]{2000A&A...361L..25P,2001ApSSS.277..543K,2012ApJ...757...15H,2014MNRAS.441L...1S}, enormous Lyman $\alpha$ nebulae \citep[e.g.,][]{2018A&A...620A.202A,2018ApJ...861L...3C,2021ApJ...922..236L,2022A&A...658A..77N} and quasars \citep[e.g.,][]{2005ApJ...626...44S,2017A&A...606A..23B,2021arXiv210909754G}.

Given the fact that larger SFR is observed in denser region at high redshifts and the predominant role of dusty star-formation activity at high redshifts, it is widely adopted to search for protoclusters either through overdensities of dusty star-forming galaxies (DSFGs) or using bright DSFGs as a signpost. DSFG can be selected as bright far-infrared (FIR) or submillimeter detections. Their star-formation activity is obscured by dust which absorbs UV/optical photons and re-emits them in the longer wavelength range. One efficient method to find overdensity of DSFGs is using large FIR/submillimeter surveys covering a wide area such as surveys with \textit{Planck} and \textit{Herschel}. The total flux density within the beam of a low-resolution instrument is usually a combination of multiple sources located within the same beam. Follow-up high-resolution imaging can help confirm the overdense nature for regions with exceedingly large flux densities \citep[e.g.,][]{2005MNRAS.358..869N}. For example, by observing \textit{Planck} pre-selected protocluster candidates through \textit{Herschel}, a number of protoclusters have been confirmed \citep[e.g.,][]{2014MNRAS.439.1193C, 2016A&A...585A..54F, 2018MNRAS.476.3336G,2018A&A...620A.198M}. A famous protocluster selected in this way is SPT2349-56 at $z=4.3$ which was first found in the 2,500 deg$^2$ South Pole Telescope (SPT) submillimeter survey. Follow-up Atacama Large Millimeter/submillimeter Array (ALMA) observations reveal that this protocluster consists of at least 14 gaseous galaxies in a region of 130 kpc diameter, and is very likely to eventually grow into one of the most massive structures in the local universe \citep{2018Natur.556..469M,2020MNRAS.495.3124H}. This method of targeting bright sources detected in low-resolution single dish surveys can be very effective at pinpointing the most intense star-forming cores of protoclusters, which may be evolutionarily connected to the bright cluster galaxies (BCGs) in the local Universe \citep[e.g.,][]{2021MNRAS.502.1797R}.

The success of tracing protoclusters through DSFGs naturally leads us to ask whether the most extreme DSFG live in the densest regions and trace the most massive protoclusters. Hyper luminous infrared galaxies (HLIRGs) are extremely luminous in the infrared (IR) band (rest-frame 8-1000 $\mu$m), reaching total IR luminosity $L_{IR}> 10^{13} L_{\odot}$ \citep{2000MNRAS.316..885R}. This huge amount of IR luminosity implies a SFR of a few thousand $\mathrm{M_{\sun}}\,\mathrm{yr^{-1}}$. In addition, active galactic nucleus (AGN) activity can also contribute significantly to their extreme IR luminosity. %Whether HLIRGs are starburst dominated or AGN dominated is still debated and could vary significantly on a source by source basis. By fitting submillimeter spectra of 11 HLIRGs up to $z\sim 1.1$, \citet{2002MNRAS.335.1163F} found an AGN contribution of 20-80$\%$ with a mean value of 35$\%$. 
%\citet{2003MNRAS.338L..19W} observed two low-z HLIRGs with \textit{XMM-Newton} and revealed their Compton-thick nature of AGN activity, with line of sight column density $N_{H}>1.5\times 10^{24} cm^{-2}$. \citet{2018MNRAS.479L..91S} approached this question by investigating the infrared luminosity function of galaxies and the X-ray luminosity function of AGN at $z\sim 1-2$. They found an increased AGN dominated sources as total IR luminosity increases. Even dominated by AGN activity, it is still highly likely that these HLIRGs may live in dense regions as elaborated in Section \ref{data3}.
HLIRGs are expected to live in overdense environment as they are extreme DSFGs and/or show vigorous black hole accretion. Therefore, they are good targets to search for protoclusters at higher redshifts. To date, there are only a few studies focused on the environment of HLIRGs due to their rarity. \citet{2004MNRAS.349..518F} studied six HLIRGs selected using data from the Infrared Astronomical Satellite (IRAS) at $0.44<z<1.55$ and evaluated their environment through the amplitude of the spatial cross-correlation function, based on a quantitative measurement presented in \citet{1979MNRAS.189..433L}. They found a larger average clustering level of their HLIRG sample compared with local IR luminous galaxies. \citet{2014MNRAS.443..146J} selected 10 hot dust-obscured galaxies (hot DOGs) that are detected in WISE 12 or 22 $\mu m$ but undetected in 3.4 and 4.6 $\mu m$, six of which are HLIRGs. They found a factor of $\sim$3 overdensity of SMGs around these hot DOGs compared with 
blank-field submillimeter surveys. \citet{2015ApJ...804...27A} counted galaxies having red IRAC colors within 1\arcmin of 90 HLIRGs selected from WISE. They found that the number of red IRAC sources is significantly higher than those around random pointings. 

Following the largest HLIRG sample constructed in \citet{2021A&A...648A...8W} and further studied in \citet{2021A&A...654A.117G}, in this work we investigate whether they live in dense regions as expected. The structure of this paper is as follows. We describe our data in Section \ref{data}. We first introduce our HLIRG sample, then describe the deblended \textit{Herschel} 250 $\mu$m catalogs and the deep IRAC-selected photometric redshift catalogs used to search for HLIRG neighbours. We also include a quasar sample and a SMG sample for cross-checks. We study the number of HLIRG neighbours within different projected separations as well as spatial volumes and compare with that around randomly chosen sources in Section \ref{re}. In addition, we discuss the influence due to photometric redshift uncertainty and survey depth. We also adopt a Friends of Friends (FOF) algorithm to search for overdense regions that are associated with HLIRGs. We summarize our findings and provide a list of 30 most promising protocluster candidates for future observations in Section \ref{conclu}. Throughout this paper, we assume a flat $\Lambda$CDM universe with $\Omega_{\text{M}} = 0.286$ and $H_{0}=69.3 \,\rm km s^{-1} Mpc^{-1}$ \citep[Nine-year Wilkinson Microwave Anisotropy Probe (WMAP) results;][]{2013ApJS..208...19H}. Unless otherwise stated, we adopt a \citet{1955ApJ...121..161S} initial mass function (IMF). For magnitudes, we use standard AB magnitude system.

\section{Data and Method}\label{data}
In this section, we introduce the various samples used in this work. We first describe our HLIRG sample in Section \ref{data1}. They are selected in \citet{2021A&A...648A...8W} and further studied in \citet{2021A&A...654A.117G}. In order to search for neighbours of HLIRGs and study their distributions, we need deep source catalogs which trace the galaxy density field. Source catalogs without redshift information can help find neighbours within certain projected separations, and source catalogs containing redshift information can assist in seeking neighbours within certain spatial volumes. We first introduce the deblended \textit{Herschel} source catalogs in Section \ref{data2} used to search for SFGs around HLIRGs within different projected separations. We then describe the deep IRAC-selected source catalogs and their good-quality photometric redshifts in Section \ref{data3} (referred as deep photo-z catalogs hereafter) to find HLIRG neighbours that reside within different spatial volumes. We characterize the overdense nature of HLIRG environment by comparing the distributions of HLIRG neighbours with that of random galaxy neighbours. In order to validate our method, we make use of a quasar sample and an SMG sample which have estimates of host dark matter halo mass through clustering studies. We briefly summarize our samples in Section \ref{data6}.

\begin{figure}[htbp]
    \centering
    \begin{subfigure}[b]{.5\textwidth}
        \includegraphics[width=\textwidth]{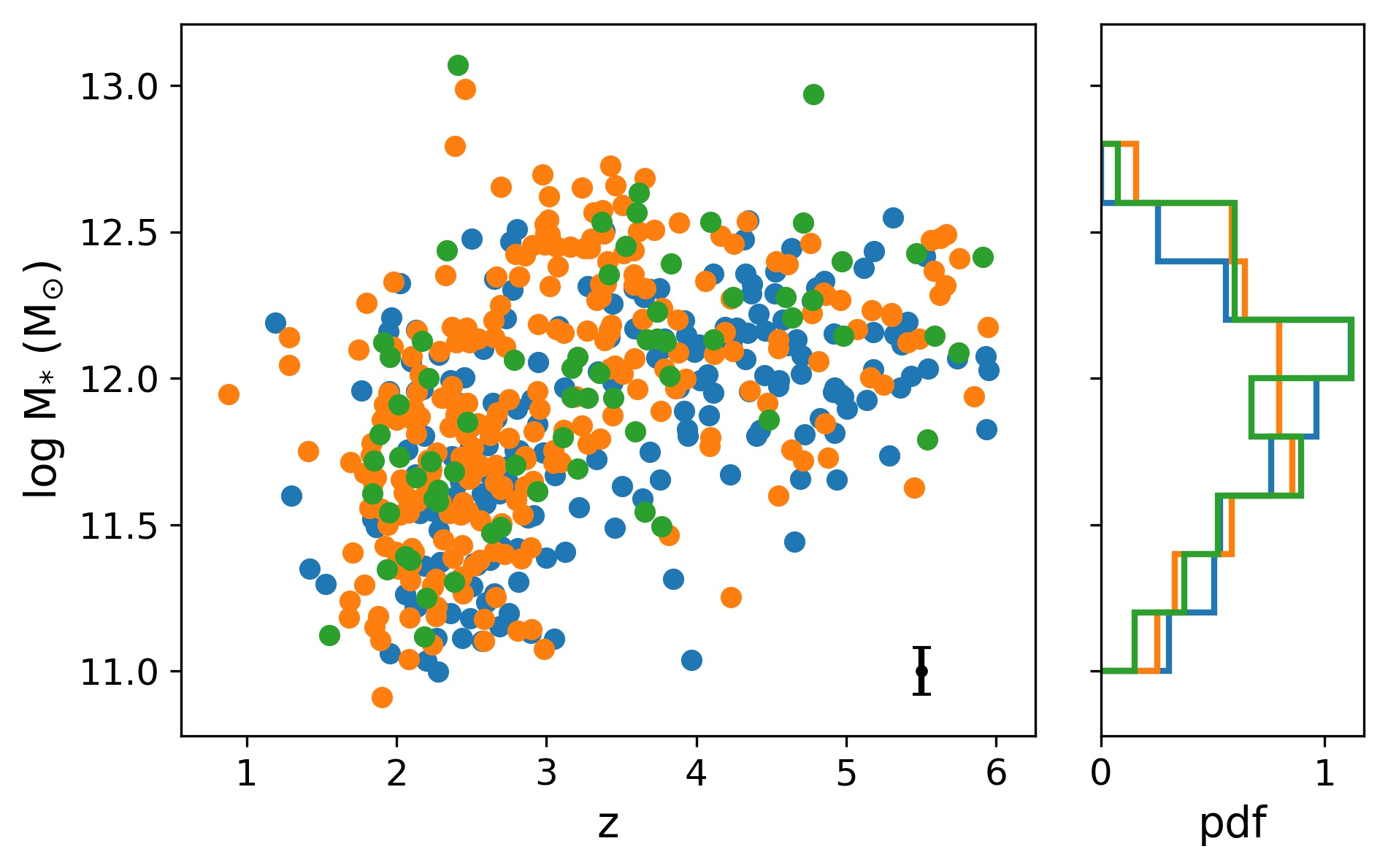}
    \end{subfigure}
    \begin{subfigure}[b]{.5\textwidth}
        \includegraphics[width=\textwidth]{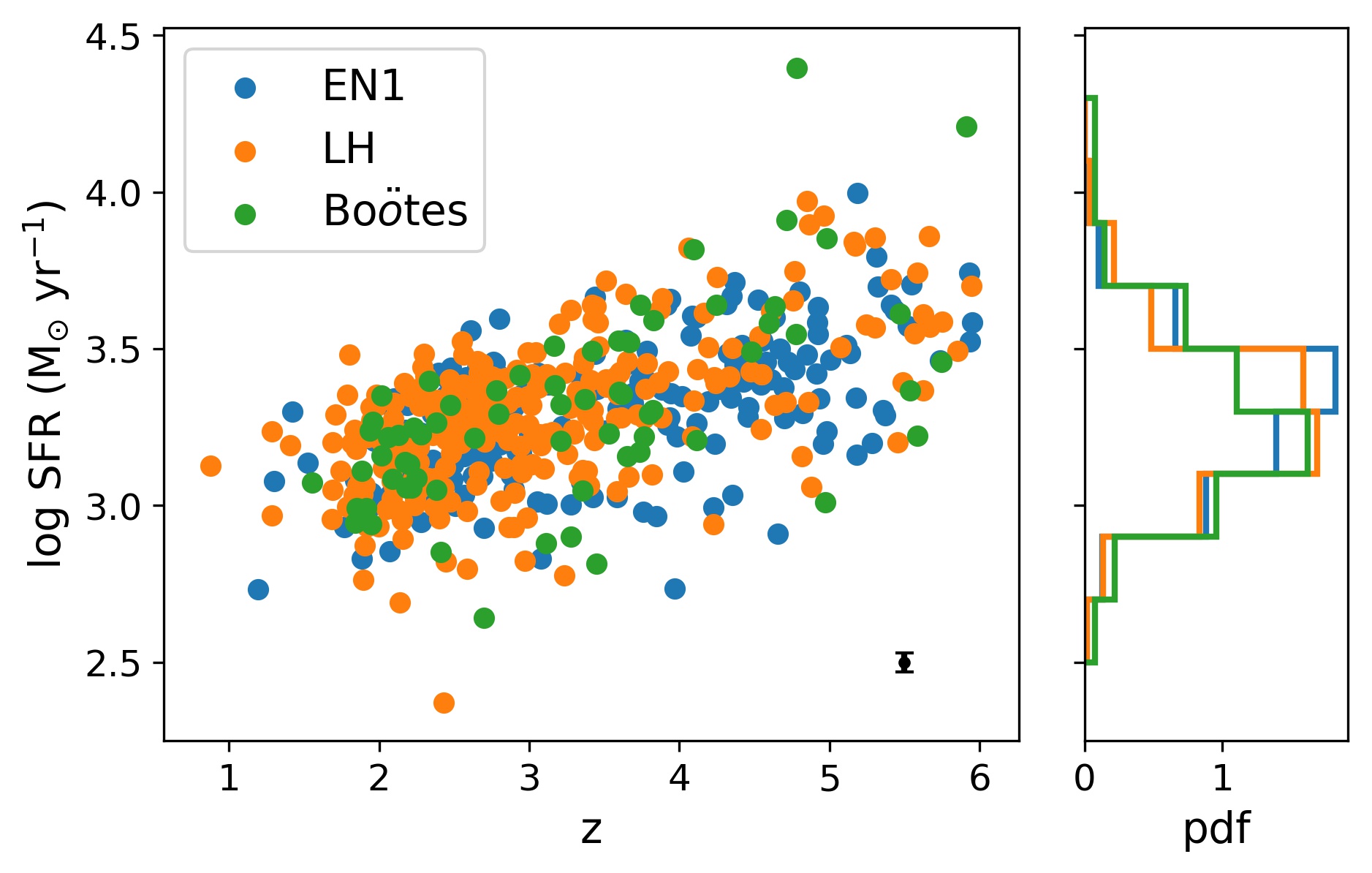}
    \end{subfigure}
    \begin{subfigure}[b]{.5\textwidth}
        \includegraphics[width=\textwidth]{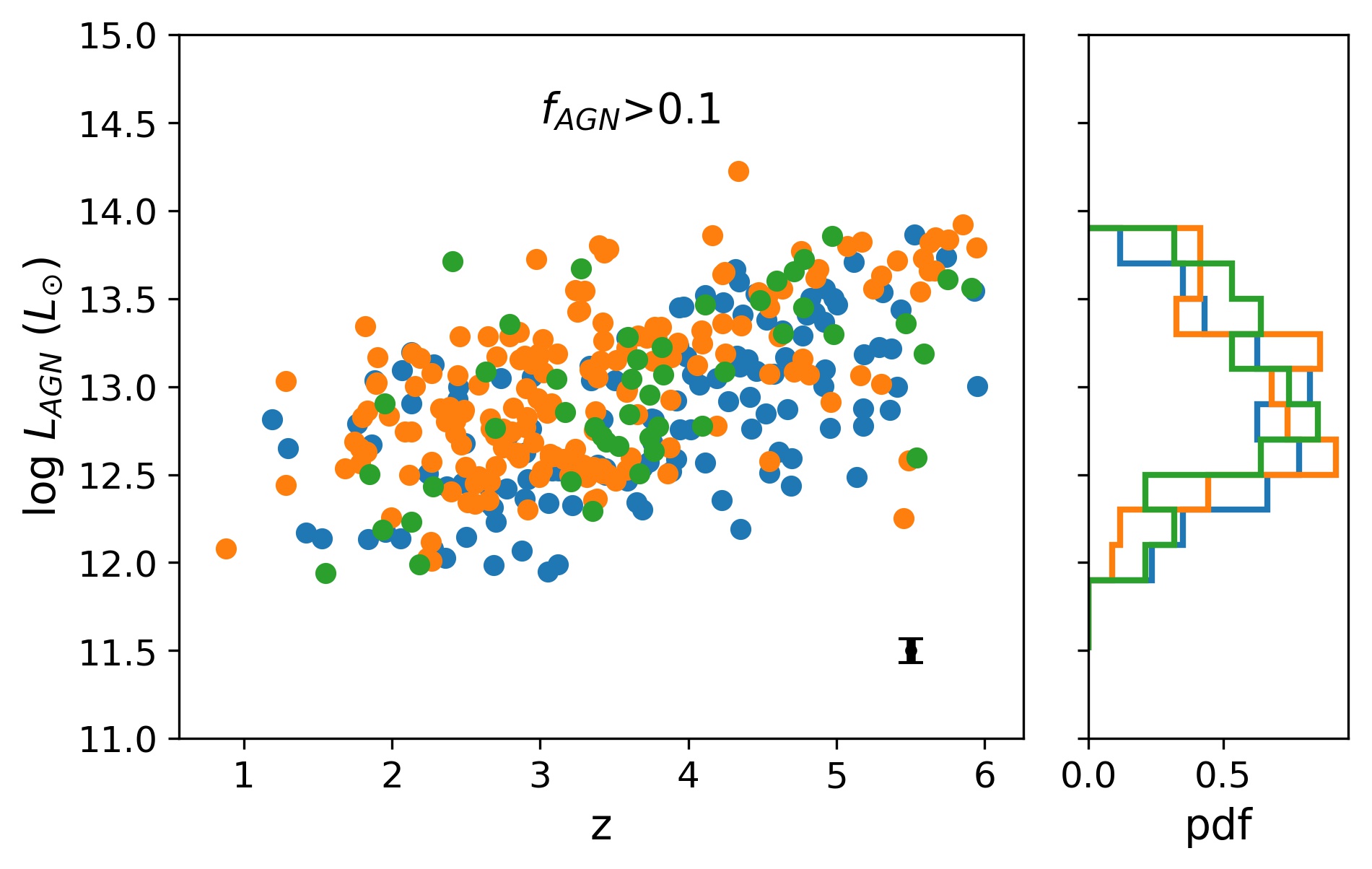}
    \end{subfigure}
\caption{Distributions of stellar mass (upper), SFR (middle) and AGN luminosity (bottom; only for HLIRGs with AGN fraction>0.1) as a function of redshift for all HLIRGs. The normalized histograms are also inserted. The median uncertainty is indicated in the right bottom corner of each panel.}
\label{allhlirg}
\end{figure}

\subsection{The HLIRG sample}\label{data1}
Our HLIRG sample is first identified in \citet{2021A&A...648A...8W}, based on a parent sample of IR luminous galaxies selected from the \textit{Herschel} blind source catalogs. These blind \textit{Herschel} source catalogs are then cross-matched with the Low Frequency Array (LOFAR) 150 MHz source catalog, specifically the LOFAR Two-metre Sky Survey (LoTSS) Deep Fields First data release \citep{2021A&A...648A...4D, 2021A&A...648A...3K, 2021A&A...648A...2S, 2021A&A...648A...1T} in three deep fields, Bo$\rm \ddot{o}$tes, ELAIS-N1 (EN1) and Lockman-Hole (LH). The cross-matching between \textit{Herschel} and LOFAR catalogs is secure thanks to the well-known tight correlation between the far-IR and radio \citep[FIRC; e.g.,][]{2017MNRAS.469.3468C, 2019A&A...631A.109W}. The angular resolution ($\sim 6\arcsec$), positional accuracy (0.2\arcsec) and the superb sensitivity (average 71 $\upmu$Jy beam$^{-1}$) of LOFAR imaging help to associate the multi-wavelength counterparts of these radio sources. Only 2-3\% of LOFAR sources do not have multi-wavelength counterparts in these three fields \citep[see][]{2021A&A...648A...3K}. The extraction of multi-wavelength photometry in the LOFAR deep fields and the process of cross-matching between the LOFAR radio sources and their multi-wavelength counterparts are fully described in \citet{2021A&A...648A...3K}. As our blind \textit{Herschel} sources have already been matched to the LOFAR radio sources, it is then straightforward to find their multi-wavelength counterparts.

We selected \textit{Herschel} blind sources with 250 $\mu$m fluxes above 45, 35 and 40 mJy in Bo$\rm \ddot{o}$tes, EN1 and LH respectively. At these flux density cuts $>90\%$ \textit{Herschel} sources are matched with LOFAR sources. The majority of \textit{Herschel} sources are matched with only one LOFAR source (the unique sample) and around 16\% are matched with at least two LOFAR sources (the multiple sample). We deblended the \textit{Herschel} 250, 350 and 500 $\mu$m flux densities for the multiple sample using XID+, a probabilistic de-blender tool \citep{2017MNRAS.464..885H} and taking advantage of the aforementioned FIRC to calculate flux density priors. For more details we refer the reader to \citet{2021A&A...648A...8W}.

In \citet{2021A&A...648A...8W} we used the spectral energy distribution (SED) fitting code Code Investigating GALaxy Emission \citep[CIGALE][]{2005MNRAS.360.1413B, 2009A&A...507.1793N} to obtain physical properties of these IR luminous galaxies such as the total IR luminosity. We then further analyzed the HLIRGs (69, 198, and 259 HLIRGs in Bo$\rm \ddot{o}$tes, EN1, and LH respectively) in \citet{2021A&A...654A.117G}, adopting two different SED fitting codes CIGALE and CYprus Models for Galaxies and their NUclear Spectral \citep[CYGNUS;][]{1995MNRAS.273..649E, 2000MNRAS.313..734E, 2003MNRAS.343..322E, 2013MNRAS.436.1873E} and various AGN models \citep{1995MNRAS.273..649E,2006MNRAS.366..767F, 2012MNRAS.420.2756S}, to explore in detail how their physical properties depend on model assumptions. The distributions of stellar mass estimates, SFR estimates and AGN luminosity estimates as a function of redshift is shown in Figure \ref{allhlirg}. We find that HLIRGs are extremely massive (with a median stellar mass of $10^{12}\,\mathrm{M_{\sun}}$) and have a co-moving volume density higher than what is expected from previous studies of the global stellar mass functions. They also undergo active star-forming activities with a median SFR of $10^{3.3}\, \mathrm{M_{\sun}}\,\mathrm{yr^{-1}}$. Moreover, many HLIRGs are associated with vigorous AGN activity. There are $30-50\%$ HLIRGs that have AGN fraction $>0.3$ depending on which SED fitting code and AGN model being used.

\subsection{The deblended \textit{Herschel} 250 $\mu$m source catalogs}\label{data2}
To explore whether HLIRGs live in dense environments such as protoclusters with intense on-going star-formation activity, we need deep catalogs of SFGs. The Herschel Extragalactic Legacy Project \citep[HELP;][]{2019MNRAS.490..634S} is a large program which combines and homogenises a wide range of multi-wavelength surveys in fields observed by \textit{Herschel}. HELP presents a catalog of around 170 million sources covering 1270 deg$^2$ selected in optical-near IR (NIR) wavelength range.

We use the deblended \textit{Herschel} 250 $\mu$m source catalogs (DMU26) from HELP to search for star-forming neighbours around HLIRGs. The blind \textit{Herschel} source catalogs we used to build the parent sample of IR luminous galaxies are selected by finding peaks in the matched filtered (MF) maps \citep{2011MNRAS.411..505C} and thus limited to bright sources. Therefore, they are not suitable as tracers of the general SFG population. In contrast, the deblended catalogs are generated by deblending the \textit{Herschel} maps based on positions of known sources in the 24 $\mu$m maps with higher resolution. They are therefore deeper than the blind \textit{Herschel} source catalogs and thus better for studying the environment of HLIRGs. We use the deblended 250 $\mu$m source catalogs to search for DSFGs around HLIRGs within different projected separations and compare with those around random positions. In order to remove less reliable detections of low significance, we apply a flux density cut of 10 mJy, resulting in 42\,406, 42\,694 and 41\,651 sources in Bo$\rm \ddot{o}$tes, EN1 and LH respectively.

\begin{table*}[htbp]
   \caption{Statistics of the HLIRGs, quasars, SMGs, the deblended 250 $\mu$m sources and deep IRAC-seletced photo-z catalogs in the three fields. Since all HLIRGs are all above $z=1$ (only one HLIRG is at $z<1$), we select sources at $z>1$ in all catalogs. No redshift cut to the HELP 250 $\mu$m deblended catalogs has been made as we only use them in the surface density analysis.}
    \centering
    \begin{tabular}{ccccccc}
    
    \hline

        field&limit area (deg$^2$)&HLIRG&quasar&SMG&deep photo-z catalog&deblended 250 $\mu$m catalog\\
        \hline
        Bo$\rm \ddot{o}$tes&5.06&31&235&--&150\,261&19\,025\\
        EN1&5.75&85&272&--&161\,845&16\,255\\
        LH&6.5&86&346&93&130\,907&14\,593\\
         \hline
         \multicolumn{7}{l}{Note: SMGs only locate in the LH field.}
    \end{tabular}
    \label{field}
\end{table*}

%The \textit{Herschel} Multi-tiered Extragalactic Survey (HerMES) is a legacy program using SPIRE instrument \citep{2010A&A...518L...3G} on board of \textit{Herschel}, in order to probe star-forming galaxies in a wavelength range at which their SED peak, representing the re-emitted energy by dust from star-formation activity \citet{2012MNRAS.424.1614O}. This survey provides a catalog with over 340,000 galaxies detected with $>5 \sigma$ significance at 250$\mu$m in several fields, covering a total area of $\sim$ 380 deg$^2$. Together with other multi-wavelength photometry, HerMES data can help better constrain the total amount of IR luminosities and benefit the study of galaxy evolution. The 5-$\sigma$ total noise combining instrument noise and confusion noise is 28 mJy for EN1 and LH fields and 35 mJy for Bo$\rm \ddot{o}$tes field \citep[see ][]{2021A&A...648A...8W}. 

\begin{figure}[htbp]
    \centering
    \begin{subfigure}[b]{.5\textwidth}
        \includegraphics[width=\textwidth]{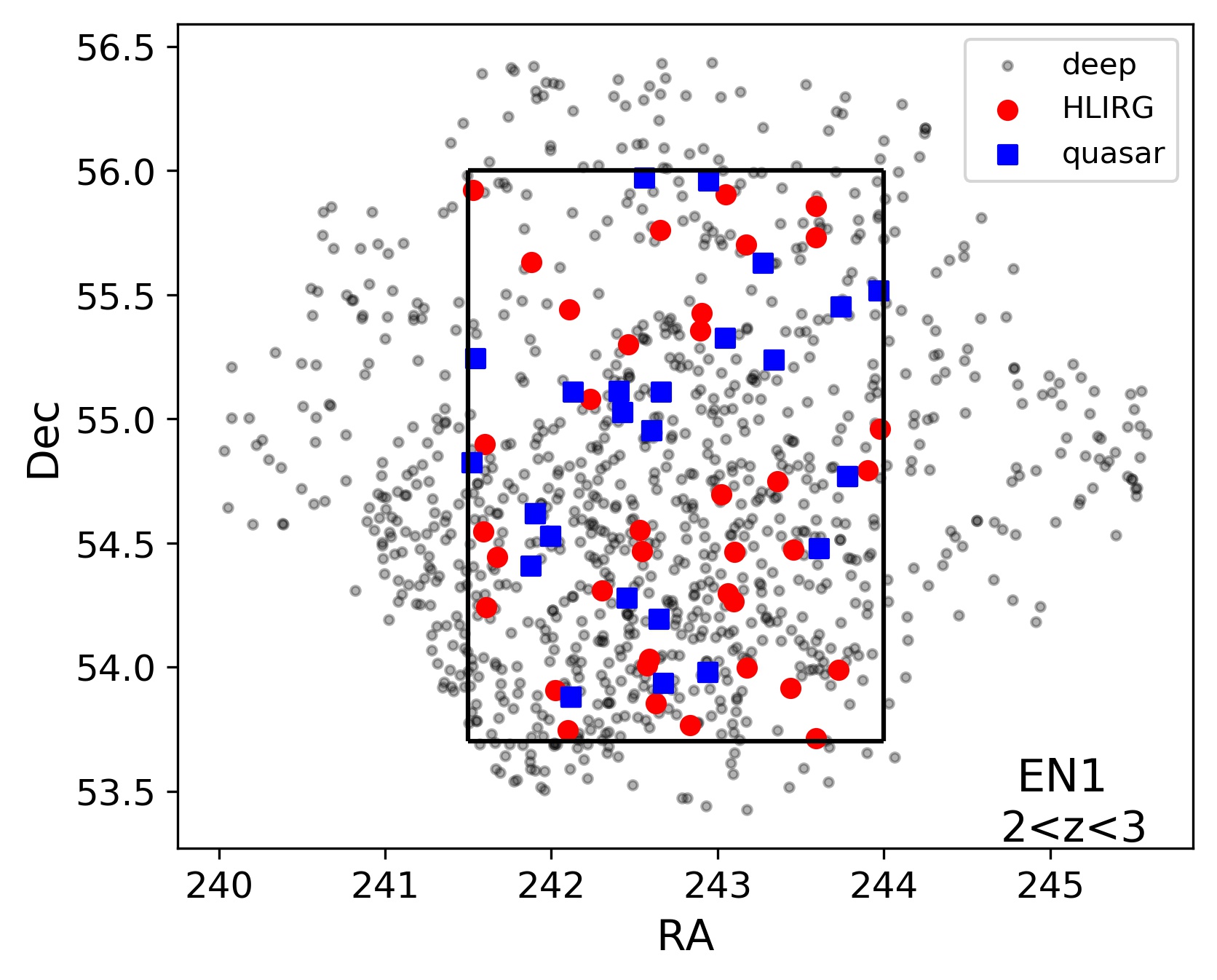}
    \end{subfigure}
    \begin{subfigure}[b]{.5\textwidth}
        \includegraphics[width=\textwidth]{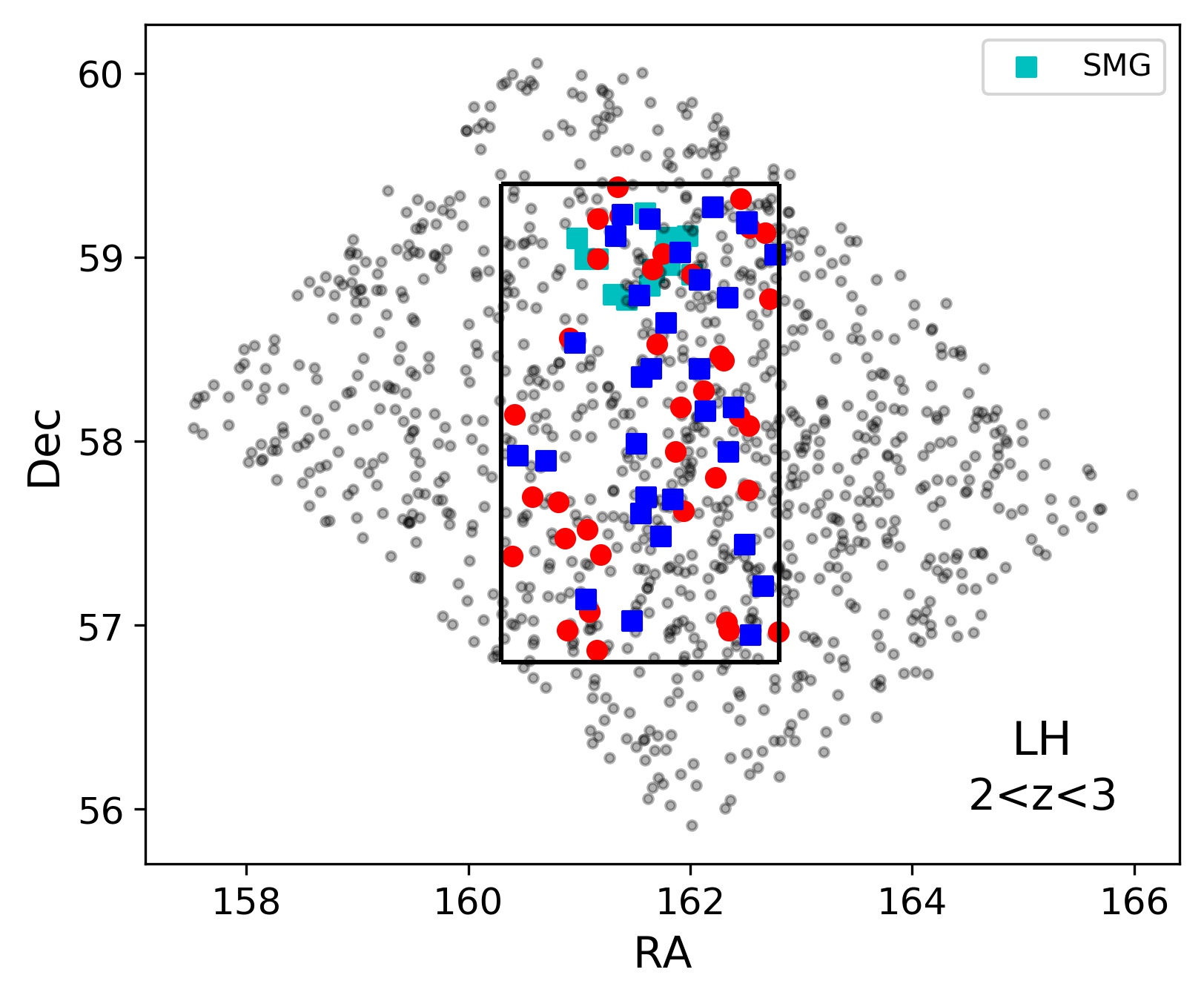}
    \end{subfigure}
    \begin{subfigure}[b]{.5\textwidth}
        \includegraphics[width=\textwidth]{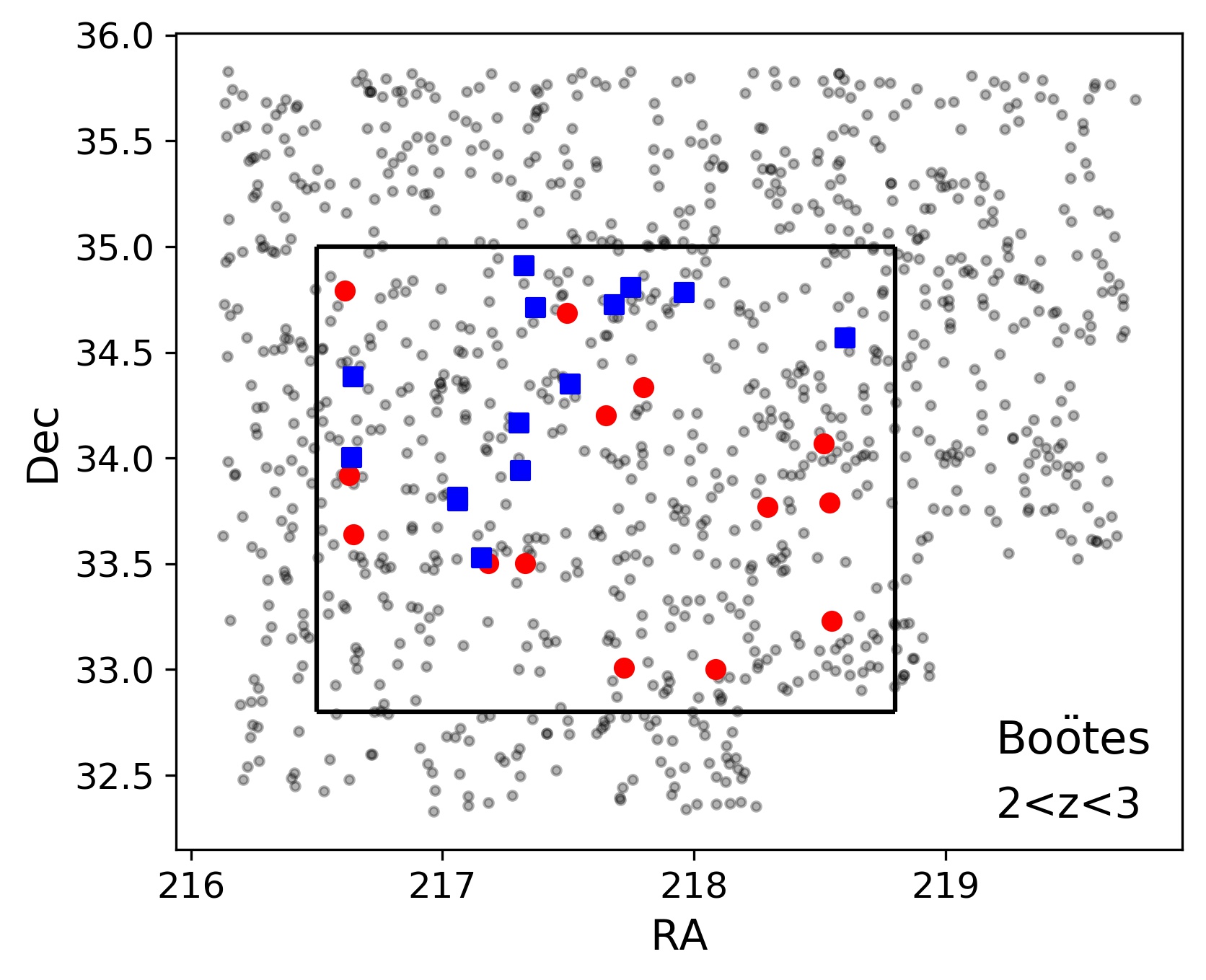}
    \end{subfigure}
\caption{Sky distributions of the various samples used in this study. Black dots are 1000 randomly selected sources from the deep IRAC-selected photo-z catalogs at $2<z<3$ for clarity. We require all samples to be located inside the central black boxes in order to avoid searching for neighbours outside of the sky area coverage. HLIRGs at $2<z<3$, redshift-matched quasars and SMGs (only in the LH field) are plotted as red circles, blue squares and cyan squares respectively.}
\label{sky}
\end{figure}

\subsection{Deep photo-z source catalog}\label{data3}
In order to find neighbours of HLIRGs in a certain spatial volume and characterize their environment in 3D, we need deep source catalogs that contain good-quality redshift information. SED fitting to the multi-wavelength photometry can help determine galaxy properties such as stellar mass for these sources.

We use deep source catalogs in the three deep fields compiled in \citet{2021A&A...648A...3K} to search for neighbours around our HLIRG sample. The authors generated deep source catalogs for the LH and EN1 fields
and combined existing catalogs for the Bo$\rm \ddot{o}$tes field. Source detection is carried out using SExtractor \citep{1996A&AS..117..393B} and multi-wavelength photometry is measured first through forced, matched aperture then corrected to total fluxes using aperture corrections across all 20 and 16 bands for EN1 and LH fields respectively. %They then ran radio-optical cross identification successfully for over 97 $\%$ LOFAR radio sources based on the likelyhood ratio method and visual classification. 

Photometric redshifts of the deep source catalogs are determined in \citet{2021A&A...648A...4D}, adopting a hybrid method that combines template fitting
and machine learning. They achieve a good agreement when comparing with spectroscopic redshifts, with a 1.6-2\% scatter from spectroscopic redshifts for galaxies and a 1.5-1.8\% of outlier fractions (defined as |$\Delta$z|/(1+spec-z)>0.15).

We first apply quality flags in the three source catalogs to eliminate unreliable and duplicate sources. As elaborated in \citet{2021A&A...648A...3K}, the FLAG\_CLEAN flag indicates regions masked by bright stars and FLAG\_OVERLAP flag indicates multiwavelength coverage in multiple surveys. For Bo$\rm \ddot{o}$tes, an additional flag FLAG\_DEEP represents duplicate sources in the $I$-band catalog of \citet{2008ApJ...682..937B}. We then require the signal-to-noise (S/N) ratio in the \textit{Spitzer} IRAC 3.6 $\mu$m band to be above 3$\sigma$ to further reduce spurious sources, reaching a limiting magnitude of 23.63. Since our HLIRGs are all at high redshifts beyond $z=1$, 3.6 $\mu$m is close to rest-frame NIR band whose emission comes mostly from evolved stellar populations. Hence the 3.6 $\mu$m emission provides a good indicator of stellar mass \citep[e.g.,][]{1998MNRAS.297L..23K, 2001MNRAS.326..255C} and it suffers less dust obscuration. After applying the quality flags and S/N cuts, the total number of galaxies above $z=1$ reduces to 331\,032, 275\,672, and 369\,729 sources in the Bo$\rm \ddot{o}$tes, EN1 and LH fields respectively.

\begin{table}[htbp]
    \caption{The number of HLIRGs, redshift-matched quasars and SMGs summed over the three fields. We select the same number of quasars in each redshift bin as the number of HLIRGs. Numbers in parenthesis are the number of HLIRGs in the LH field. We select SMGs with half the number of HLIRGs due to small number statistics of the SMG sample.}
    \centering
    \begin{tabular}{cccc}
    \hline
    redshift bin&HLIRG&quasar&SMG (LH field)\\
    \hline
    2.0-2.2&13(7)&13&3\\
    2.2-2.4&18(8)&18&4\\
    2.4-2.6&19(8)&19&4\\
    2.6-2.8&22(6)&22&3\\
    2.8-3.0&15(7)&15&3\\
    3.0-3.2&8(6)&--&3\\
    3.2-3.4&11(5)&--&2\\
    3.4-3.6&11(6)&--&3\\
    3.6-3.8&13(4)&--&2\\
    3.8-4.0&8(2)&--&1\\

    \hline
    \end{tabular}
    
    \label{number}
\end{table}

\subsection{Quasar sample}\label{data4}

Quasars are among the brightest galaxies that are powered by accretion onto the central super massive black holes \citep{1964ApJ...140..796S,1969Natur.223..690L}. %They are popular targets in studying the large-scale structures (LSS) and its evolution because they are easy to be selected due to their bright luminosity. 
Since the advent of large sky surveys such as the Sloan Digital Sky Survey \citep[SDSS;][]{2000AJ....120.1579Y} and the 2dF Quasi-Stellar Object (QSO) redshift survey \citep[2QZ;][]{2004MNRAS.349.1397C}, it has long been well established that quasar clustering becomes stronger as redshift increases \citep{2004MNRAS.355.1010P,2005MNRAS.356..415C,2006ApJ...638..622M,2007AJ....133.2222S}. They are expected to be hosted by dark matter halos with characteristic halo masses around a few times of $10^{12}\, \mathrm{M_{\sun}}$ with little to no dependence on quasar luminosity \citep{2005MNRAS.356..415C,2007ApJ...658...85M,2007AJ....133.2222S,2009ApJ...697.1634R,2012MNRAS.424..933W,2015MNRAS.453.2779E,2015MNRAS.449.4476W}. According to the well studied relationship between stellar mass and halo mass \citep[SMHM relationship;][]{2010MNRAS.404.1111G,2010ApJ...710..903M,2013ApJ...770...57B,2013MNRAS.431..648W,2015ApJ...799..130R,2019MNRAS.488.3143B}, halos around a few times $10^{12}\, \mathrm{M_{\sun}}$ host galaxies with stellar masses around a few times of $10^{10}\,\mathrm{M_{\sun}}$ over a wide range of redshifts.

We use the latest SDSS sixteenth quasar catalog \citep{2020ApJS..250....8L} which consists of the largest number of spectroscopically confirmed quasars to date, with 750\,414 quasars over 7\,500 deg$^2$, reaching a 99.8\% completeness with a contamination fraction of 0.3-1.3\%. We make use of this quasar catalog due to several reasons. First, estimates of the host dark matter halo mass for quasar already exists thanks to extensive clustering analyses. Comparing the neighbours around HLIRGs with those around quasars can give us an indication of the environment of HLIRGs relative to the quasars. Moreover, these quasars partly overlap with our HLIRG sample in the redshift distributions as later described in Section \ref{data6}. %Finally, some HLIRGs are also partly powered by rapid black hole accretion rather than dusty star-formation activity alone. The enhanced clustering of quasars at similar redshift range can shed light on the overdense nature of environment in which these HLIRGs may live in.%, even though they are not extreme dusty star-forming that are expected to live in the most extremely dense environments. Although there are studies finding that optical selected quasars live in lower density environment than more powerful radio quasars, and only the most luminous ones are more clustered and live in the more massive halos \citep[e.g.,][]{2009ApJ...697.1656S, 2018PASJ...70S..32U,2019ApJ...874...85G}, we use quasar sample to validate our method, rather than directly comparing the environment of HLIRGs to that of quasars. 
In summary, the brightness of quasars, good understanding of their host halo environment, and overlapping redshift range make them a good sample to serve as a cross-check. %We use the deep IRAC-selected photo-z source catalog in Section \ref{data3} to search for neighbours around quasars and compare them with neighbours of randomly selected galaxies. Then, we compare the neighbours around HLIRGs with those around quasars in order to check whether HLIRGs live in similar or more massive halos. %If we can find a similar excess of neighbours around HLIRGs, then it strongly indicate that HLIRGs also reside in overdense regions.

\begin{figure}[htbp]

    \centering
    \begin{subfigure}[b]{.5\textwidth}
        \includegraphics[width=\textwidth]{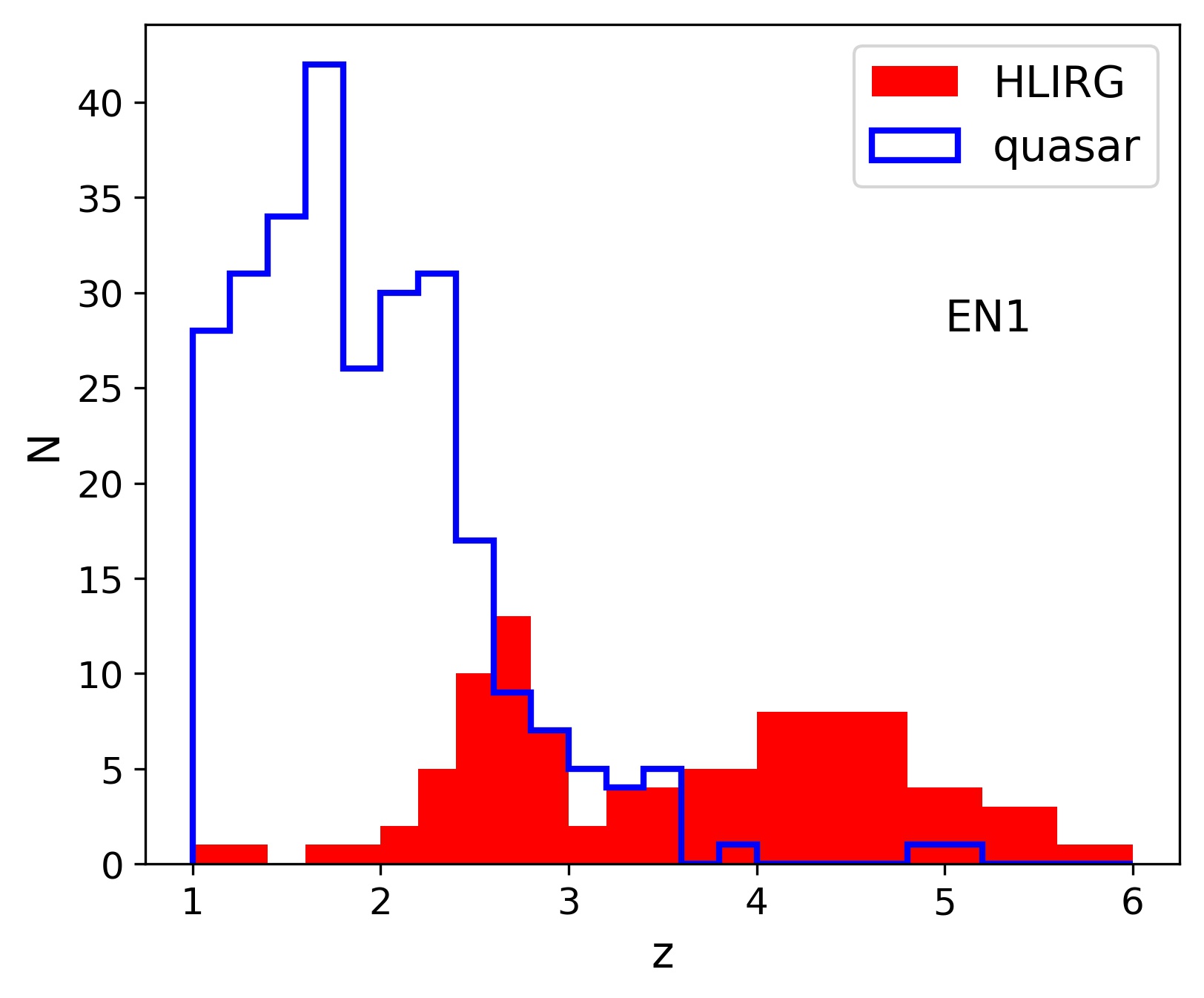}
    \end{subfigure}
    \begin{subfigure}[b]{.5\textwidth}
        \includegraphics[width=\textwidth]{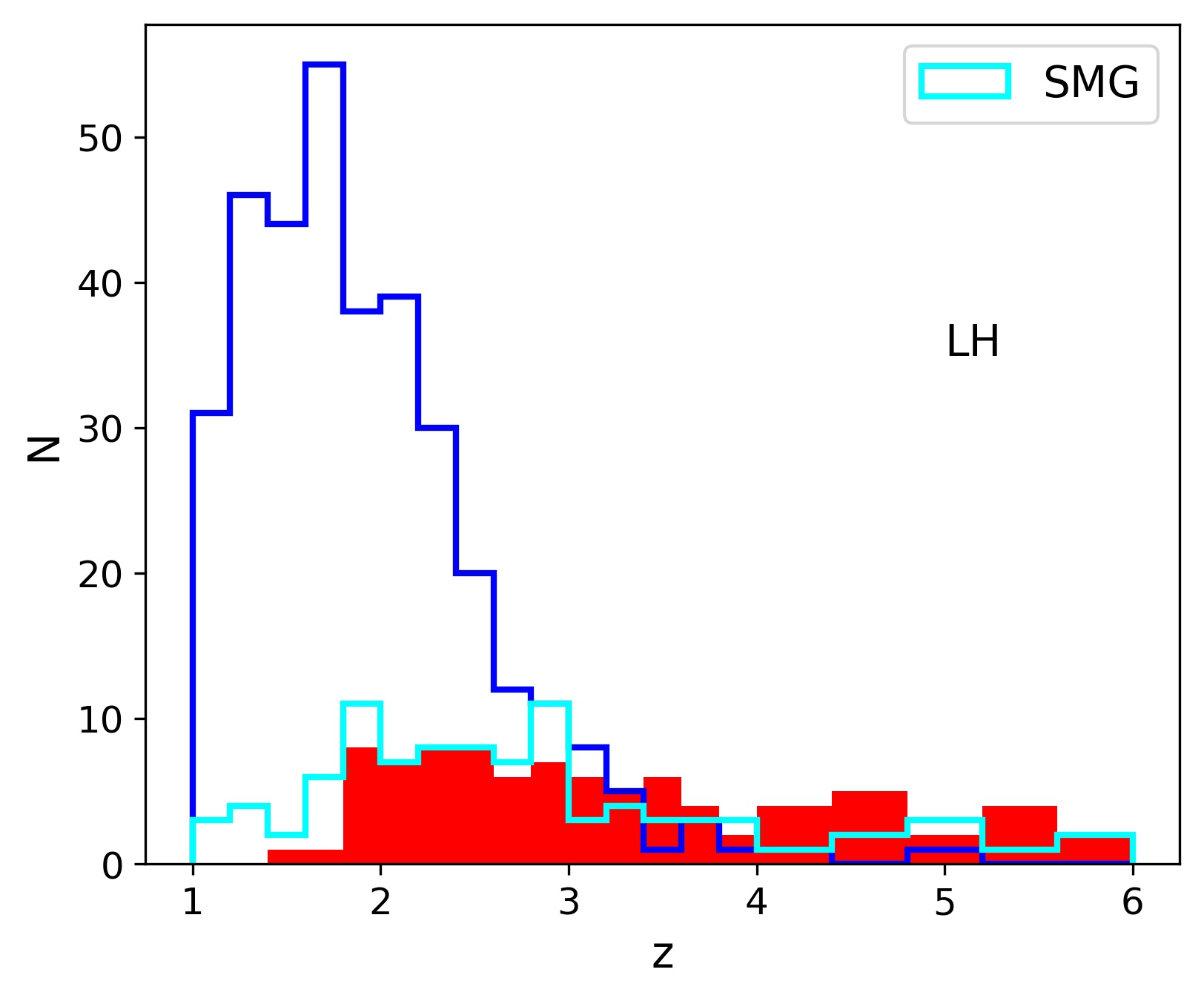}
    \end{subfigure}
    \begin{subfigure}[b]{.5\textwidth}
        \includegraphics[width=\textwidth]{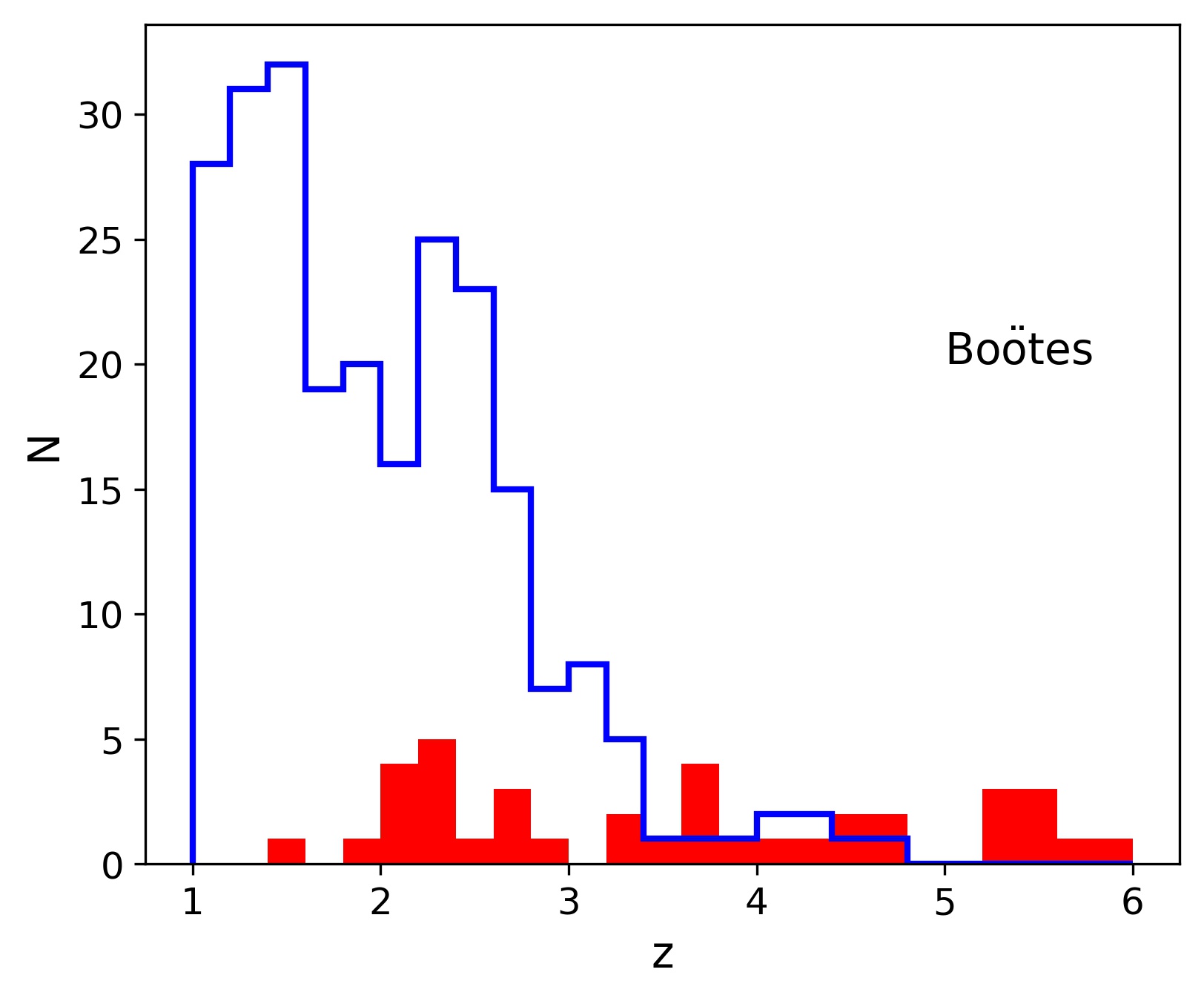}
    \end{subfigure}
\caption{Redshift distribution of our HLIRG (red), quasar (blue) and SMG (cyan) samples in the three deep fields after location restriction. Most HLIRGs are locacted at $z>2$. SMGs have a broadly similar redshift distribution as HLIRGs, while quasars only partly overlap with the HLIRGs at $2<z<3$. Due to small number statistics, we only study HLIRGs at $2<z<4$ and use quasars at $2<z<3$ for a cross-check.}
\label{z}
\end{figure}

\subsection{Sub-millimeter Galaxy sample}\label{data5}
Sub-millimeter Galaxies (SMG) are rapidly star-forming galaxies at high redshifts. Their rest-frame FIR emission from obscured star-forming activity shifts towards longer wavelengths and make them detectable in the (sub-)millimeter band. HLIRGs and SMGs are broadly similar in terms of their power source (i.e., dusty star formation, and AGN activity in some cases) although SMGs typically have IR luminosity more comparable to ultra luminous  IR galaxies (ULIRGs), reaching a few times of 10$^{12} L_{\odot}$ \citep[e.g.,][]{2005ApJ...622..772C,2012A&A...539A.155M, 2014MNRAS.438.1267S}. In addition, SMGs in general are biased towards colder dust temperatures compared to \textit{Herschel}-selected galaxies with similar IR luminosities \citep[e.g.,][]{2014MNRAS.438.1267S,2015ApJ...806..110D}.

Clustering measurements of SMGs report a strong clustering and typical host halos with $M_{halo} \sim 10^{12}-10^{13}\, \mathrm{M_{\sun}}$ \citep{2004ApJ...611..725B,2012MNRAS.421..284H,2016ApJ...820...82C,2017MNRAS.464.1380W,2021MNRAS.504..172S}. Such halos will host galaxies with stellar masses a few times of $10^{10}\, \mathrm{M_{\sun}}$ based on the aforementioned SMHM relationship at $2<z<4$ where the distribution of SMGs peaks. Bright SMGs can be used as a signpost to trace high-redshift protocluster candidates and several studies have successfully found SMGs that live in protoclusters \citep[e.g.,][]{2003ApJ...583..551S,2009ApJ...694.1517D,2014ApJ...796...84R, 2019MNRAS.490.3840C, 2021MNRAS.508.3754W}. In our work, we also select a SMG sample to compare their neighbours with random galaxy neighbours. Similar to the quasar sample described in Section \ref{data4}, by comparing HLIRG neighbours with SMG neighbours, we can find out whether HLIRGs reside in similar or more massive halos. Moreover, SMGs typically have a broadly similar redshift range (as later described in Section \ref{data6}) and number distribution as HLIRGs as they both trace dusty star-forming activity in the early universe.

The SMG sample we use is taken from \citet{2017MNRAS.465.1789G}, which contains $\sim$ 3000 SMGs above 3.5 $\sigma$ at 850 $\mu$m over $\sim 5$ deg$^2$, as part of the James Clerk Maxwell Telescope (JCMT) SCUBA-2 Cosmology Legacy Survey (S2CLS). This survey reached an average 1 $\sigma$ depth of 1.2 mJy/beam, close to the SCUBA-2 confusion limit of 0.8 mJy/beam. The catalog covers six fields in total, but only the Lockman-Hole North (LH-N) field falls into the regions we study in this paper. We cross-match the 850 $\mu$m source catalog with LOFAR radio-optical cross-identified catalog in the LH field, in order to associate SMG sources with their multi-wavelength counterparts. In contrast to the quasar sample, the SMG sample suffers from small sample size and only has photometric redshifts. Therefore, when comparing both SMG neighbours and quasar neighbours with random galaxy neighbours, we can also have an idea of the influence due to small number statistics and photometric redshift uncertainty, given the fact that both samples have been well studied through clustering analyses, and the quasar sample is larger and has spectroscopic redshifts. The effect from photometric redshift uncertainty will be further discussed in Section \ref{re3}.

\begin{figure*}[htbp]
    \centering
    \includegraphics[width=\linewidth]{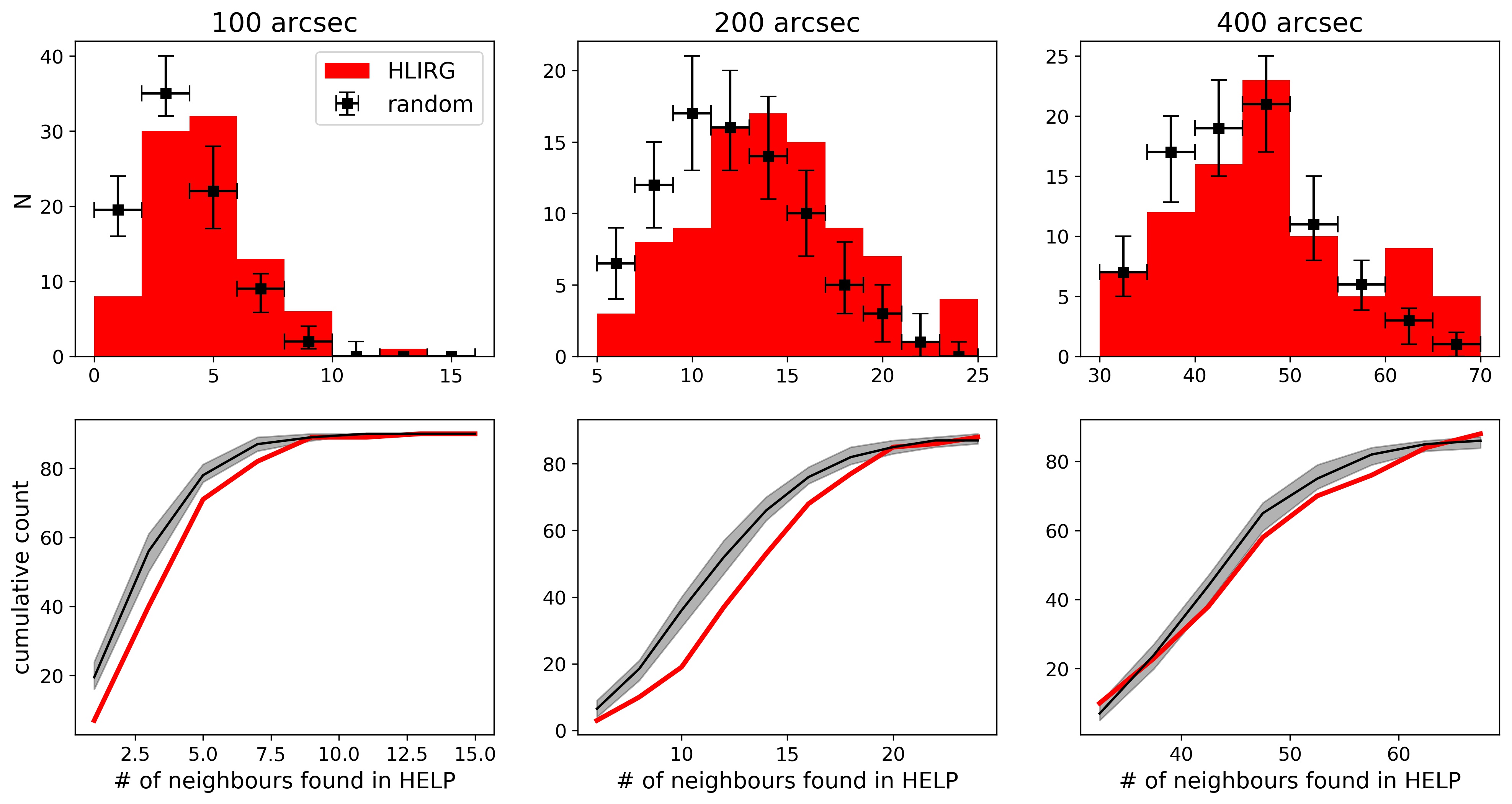}
    \caption{Upper: Number distributions of deblended 250 $\mu$m galaxies around HLIRGs within 100, 200 and 400$\arcsec$ respectively. The black symbols and errorbars are the median value and 16th/84th percentiles calculated from 100 realizations of random galaxies. Bottom: the cumulative distribution function (CDF) of the number of HLIRG neighbours and random neighbours. HLIRGs on average have more 250 $\mu$m neighbours than random positions. This difference becomes weaker as radius increases, reducing from 3.7$\sigma$ within 100$\arcsec$ to 2.7 $\sigma$ within 200$\arcsec$ and 1.6$\sigma$ within 400$\arcsec$ respectively.}
    \label{help}
\end{figure*}

\subsection{Summary}\label{data6}
Our HLIRG sample is selected from the blind \textit{Herschel} catalogs. To find star-forming galaxies around HLIRGs we take advantage of the HELP 250 $\mu$m deblended source catalogs. We also use a deep IRAC-selected source catalog with good-quality photometric redshifts in order to study the 3D environment of the HLIRGs within different spatial volumes. We also include a quasar sample (at $2<z<3$) and a SMG sample (only in the LH field) as cross-checks.

We first restrict all samples to a limited sky area within the coverage of the deblended \textit{Herschel} 250 $\mu$m source catalogs and the deep photo-z catalogs, in order to avoid searching for neighbours outside of the sky coverage. In Figure \ref{sky},the sky areas covered by the deep photo-z catalogues are illustrated using 1000 random sources over the redshift range 2<z<3 for clarity. We confine all samples to be located in the central part of the regions shown by the black boxes. The sky coverage of the deblended \textit{Herschel} 250 $\mu$m source catalogs is almost the same as that covered by the deep photo-z catalogs, therefore we adopt the same limited zones. The sky area and the total number of each sample in the three fields are listed in Table \ref{field}.

%When exploring the star-forming neighbours of HLIRGs within a certain projected separation, we study all HLIRGs in the limited zones and compare the number of their neighbours with the number of random galaxy neighbours. We also compare the number of quasar neighbours with the number of random galaxy neighbours. \textit{Herschel} observed LH-N filed with a different noise level than LH field \citep[see][]{2012MNRAS.424.1614O}, resulting a deeper source catalog in LH-N field. We compare the number of SMG neighbours with the number of neighbours around random positions located in LH-N. All HLIRGs, SMGs (only in the LH-N field), and 50 quasars randomly chosen for clarity are shown in red circles, cyan squares and blue squares respectively in panel (a) of Figure \ref{sky}. The sky coverage and total numbers of each sample in three fields are listed in Table \ref{field}

When exploring the star-forming neighbours of HLIRGs, we study all HLIRGs in the limited zones and compare the number of their neighbours with the number of random galaxy neighbours.
To characterize the 3D environment of HLIRG neighbours, we compare the distribution of HLIRG neighbours from the deep photo-z catalogs with the distribution of redshift-matched random galaxy neighbours. We also study the distribution of redshift-matched quasar neighbours and SMG neighbours as cross-checks. After restriction in position according to the sky coverage of the deep photo-z catalogs, the redshift distributions of HLIRGs, quasars and SMGs are shown in Figure \ref{z}. Our HLIRG sample lies mostly in $2<z<5$ and quasar samples are abundant below $z<3$. In spite of a small number, SMGs occupy roughly the same redshift range as HLIRGs. Based on the size of each sample, we restrict our analysis to HLIRGs at $2<z<4$, quasars at $2<z<3$, and SMGs at $2<z<4$. To obtain a redshift-matched comparison, we split our HLIRG sample into 10 redshift bins with 0.2 interval over $2<z<4$. For each redshift bin, we randomly select the same number of quasars as the number of HLIRGs. The number of the selected SMGs is half of the number of HLIRGs, due to their small sample size. The numbers of each sample in each redshift bin are listed in Table \ref{number}.
%Panel (b) of Figure \ref{sky} shows the positions of HLIRG sample, redshift-matched quasar sample and SMG sample in $2<z<4$ respectively. %Figure \ref{zbin} demonstrates the redshift distributions of our HLIRG sample, selected quasar and SMG samples in the LH field. 
%The sky area and total number of sources in each field is listed in Table \ref{field}.

\begin{figure*}[htbp]
\resizebox{\hsize}{!}{\includegraphics[width=\linewidth]{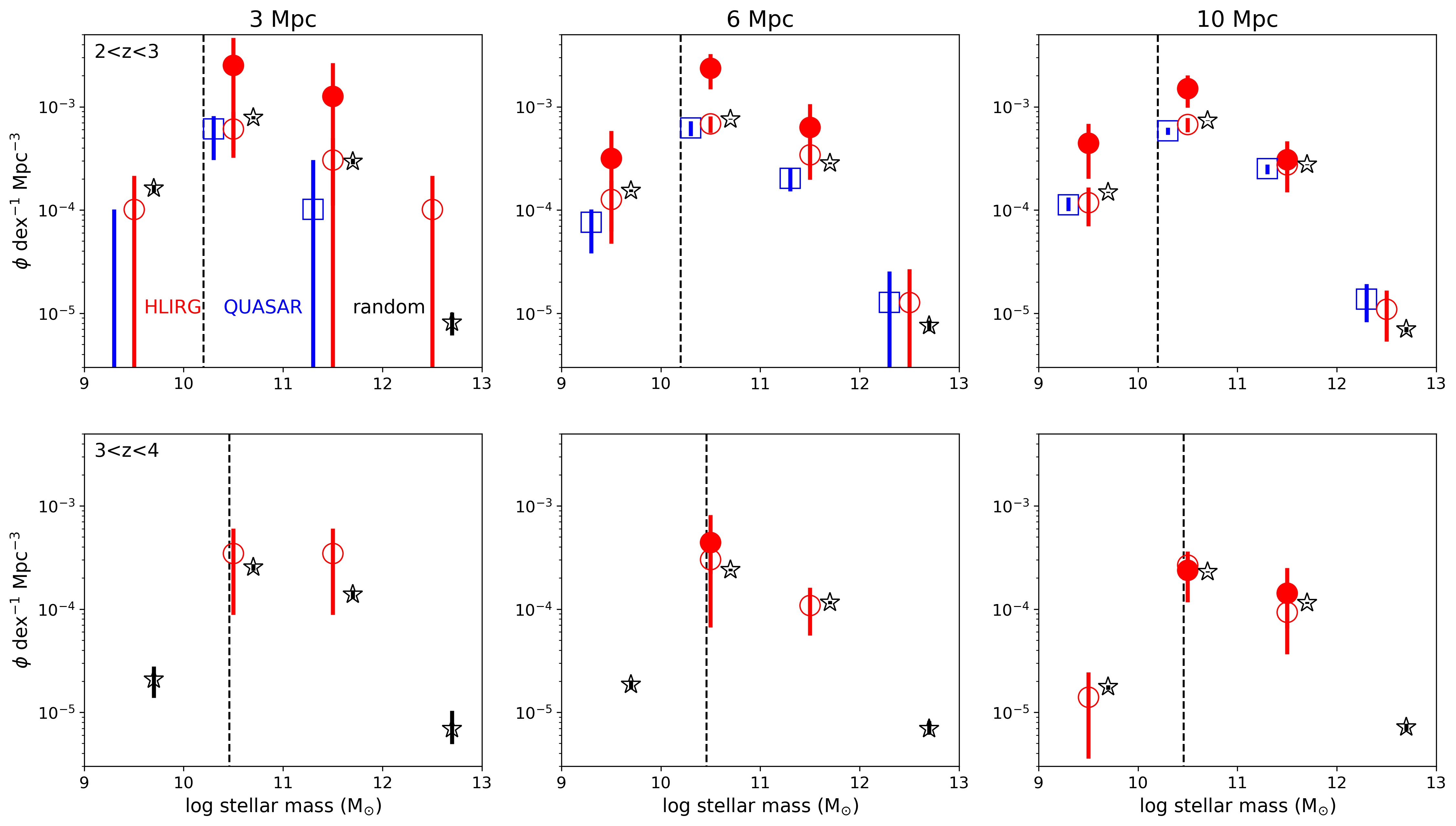}}
    \caption{The distribution of HLIRG neighbours (red empty circles), quasar neighbours (blue squares) and random galaxies
    neighbours (black stars) as a function of stellar mass. The red solid circles represent the subset of HLIRGs which are the most promising protocluster candidates (see Section \ref{conclu}). Three columns represent neighbours found within 3 Mpc, 6 Mpc and 10 Mpc respectively. Two rows display results in two redshift bins. The dashed lines are stellar mass completeness limits in each redshift bins. We combine Poisson error, photometric redshift uncertainty and standard deviation in the three fields for the HLIRG neighbours. We only consider sampling uncertainty for quasar neighbours and random neighbours. We find no excess in quasar neighbours at $2<z<3$ and a weak excess in HLIRG neighbours at $3<z<4$ compared with random neighbours. This excess disappear as searching radius increases. For the most promising protocluster candidates, we observe an enhanced excess signal.}
\label{smf}
\end{figure*} 

\begin{figure*}[htbp]
\resizebox{\hsize}{!}{\includegraphics[width=\linewidth]{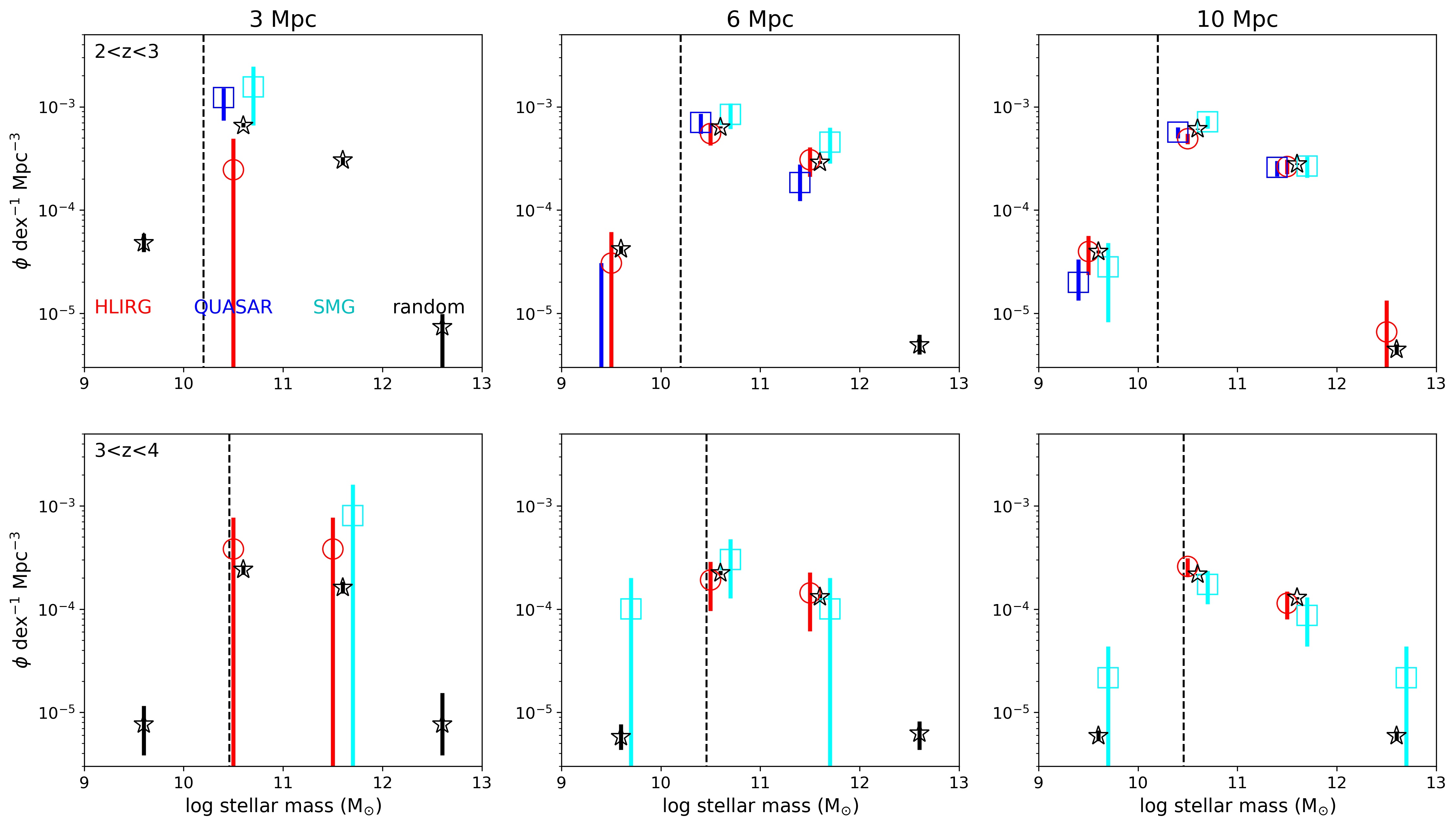}}
\caption{Similar to Figure \ref{smf}, but only in the LH filed. Neighbours of SMGs are plotted as cyan squares. We only include Poisson error for the SMG neighbours. We also find weak excess in the SMG neighbours and this excess disappears as searching radius increases.}
\label{smf_smg}
\end{figure*}

\section{Results}\label{re}
\subsection{The 2D surface distribution of 250 $\mu$m deblended star-formation galaxies around HLIRGs}\label{re1}

 High-redshift protoclusters are expected to be overdense regions with vigorous on-going star-formation activity. Therefore, it is reasonable to target overdensities of highly star-forming galaxies in order to identify protocluster candidates. This can be achieved by using high-resolution millimeter/submillimeter surveys to search for overdensity of DSFGs around protocluster candidates, which can be pre-selected as sources with exceedingly large flux density in low-resolution surveys. For example, \citet{2016A&A...596A.100P} presented a total of 2151 high-redshift sources whose flux density at 545 GHz are above 500 mJy and colors indicate $z>2$. Optical to submillimeter observations reveal that only a small fraction are strong gravitational lensed sources and the vast majority of them ($\sim 97\%$) are overdense regions of DSFGs. In order to investigate further the nature of these potential protoclusters, follow-up observations have been made. For example, SCUBA-2 observations at 850 $\mu$m \citep{2017MNRAS.468.4006M,2019MNRAS.490.3840C, 2020MNRAS.494.5985C} have found a significant overdensity of 850 $\mu$m sources compared with blank-field distributions and a rapid on-going star-formation activity. Another example is observing unresolved bright sources detected by SPT\footnote{Due to the smaller beam size of SPT (1\arcmin compared to 3\arcmin of Planck), SPT-detected bright sources are mostly lensed galaxies.}, a single-dish submillimeter telescope with a beam size of one arcminute \citep{2011PASP..123..568C}, through interferometric facilities such as ALMA \citep[e.g,][]{2020hst..prop16237C,2021MNRAS.508.3754W}. %The famous source SPT2349-56 is the brightest protocluster found in SPT survey \citep{2018Natur.556..469M,2020MNRAS.495.3124H}.

%For instance, using ALMA imaging to confirm the excess of DSFGs in protocluster candidates is a widely adopted method \citep{2014MNRAS.443..146J,2016MNRAS.461.2944A,2018Natur.556..469M,2018PASJ...70...65U,2020MNRAS.496.4358I}.
We search for deblended 250 $\mu$m sources surrounding our HLIRGs within different projected separations, and compare with that surrounding random galaxies. Those which have more star-forming neighbours than random positions are likely to be potential protoclusters and can be used as targets for future observations to distinguish real protocluster members from chance projections. We first study the total number of HLIRGs in the central confined region and count the number of their neighbours within 100$\arcsec$. We visually inspect them and remove a few HLIRGs due to their closeness to the boundary respectively, resulting in 27, 77 and 74 HLIRGs in Bo$\rm \ddot{o}$tes, EN1 and LH respectively. We randomly select the same number of random galaxies and count their neighbours within 100$\arcsec$. These randoms galaxies are required to be located at least 200$\arcsec$ away from any HLIRG, to make sure that there is no overlap between HLIRG environments and random environments. We run random selections 100 times and calculate the average distribution of the number of their neighbours. We find that HLIRGs have a median value of $13.0\pm{5.0}$ neighbours while random galaxies have a median value of $11.4\pm{0.5}$ neighbours (calculated as the mean value and the standard deviation of median number of neighbours from 100 realizations). HLIRGs live in an overdense environment at a 3.2$\sigma$ significance level. As radius increases, the difference in environment between HLIRGs and random positions becomes weaker (with 2.9$\sigma$ significance level within 200$\arcsec$) or even disappear (within 400$\arcsec$).

To reduce contaminants from low-redshift sources, we adopt the \textit{Herschel} color cut which requires $S_{350 \mu m}/S_{250 \mu m}>0.7$ and $S_{500 \mu m}/S_{350 \mu m}>0.6$ to effectively select sources at $1.5<z<3$ \citep[see][]{2016A&A...596A.100P}. The numbers of HLIRGs occupying this redshift range are 16, 39, and 45 in Bo$\rm \ddot{o}$tes, EN1 and LH respectively, and reduce to 16, 34, and 40 after visual inspection to remove objects near the boundary. In the upper panel of Figure \ref{help}, we present the number distributions of star-forming neighbours around each HLIRGs within 100$\arcsec (\sim$ 1 Mpc at $1.5<z<3$), 200$\arcsec$ and 400$\arcsec$ respectively. The black symbols and errorbars are the median values and 16th/84th percentiles calculated from 100 realizations which select the same number of random galaxies and count their neighbours. We also require no overlap between HLIRG neighbours and random neighbours by requiring random galaxies locate at least 2 times of searching radius away from every HLIRG. The cumulative distribution function (CDF) of the number of HLIRG neighbours and random neighbours are shown in the lower panel of Figure \ref{help}. We observe a clear difference in the number of star-forming neighbours within 100$\arcsec$ as HLIRGs at $1.5<z<3$ tend to have more neighbours, with a median value of $4.0\pm{2.3}$, compared with a median value of $2.9\pm{0.3}$ for random positions (calculated as the mean value and the standard deviation of median number of neighbours from 100 realizations). This difference implies that HLIRGs live in overdense regions at a 3.7$\sigma$ significance level. This difference becomes weaker as the search radius increases, reducing to 2.7 and 1.6 $\sigma$ for 200" and 400", respectively. We also adopt IRAC color cut which requires $m_{3.6\mu m, AB}-m_{4.5\mu m, AB}>-0.1$ to select sources at $z>1.3$ \citep{2008ApJ...676..206P}. The similar discrepancy in the number of star-forming neighbours to what has been found using the \textit{Herschel} color cut also supports our expectation in the overdense nature of HLIRG environments.

\subsection{Spatial distribution of IRAC-selected sources around HLIRGs}\label{re2}
To probe the small-scale environment of HLIRGs in 3D, we search for their neighbours using the deep IRAC-selected photo-z catalogs. In this section, we search for neighbours around HLIRGs, and calculate the co-moving volume density of these neighbours as a function of stellar masses. We also study the neighbours around quasars and SMGs to validate our method and to explore how they compare with HLIRG neighbours. We adopt three fixed radius, 3 Mpc, 6 Mpc and 10 Mpc \citep[typical scale of protoclusters, see e.g.,][]{2016ApJ...833..135C} to find nearby galaxies.

%We first investigate the distribution of absolute IRAC 4.5 $\mu$m magnitudes of HLIRG neighbours. This band is close to rest-frame NIR band at high redshifts. The absolute magnitudes can serve as a simple and direct measurement of the stellar mass distribution of HLIRG neighbours as NIR traces evolved stellar emissions. Figure \ref{ab} shows the distribution of HLIRG neighbours as a function of absolute IRAC 4.5 $\mu$m magnitudes, as well as comparisons with quasar neighbours and random galaxy neighbours. Figure \ref{ab_smg} shows the same distributions but compared HLIRG neighbours with SMG neighbours in the LH field. Dashed lines represent completeness limit of absolute IRAC 4.5 $\mu$m  magnitudes obtained from $M_{ab, 4.5\mu m}-z$ distribution at the higher boundary of each redshift bin.

%We then study the distribution of rest-frame K-band luminosities in HLIRG neighbours. Rest-frame K-band luminosities traces emission from evolved stars and hence provides a good estimate of stellar mass. They are less direct than absolute magnitudes because they depend on SED fitting method to derive their values. However, they are advantageous than absolute magnitudes of one specific band as the latter measures different part of rest-frame spectrum for galaxies at different redshifts.  

Following \citet{2021A&A...648A...8W} and \citet{2021A&A...654A.117G}, we use CIGALE to estimate stellar mass for the deep photo-z source catalogs. We use a delayed-$\tau$ plus a starburst star formation history (SFH) and \citet{2003MNRAS.344.1000B} single stellar populations (SSPs), with a \citet{1955ApJ...121..161S} IMF and solar metallicity. We adopt a double power-law dust attenuation law based on \citet{2000ApJ...539..718C} and dust emission models from \citet{2014ApJ...780..172D}. We use the \citet{2006MNRAS.366..767F} AGN models but include fewer parameters for simplicity. We refer to \citet{2021A&A...654A.117G} for a detailed description of the parameter space. %Figure \ref{rest} shows the distribution of rest-frame K-band luminosities of HLIRG neighbours compared with redshift-matched quasar neighbours and random galaxy neighbours. Figure \ref{rest_smg} shows the same distributions but only for HLIRGs in the LH field and SMGs. Dashed lines represent completeness limit of rest-frame K band luminosities obtained from $L_{K, rest-frame}-z$ distribution at the higher boundary of each redshift bin.

%Finally we study the stellar mass distribution of HLIRG neighbours. Stellar mass estimates characterize how different HLIRG neighbours compared with quasars, SMGs and random galaxies directly. However, they depend on many factors such as the accuracy of redshifts and multiwavelength photometry, the selection of SED fitting models and parameter space and so on. Consequently, they are less reliable and suffer large uncertainties. As mentioned above, stellar masses are estimated through SED fitting.

Figure \ref{smf} shows the distribution of stellar mass estimates of HLIRG neighbours as well as redshift-matched quasar neighbours and random galaxy neighbours. Figure \ref{smf_smg} shows the same distribution but only for HLIRGs in LH and redshift-matched SMGs. Dashed lines represent the stellar mass completeness limit calculated using the method described in \citet{2010A&A...523A..13P}. Briefly, we calculate the limiting stellar mass $M_{limit}$ for each galaxy which is the value a galaxy will have if its apparent 3.6 $\mu$m magnitude equals to the limiting magnitude of the entire catalog. Then, we select the 20\% of the faintest galaxies in each redshift bin and assign the completeness limit as the $M_{limit}$ value below which 90\% of these faint galaxies lie. The stellar masses completeness limit derived in this way is $10^{10.2}\, \mathrm{M_{\sun}}$ and $10^{10.5}\, \mathrm{M_{\sun}}$ in $2<z<3$ and $3<z<4$ respectively.

In Figures \ref{smf} and \ref{smf_smg}, the three columns represent neighbours found within different radius as indicated at the top. The two rows display two redshift bins as indicated in the upper-left corner. For the uncertainty of the HLIRGs neighbours, we combine the Poisson error, standard deviation in three fields (for Figure \ref{smf} only) and uncertainty from photometric redshifts in quadrature. The latter comes from the standard deviation from 100 realizations. In every realization the photometric redshift of each HLIRG is randomly determined from a Gaussian distribution with a mean value of its photometric redshift $z$-photo and a standard deviation of (1+$z_{\rm photo})\times$ 0.15 (typical boundaries used for counting outliers in photometric redshift fitting, see Section \ref{data3}). Data points and lower/upper boundaries for the quasar neighbours and random neighbours are median value and 16th/84th percentiles respectively of 100 realizations. In each realization we randomly selected redshift-matched quasars and random positions to take uncertainty due to sampling into account. We do not consider Poisson error, standard deviation in three fields or uncertainty from photometric redshifts, because uncertainty due to random sampling dominates. For the uncertainty of SMG neighbours, we only consider Poisson errors due to their small sample size. %We cannot obtain enough SMGs in some redshift intervals after shifting their redshifts when simulate uncertainty due to photometric redshift.

As Figures \ref{smf} and \ref{smf_smg} show, there is a weak excess of HLIRG neighbours at $3<z<4$ compared with random positions. This excess of HLIRG neighbours become weaker as searching radius increases, implying that large-scale environment of HLIRGs is no different from random positions. The environment of quasars and SMGs within different scales is similar or marginally denser than random positions. We attribute the lack of strong excess when compared with random positions to the selection criterion of the deep photo-z catalogs we use. As explained in Section \ref{data4} and \ref{data5}, quasars are typically hosted in dark matter halos around a few times $10^{12}\, \mathrm{M_{\sun}}$, while SMGs are typically hosted in slightly more massive halos with masses around $10^{13}\, \mathrm{M_{\sun}}$. Such halos will host galaxies with stellar masses a few time of $10^{10}\, \mathrm{M_{\sun}}$. The completeness limits of our deep photo-z catalogs reach $10^{10.2}\, \mathrm{M_{\sun}}$ and $10^{10.5}\, \mathrm{M_{\sun}}$ in $2<z<3$ and $3<z<4$ respectively after applying a 3$\sigma$ cut at 3.6 $\mu$m. It is expected that the environments of SMGs and quasars are similar to the environment of these massive random positions. We will discuss further about the influence of survey depth in Section \ref{re4}. In addition, there are some studies pointing out that quasars are not good beacons of overdense structures \citep[e.g.,][]{2012ApJ...752...39T,2013MNRAS.436..315F,2013ApJ...773..178B,2018PASJ...70S..32U}.

In order to test whether the stellar mass estimates are reliable, we carry out several SED fitting runs with different models and parameters. For example, we tried different SFHs, including double power-laws and delayed-$\tau$ models without a later burst, and dust attenuation models with different slopes. We find these changes do not introduce significant systematic difference to the stellar mass estimates.  We also adopt a machine learning random forest (RF) method to estimate stellar mass in Appendix \ref{RF}. When comparing HLIRG neighbours with random galaxy neighbours as a distribution of RF derived stellar mass, we find a similar picture to the one presented in Figures \ref{smf} and \ref{smf_smg}.

\subsection{Search for clusters using Friend of Friend algorithm}\label{re5}
%In the previous sections we investigate the entire sample of HLIRGs, finding that HLIRGs typically have more star-forming neighbours within 100$\arcsec$ and a weaker excess in spatial neighbours. 
In this section, we take advantage of the method used in \citet{2021MNRAS.506.6063H}, which discovered 88 cluster candidates that consist of 4390 member galaxies at $z < 1.1$ in the 5.4 deg2 AKARI North Ecliptic Pole (NEP) field, based on photometric redshifts. The authors calculated the local density for each galaxy through the angular separation to the 10th nearest neighbour \citep{2021MNRAS.507.3070S}, then adopted a FOF algorithm to overdense galaxies in order to determine cluster members. The FOF algorithm is improved in that it considers neighbour number as a criterion for grouping friends so that the chain-shaped linking is avoided. Also, using redshift-dependent linking length makes the FOF algorithm valid at higher redshift.

We make use of this method to find (proto)cluster candidates in deep IRAC-selected photo-z catalogs after both quality flag cuts and 3.6 $\mu$ m detection significance cuts are applied (see Section \ref{data3}) to reduce spurious sources which could influence the search for overdense regions. The Normalized Median Absolute Deviation (NMAD) of photometric redshift uncertainty are 0.022, 0.029, and 0.046 for Bo$\rm \ddot{o}$tes, EN1, and LH fields, respectively. There are 73\,974, 61\,281, and 72\,639 overdensities selected as having local density greater than 2 in three fields respectively. Then we apply the FOF algorithm on the overdensities. The projected linking length is determined by fitting the distances to the tenth neighbour for all galaxies. Having at least 10 friends within the linking length is required to continue making new friends. Clusters are selected as having more than 30 members. We find 52, 49 and 57 clusters above z > 1 in Bo$\rm \ddot{o}$tes, EN1 and LH fields respectively with 1, 2, 1 of them hosting HLIRGs respectively. 

The majority of our HLIRG sample are not associated with (proto)cluster candidates identified by FOF algorithm. One explanation could be due to the missing overdense regions traced by the IRAC-selected photo-z catalogs. Dust-obscured overdense regions (i.e, 250 $\mu$m bright overdense regions in Section \ref{re1}) may have low S/N ratios or be totally missed in the photo-z catalogs. For example, \citet{2019ApJ...887..214K} found that at $z\sim 3.8$, protoclusters may host obscured AGNs missed by optical selection. Therefore, the potential overdense environment which host HLIRGs cannot be identified by the FOF algorithm. Another explanation could be that HLIRGs reside in various environments and only a small fraction of them are hosted by protoclusters. We will discuss further in Section \ref{dis3}.

\begin{figure}
\includegraphics[width=\linewidth]{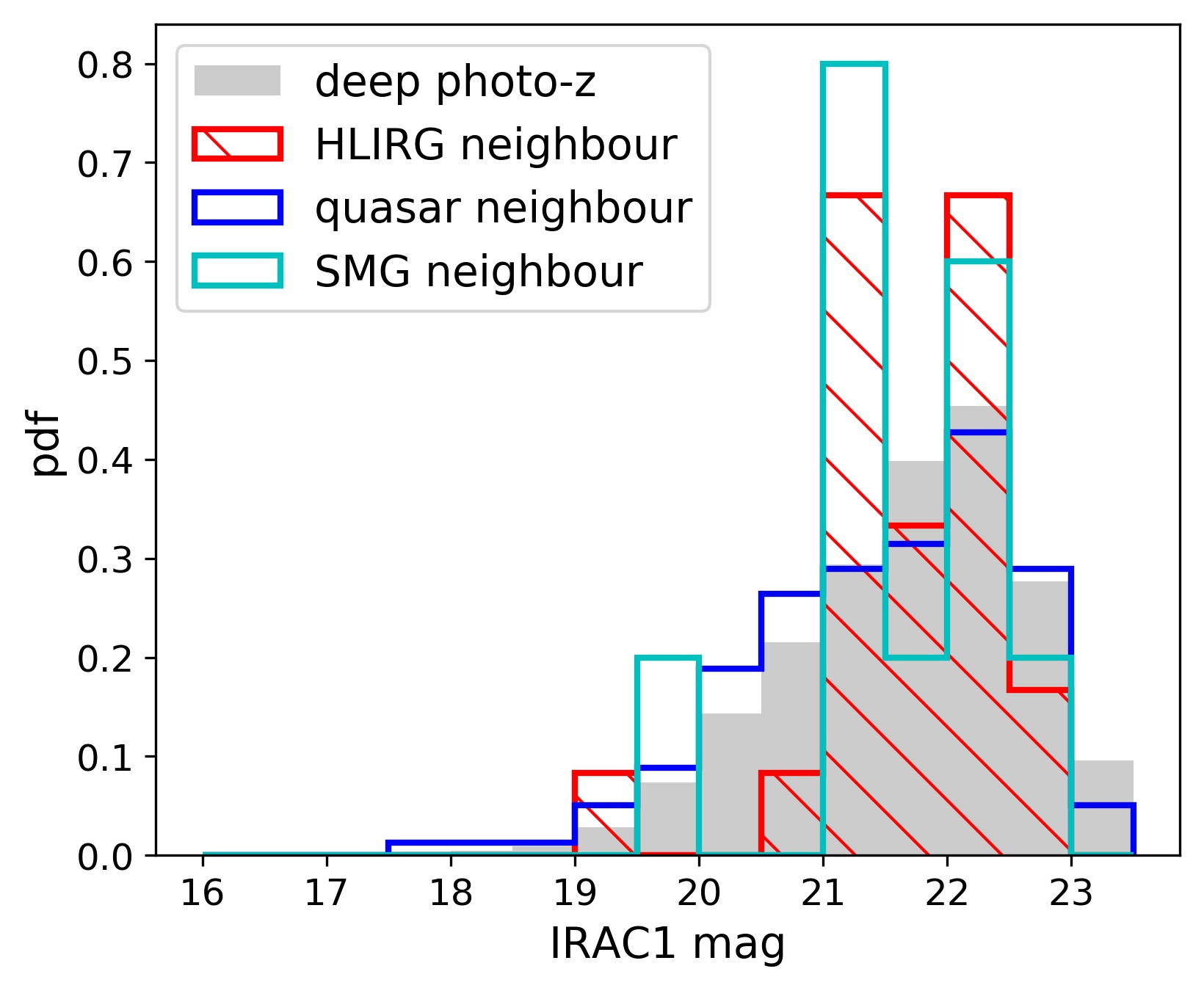}
\caption{The normalized 3.6 $\mu$m magnitudes distribution of the deep IRAC-selected photo-z catalogs (black), HLIRG neighbours (red hatched), quasar neighbours (blue) and SMG neighbours (cyan; only in the LH field) within 3 Mpc. We find that HLIRG neighbours are predominantly faint in the 3.6 $\mu$m band.}
\label{irac1}
\end{figure}

\begin{figure*}[htbp]
\resizebox{\hsize}{!}{\includegraphics[width=\linewidth]{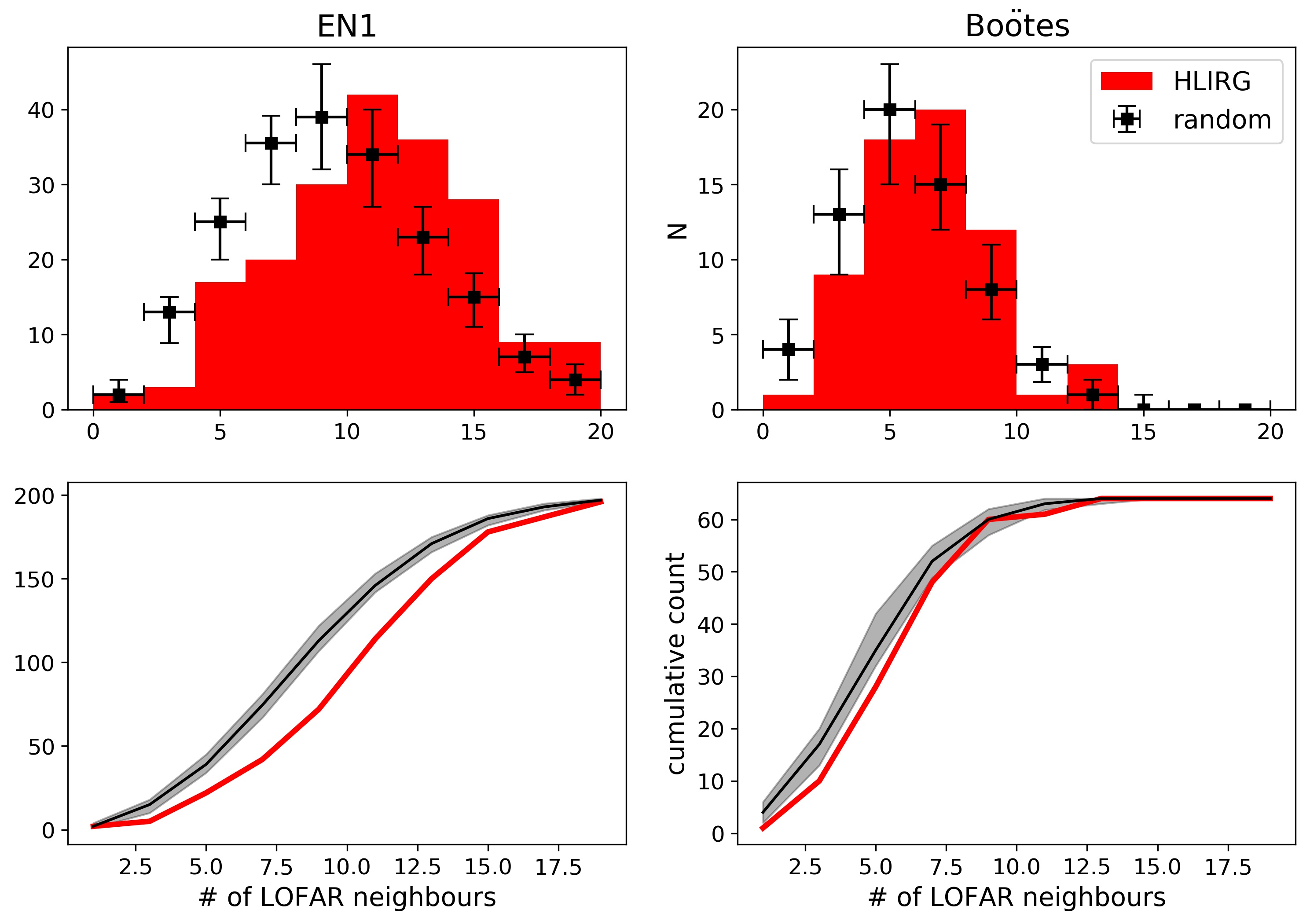}}
\caption{Number distribution of LOFAR-detected neighbours around HLIRGs within 100$\arcsec$ compared with that around random positions in Bo$\rm \ddot{o}$tes and EN1 fields respectively. The minimum 150 MHz flux densities are 0.15 and 0.07 mJy respectively. We find a more significant difference (4.7$\sigma$ in comparison with 1.9$\sigma$) between HLIRG neighbours and random neighbours in the EN1 field which is the deepest LOFAR field.}
\label{lofar}
\end{figure*}

\section{Discussion}
\subsection{The effect of survey depth}\label{re4}
We do not find a strong excess in the quasar neighbours and SMG neighbour compared to random neighbours. The main explanation is the similar halo mass range that host quasars, SMGs and random galaxies within our deep photo-z catalogs. In contrast, our HLIRGs are ultra-massive and thus are expect to reside in dark matter halos with masses in a range of a few times of $10^{13}\, \mathrm{M_{\sun}}$ to $10^{14}\, \mathrm{M_{\sun}}$ at $2<z<3$ where most our HLIRG lie. Such extremely massive halos should host a large number of member galaxies. However, we only observe a relatively weak excess in HLIRG spatial neighbours compared to random neighbours.

We attribute this lack of excess to the fact that some HLIRG neighbours are not detected due to the limited survey depth. We select the deep photo-z catalogs requiring 3-$\sigma$ detection in the 3.6$\mu$m band which is sensitive to emission from evolved stars. Some star-forming neighbours of HLIRGs may be too dusty to be selected. Therefore, they are faint and of low significance in the 3.6$\mu$m band, resulting in non-detection in the deep photo-z catalogs. In Figure \ref{irac1} we display the normalized probabilistic distribution of 3.6 $\mu$m magnitudes for all sources in the deep photo-z catalogs, as well as neighbours of HLIRGs, quasars and SMGs (only in the LH field). We find that HLIRG neighbours and SMG neighbours typically have fainter 3.6 $\mu$m magnitudes. It is highly possible that some neighbours are too faint in the 3.6 $\mu$m bands and are not included in our deep photo-z catalogs after applying a 3$\sigma$ significance cut. We may need deeper catalogs in order to find these fainter neighbours and study the environment of HLIRGs.We anticipate that with deeper catalogs, more HLIRG neighbours will be revealed relative to the random neighbours, leading to a stronger excess signal in the HLIRG environments when comparing with random positions.

We also explore the effect due to survey depth by seeking LOFAR-detected neighbours around HLIRGs. The sensitivity of LOFAR radio survey varies in different fields, reaching 32, 20 and 22 $\upmu$Jy beam$^{-1}$ in Bo$\rm \ddot{o}$tes, EN1 and LH fields respectively. %We search for LOFAR detected neighbours within 100$\arcsec$ around 64, 198 and 185 HLIRGs in three fields respectively. We keep most of HLIRGs in the Bo$\rm \ddot{o}$tes and LH fields and all HLIRGs in the EN1 fields as LOFAR observations cover a large sky area. We require a 3$\sigma$ cut in signal-to-noise ratio of 150 MHz detection and confine all sources to locate in the central 14, 24, and 18 degree$^2$ respectively. The median 150 MHz flux densities are 0.55, 0.31 and 0.39 mJy respectively. The surface number densities of LOFAR detected sources are 1668, 1945, and 1559 degree$^{-2}$ respectively. In each field we follow the same procedure as described in Section \ref{re1}, searching for neighbours around HLIRGs and counting neighbours around random positions. The distribution of the number of LOFAR-detected neighbours around HLIRGs in three fields are shown in Figure \ref{lofar}.   
We search for LOFAR-detected neighbours within 100$\arcsec$ around 64 and 198 HLIRGs in Bo$\rm \ddot{o}$tes and EN1 respectively as these two fields are the most different: After requiring a 3$\sigma$ cut in the signal-to-noise ratio of the 150 MHz detections and confining all sources to be located in the central 14 and 24 degree$^2$ respectively, the minimum 150 MHz flux densities are 0.16 and 0.07 mJy respectively. We keep most of HLIRGs in the Bo$\rm \ddot{o}$tes field and all HLIRGs in the EN1 field as LOFAR observations cover a large sky area. In each field we follow the same procedure as described in Section \ref{re1}, searching for neighbours around HLIRGs and counting neighbours around random positions. The distribution of the number of LOFAR-detected neighbours around HLIRGs in two fields are shown in Figure \ref{lofar}. 

As Figure \ref{lofar} shows, the difference in the abundance between HLIRG neighbours and random neighbours is larger in the EN1 field, at a 4.7$\sigma$ significance level compared to 1.9$\sigma$ significance level in the Bo$\rm \ddot{o}$tes field. When requiring the same depth of LOFAR detections with 150 MHz flux density $>0.5$ mJy, this discrepancy in different fields disappears, reaching 1.0$\sigma$ and 0.9$\sigma$ significance levels respectively. We hypothesize that deeper surveys can play an important role in studying small-scale environments as they can include fainter and less massive galaxies which trace more typical environments with lower host halo mass values. The excess signal may become more significant when comparing HLIRG environments with these less massive environments. %Although we investigate the abundance of 2D neighbours of HLIRGs and do not make any redshift constraints (see Section \ref{re1}) to reduce chance projections, this analysis provides a hint on how survey depth may influence the study of 3D environments.

%As Figure \ref{lofar} shows, the difference in the abundance between HLIRG neighbours and random neighbours is larger in the EN1 field, at a 4.7$\sigma$ significance level compared with 2.1$\sigma$ and 1.8$\sigma$ significance levels in the LH and Bo$\rm \ddot{o}$tes fields respectively. In addition, this large difference in the EN1 fields disappears when we require the same survey depth in all three fields. The different levels of excess in different fields suggests that deeper surveys play an important role in studying small-scale environments. Some faint HLIRG neighbours may not be uncovered in shallower surveys and thus resulting in smaller differences when compared with random positions. Although we do not make any redshift constraints like Section \ref{re1} and thus include more chance projections of low-redshift galaxies when counting LOFAR-detected neighbours, this analysis provide a hint on how survey depth may influence the study of environments.

\subsection{The need for spectroscopic follow-up}\label{re3}

%We only find weak excess of HLIRG neighbours compared to random galaxy neighbours in terms of distribution of stellar mass estimates. \sout{While in contrast, quasar neighbours display a significant excess at the massive end. A weak excess also exists in the SMG neighbours which is inconsistent with the fact that both SMGs and quasars have been reported being clustered and SMGs reside in slightly more massive halos. Apart from small sample size, this weak excess in HLIRG neighbours and SMG neighbours compared with random neighbours could be partly attributed to the fact that quasars have spectroscopic redshifts while HLIRGs and SMGs only have photometric redshifts. We expect that photometric redshift uncertainty would dilute the overdense signal and explore this factor by adding an artificial redshift uncertainty to the spectroscopic redshifts of quasars. The redshift of each quasar is randomly drawn from a Gaussian distribution with a mean value of its spectroscopic redshift spec-z and a standard deviation value of $(1+\rm spec-z) \times 0.05$. The data points and uncertainty of quasar neighbours are the median value and 16th/84th percentiles from 100 realizations in which new redshift-matched quasars are selected based on their new redshifts. We only compare the stellar mass density of quasar neighbours and random galaxies neighbours within 3 Mpc since this small-scale environment shows the most distinctions. As Figure show, the excess of quasar neighbours at the massive end displayed above vanish completely.} 

Another potential reason for the relatively weak excess of spatial HLIRG neighbours compared with random neighbours is due to photometric redshift uncertainty. At a given sky position, a HLIRG at $2<z<4$ will move into a new location that is $70-150$ Mpc away which is beyond the scale of its host halo when its redshift is shifted with a small value of 0.1. Also true neighbours of HLIRGs which live in the same overdense structure could be moved outside of the searching radius due to photometric redshift uncertainty. %It is highly possible that no nearby galaxies can be found around this new location assuming that galaxies are clustered around real locations. Inspired by this difference, we may obtain a more significant excess if we have spectroscopic redshifts of HLIRGs.

%\begin{figure}
%\includegraphics[width=\linewidth]{test_quasar_specz.jpg}
%\caption{Stellar mass density of new quasar neighbours (blue solid square) and random galaxy neighbours (black stars) within 3 Mpc. The blue open squares are data points without any shifting in redshifts (i.e., the corresponding data points in Figure \ref{smf}).}
%\label{test_specz}
%\end{figure}

 It is challenging to obtain spectroscopic redshifts for HLIRGs due to severe dust obscuration. So far, we have managed to obtain spectroscopic redshifts for a few HLIRGs. First, there are 28 HLIRGs (6, 8, and 14 in Bo$\rm \ddot{o}$tes, EN1, and LH respectively) that have spectroscopic redshifts collected and compiled in the deep source catalogs \citep{2021A&A...648A...4D}. Second, two HLIRGs, LH\_16049 and LH\_23905 are observed by Northern Extended Millimeter Array (NOEMA) and their spectroscopic redshifts are consistent with photometric redshifts. LH\_16049 has a $z_{\rm phot}$ of 2.47 and a $z_{\rm spec}$ of 2.464. LH\_23905 has a $z_{\rm phot}$ of 2.66 and a $z_{\rm spec}$ of 2.558 (private communications with Sharon Chelsea and Axel Weiss). These two HLIRGs also demonstrate the good quality of photometric redshifts. We also obtained SUBARU observations for an extreme starburst candidate EN1\_19770. Details of the observation campaign and data processing are described in the Appendix \ref{ob}. We retrieve a spectroscopic redshift of 3.74 which is in good agreement with the photometric redshift estimate of 3.91 in \citet{2021A&A...648A...4D}.
 
\subsection{Are HLIRGs good tracers of protoclusters?}\label{dis3}

We find that HLIRGs in general have more bright \textit{Herschel} sources (with $250\mu$m flux density above 10 mJy) within 100$\arcsec$ than random positions at a $3.7\sigma$ significance level. However, when seeking 3D neighbours with IRAC-selected photo-z catalog, we only find relatively weak excess compared to random positions. Apart from the influence due to survey depth and photometric redshift uncertainty, another potential explanation could be attributed to HLIRG themselves. Are HLIRGs good tracers of overdense regions such as protoclusters?

From Figure \ref{help} we find that around 68-87\% of our HLIRGs (11 of 16, 23 of 34, 35 of 40 in Bo$\rm \ddot{o}$tes, EN1 and LH fields respectively) have more \textit{Herschel}-detected neighbours within 100$\arcsec$ than the median number of neighbours of random positions while the remaining 13-32\% have fewer \textit{Herschel}-detected neighbours. We select 30 HLIRGs as the most promising protocluster candidates based on the number of \textit{Herschel} neighbours and FOF algorithm (see Section \ref{conclu}) and plot the stellar mass distribution of their neighbours as red solid circles in Figure \ref{smf}. We observe a slightly enhanced excess compared to the stellar mass distribution of neighbours of all HLIRGs. We conclude that only some HLIRGs are more likely than others to reside in overdense regions, both in 2D and 3D environments. We note here that this conclusion is reached under the above-mentioned limitations of survey depth and photometric redshift uncertainty. HLIRGs that show overdensity features under such limitations can be used as the most promising candidates to identify protoclusters in follow-up observations given that spectroscopic campaigns are time-consuming (see next section). However, the possibility that HLIRGs showing no signs of overdensity in the current study may actually  reside in overdense regions cannot be ruled out completely.

\citet{2015MNRAS.452..878M} investigated a sample of mock SMGs from the \textit{Bolshoi} cosmological simulation \citep{2013AN....334..691R,2011ApJ...740..102K,2016MNRAS.462..893R} and found that SMGs are incomplete tracers of the most-massive structures as the most majority of the most-massive structures do not host SMGs. Since the enhanced starbursting mode is short-lived, SMGs can only be observed during a relatively short period. Their rarity results in some of the most-massive structures hosting no SMGs. Moreover, due to the cosmic downsizing, the most-massive structures are expected to cease star-formation activity at an earlier epoch. \citet{2015MNRAS.452..878M} also found that dark matter halos at $z<2.5$ are less likely than their high-redshift counterparts to host SMGs. This is broadly consistent with our findings that HLIRGs at $3<z<4$ show a relatively stronger excess than HLIRGs at $2<z<3$ when comparing with random positions.

Observationally,\citet{2015MNRAS.453..951C} carried out observational campaign for one of the most luminous SMGs HS1700.850.1 at $z=2.82$ and concluded that it resides in relatively voids. \citet{2015ApJ...810..130L} studied the environment around an extreme luminous IR galaxy HFLS3 at $z=6.34$ and found no evidence for an overdensity of bright sources.\footnote{HFLS3 is at much higher redshift than our HLIRG sample. Both \citet{2015MNRAS.453..951C} and \citet{2015ApJ...810..130L} searched for overdensity of LBGs which may not be a good tracer of dusty overdensities at high redshifts.} Similarly, \citet{2012ApJ...752...39T} studied a sample of 15 most luminous QSOs at $z \sim 2.7$. They found that these extremely luminous QSOs reside in host halos with masses similar to their less luminous counterparts. The possibility that extreme luminous SMGs, HLIRGs or QSOs are merely rare events and do not necessarily trace the most massive structures can not be ruled out. 

HLIRGs maybe a more incomplete tracers as being rarer and more extreme than typical SMGs. This is supported by the (proto)clusters found using FOF algorithm. Of all (proto)clusters candidates identified, $>96\%$ of them do not host one of our HLIRGs. Nonetheless, the reason why some HLIRGs do not show an overdense nature of MPC-scale environment is beyond the scope of this paper. Our work aims to study the environment of the largest HLIRG sample and select the most promising protocluster candidates that benefit further research.

\section{Conclusions}\label{conclu}

\begin{table*}[htbp]
    \caption{A list of 30 HLIRGs which have been selected as the most promising signposts of potential protoclusters. These candidates are ranked by possibility of residing in overdense regions, accounting for factors such as the number of \textit{Herschel}-detected neighbours, spatial neighbours found in the deep photo-z catalogs, the quality of their redshifts (specscopically confirmed or the reduced $\chi^{2}$ in the fitting for photometric redshifts), whether found in FOF algorithm and the q value in FIRC (defined as log$(\frac{L_{IR}/(3.75\times10^{12}Hz)}{L_{150MHz}})$; smaller values indicating more radio-loud).} 
    \centering
    \begin{tabular}{cccccccc}
    \hline
    LOFAR id&field&RA&DEC&redshift&reduced $\chi^{2}$ for $z_{phot}$&FOF&q value\\
    \hline
    \multicolumn{8}{c}{spectroscopic redshift}\\
    \hline
    21243&LH&162.33080&57.39226&1.68&--&F&1.63\\
    19726&LH&162.53003&58.08506&2.84&--&T&1.68\\

\hline
\multicolumn{8}{c}{EN1}\\
\hline

40215&EN1&242.58979&54.03271&2.29&0.27&F&1.28\\
37547&EN1&242.83600&53.76636&2.90&0.26&F&1.34\\
38464&EN1&242.75260&56.52186&2.50&0.37&F&1.75\\
34758&EN1&243.09614&54.26354&2.45&0.48&F&2.16\\
54459&EN1&241.18199&54.34068&2.69&0.32&F&1.49\\
44576&EN1&242.15551&56.47998&2.45&0.43&F&1.63\\
60669&EN1&240.40852&54.98878&2.40&0.48&F&1.72\\
40404&EN1&242.57690&54.00540&2.59&0.13&F&1.75\\
50542&EN1&241.61670&54.22248&4.92&0.32&T&1.65\\
46515&EN1&242.00465&54.43105&3.69&0.23&F&1.57\\
23891&EN1&244.16897&55.90912&4.11&1.54&F&0.94\\
39610&EN1&242.64656&54.60312&4.72&0.55&F&1.98\\
35782&EN1&243.00559&54.72401&4.02&0.17&F&1.71\\

\hline
\multicolumn{8}{c}{LH}\\
\hline
20490&LH&162.44004&58.13847&2.49&0.48&F&1.28\\
27801&LH&161.50650&59.87516&2.02&0.34&F&1.72\\
23594&LH&162.03957&56.36271&2.05&0.31&F&1.54\\

21892&LH&162.26817&58.46467&2.37&0.29&F&1.58\\
41471&LH&159.46355&59.17206&2.10&0.26&F&1.85\\

20655&LH&162.41359&57.91304&3.08&0.21&F&1.25\\
13561&LH&163.39872&58.22669&4.06&0.12&F&1.21\\
39722&LH&159.81428&58.54783&3.59&0.37&F&1.67\\
23157&LH&162.11348&58.56543&3.42&1.10&F&1.52\\
23101&LH&162.11968&58.42583&3.65&0.63&F&1.74\\

16853&LH&162.86970&56.62488&3.93&2.19&T&1.54\\

\hline
\multicolumn{8}{c}{Bo$\rm \ddot{o}$tes}\\
\hline
13499&Bo$\rm \ddot{o}$tes&218.59504&35.48697&2.18&1.22&T&1.31\\
26351&Bo$\rm \ddot{o}$tes&217.12640&35.52119&4.25&0.06&F&1.11\\
30484&Bo$\rm \ddot{o}$tes&216.57340&33.02774&3.83&0.06&F&1.52\\
13358&Bo$\rm \ddot{o}$tes&218.59538&33.50608&5.54&0.01&F&1.52\\

    \hline
\multicolumn{8}{p{.8\textwidth}}{Note: EN1\_23891 is 18.4$\arcmin$ away from the protocluster candidate PCCS1 857 G085.48+43.36 in \citet{2018MNRAS.476.3336G}, in spatial volume 
of 7.78 Mpc at $z=4.11$. Bo$\rm \ddot{o}$tes\_13499 is 4.39$\arcmin$ away from the protocluster candidate PLCKERC545 G060.36+66.56 in \citet{2018MNRAS.476.3336G}, in spatial volume of 2.23 Mpc at $z=2.18$.}

    \end{tabular}
    
    \label{list}
\end{table*}

In this work, we probe the Mpc-scale environment of the largest sample of HLIRGs in three fields, in order to investigate whether they generally trace overdense regions as expected. We first search for \textit{Herschel}-detected star-forming neighbours around HLIRGs and compare with that around random positions. We apply a \textit{Herschel} color cut to reduce chance projections. We find that on average HLIRGs do indeed have more bright (with 250 $\mu$m flux density $>10$ mJy) star-forming neighbours at a 3.7$\sigma$ significance level with a median value of $4.0\pm{2.3}$ in their close vicinity (100$\arcsec$) while the median number of random neighbours is $2.9\pm{0.3}$. 
%\textbf{In addition, HLIRG neighbours are more likely to have brighter flux densities than random neighbours. HLIRGs have more bright neighbours with 250 $\mu$m flux density $>10$ mJy than random positions at a $3.2\sigma$ significance level indicating that HLIRGs may live in dense regions with intense on-going star-formation activities.}

We then search for neighbours in 3D around HLIRGs using deep IRAC-selected photo-z catalogs. We include a quasar sample and a SMG sample as cross-checks because clustering analyses of these two samples have derived host dark matter halo masses to be around around a few times of 10$^{12} \, \mathrm{M_{\sun}}$ and 10$^{13} \, \mathrm{M_{\sun}}$ respectively. They also (partly) overlap in redshifts with the HLIRGs. We study the comoving volume density of HLIRGs neighbours as a function of stellar mass and compare with that of random neighbours. We find the environment of quasars and SMGs are similar or marginally denser than the environment of random positions, mainly due to the fact that random galaxies in the IRAC-selected photo-z catalogs live in similar massive halos based on their stellar mass estimates. 

We only find relatively weak excess of HLIRG 3D neighbours compared with random neighbours. This is due to a number of factors such as the influence of survey depth, photometric redshift uncertainty which dilutes the excess signal, small number statistics, and so on. The IRAC-selected photo-z catalogs may not be able to uncover all HLIRG neighbours as they are predominantly faint at 3.6 $\mu$m. We investigate the influence due to survey depth by seeking LOFAR-detected neighbours. In the deepest EN1 field, HLIRGs have more LOFAR-detected neighbours at a 4.7 $\sigma$ significance level while in the shallowest Bo$\rm \ddot{o}$tes field the excess signal is reduced to 1.9$\sigma$ significance level. After applying a higher flux density cut in both fields, the significance levels drop to 1.0$\sigma$ and 0.9$\sigma$ respectively, which is in agreement with our expectation. In terms of the influence of photometric redshift uncertainty, a small displacement (0.1 at $2<z<4$) in redshift will result in a change of tens to hundreds of Mpc in location where is far away from host halos and thus no neighbours may be found. We expect that with spectroscopic redshifts, we would be able to observe a much stronger excess of massive neighbours around our HLIRGs compared with random positions. So far we have managed to assemble 31 spectroscopic redshifts for our HLIRG sample from literature and new optical/submillimeter observations. We also find that HLIRGs at $3<z<4$ show stronger excess when comparing with random positions than HLIRGs at $2<z<3$, suggesting cosmic downsizing, as the most massive halos may stop star-forming activity at $2<z<3$ and hence cannot be traced by extreme SFR.  

%We also explore the effect due to survey depth by seeking LOFAR-detected neighbours around HLIRGs. We observe a more significant excess of HLIRG neighbours compared with random neighbours in the EN1 field where the median flux is the smallest and the surface density is the largest among three fields, implying that deeper surveys are needed in studying small-scale environments. HLIRG neighbours typically have fainter 3.6 $\mu$m magnitudes, indicating dustier nature. These galaxies may be missed by our deep photo-z catalogs which are based on optical-NIR detections.

Finally, we present a list of 30 HLIRGs in Table \ref{list} that are ranked by the degree of overdensity in its Mpc-scale environment. We also display the cumulative distribution of the 250 $\mu$m flux densities of their neighbours within 100$\arcsec$ as well as the overdensity parameters measured at different radius in Figure \ref{fig:all}. We take into account factors such as the quality of their spectroscopic redshifts, the number of \textit{Herschel} detected neighbours, whether or not being associated with protocluster candidates found in the FOF algorithm and so on. Two candidates are close to the protocluster candidates identified in \citet{2018MNRAS.476.3336G}. The authors cross-matched \textit{Herschel} catalogs to \textit{Planck} compact sources and selected candidates that have at least 3$\sigma$ overdensities in either 250, 350, or 500 $\mu$m sources. 

Unlike previous studies based on bright unresolved sources detected in \textit{Planck} or SPT survey that have been pre-selected as protocluster candidates, our work first select the most extreme dusty star-forming galaxies based on \textit{Herschel} blind survey and then investigate their environment. We find on average HLIRG have more star-forming neighbours in their close vicinity. Those having more bright neighbours are very likely to be hosted in extreme massive halos and thus potentially live in overdense environments such as protoclusters. By targeting these most promising protocluster candidates with follow-up observations, we can spectroscopically confirm them and study their physical properties such as the host halo mass, total SFR, gas mass, gas depletion time and star-formation efficiency to probe the evolution path of massive halos in an early universe.

\section*{Acknowledgement}
We thank Scott Trager for his help with the Subaru FOCAS spectrum analysis for the HLIRG EN1\_19770. We thank Matthieu Bethermin and Giulia Rodighiero for their enlightening comments and suggestions.

\bibliographystyle{aa}
\typeout{}
\bibliography{env.bib}

\begin{thebibliography}{134}
\expandafter\ifx\csname natexlab\endcsname\relax\def\natexlab#1{#1}\fi

\bibitem[{{Arrigoni Battaia} {et~al.}(2018){Arrigoni Battaia}, {Chen},
  {Fumagalli}, {Cai}, {Calistro Rivera}, {Xu}, {Smail}, {Prochaska}, {Yang}, \&
  {De Breuck}}]{2018A&A...620A.202A}
{Arrigoni Battaia}, F., {Chen}, C.-C., {Fumagalli}, M., {et~al.} 2018, \aap,
  620, A202

\bibitem[{{Assef} {et~al.}(2015){Assef}, {Eisenhardt}, {Stern}, {Tsai}, {Wu},
  {Wylezalek}, {Blain}, {Bridge}, {Donoso}, {Gonzales}, {Griffith}, \&
  {Jarrett}}]{2015ApJ...804...27A}
{Assef}, R.~J., {Eisenhardt}, P.~R.~M., {Stern}, D., {et~al.} 2015, \apj, 804,
  27

\bibitem[{{Ba{\~n}ados} {et~al.}(2013){Ba{\~n}ados}, {Venemans}, {Walter},
  {Kurk}, {Overzier}, \& {Ouchi}}]{2013ApJ...773..178B}
{Ba{\~n}ados}, E., {Venemans}, B., {Walter}, F., {et~al.} 2013, \apj, 773, 178

\bibitem[{{Balmaverde} {et~al.}(2017){Balmaverde}, {Gilli}, {Mignoli},
  {Bolzonella}, {Brusa}, {Cappelluti}, {Comastri}, {Sani}, {Vanzella},
  {Vignali}, {Vito}, \& {Zamorani}}]{2017A&A...606A..23B}
{Balmaverde}, B., {Gilli}, R., {Mignoli}, M., {et~al.} 2017, \aap, 606, A23

\bibitem[{{Behroozi} {et~al.}(2019){Behroozi}, {Wechsler}, {Hearin}, \&
  {Conroy}}]{2019MNRAS.488.3143B}
{Behroozi}, P., {Wechsler}, R.~H., {Hearin}, A.~P., \& {Conroy}, C. 2019,
  \mnras, 488, 3143

\bibitem[{{Behroozi} {et~al.}(2013){Behroozi}, {Wechsler}, \&
  {Conroy}}]{2013ApJ...770...57B}
{Behroozi}, P.~S., {Wechsler}, R.~H., \& {Conroy}, C. 2013, \apj, 770, 57

\bibitem[{{Bertin} \& {Arnouts}(1996)}]{1996A&AS..117..393B}
{Bertin}, E. \& {Arnouts}, S. 1996, \aaps, 117, 393

\bibitem[{{Blain} {et~al.}(2004){Blain}, {Chapman}, {Smail}, \&
  {Ivison}}]{2004ApJ...611..725B}
{Blain}, A.~W., {Chapman}, S.~C., {Smail}, I., \& {Ivison}, R. 2004, \apj, 611,
  725

\bibitem[{{Brown} {et~al.}(2008){Brown}, {Zheng}, {White}, {Dey}, {Jannuzi},
  {Benson}, {Brand}, {Brodwin}, \& {Croton}}]{2008ApJ...682..937B}
{Brown}, M. J.~I., {Zheng}, Z., {White}, M., {et~al.} 2008, \apj, 682, 937

\bibitem[{{Bruzual} \& {Charlot}(2003)}]{2003MNRAS.344.1000B}
{Bruzual}, G. \& {Charlot}, S. 2003, \mnras, 344, 1000

\bibitem[{{Burgarella} {et~al.}(2005){Burgarella}, {Buat}, \&
  {Iglesias-P{\'a}ramo}}]{2005MNRAS.360.1413B}
{Burgarella}, D., {Buat}, V., \& {Iglesias-P{\'a}ramo}, J. 2005, \mnras, 360,
  1413

\bibitem[{{Cai} {et~al.}(2016){Cai}, {Fan}, {Peirani}, {Bian}, {Frye},
  {McGreer}, {Prochaska}, {Lau}, {Tejos}, {Ho}, \&
  {Schneider}}]{2016ApJ...833..135C}
{Cai}, Z., {Fan}, X., {Peirani}, S., {et~al.} 2016, \apj, 833, 135

\bibitem[{{Cai} {et~al.}(2018){Cai}, {Hamden}, {Matuszewski}, {Prochaska},
  {Li}, {Cantalupo}, {Arrigoni Battaia}, {Martin}, {Neill}, {O'Sullivan},
  {Wang}, {Moore}, \& {Morrissey}}]{2018ApJ...861L...3C}
{Cai}, Z., {Hamden}, E., {Matuszewski}, M., {et~al.} 2018, \apjl, 861, L3

\bibitem[{{Calistro Rivera} {et~al.}(2017){Calistro Rivera}, {Williams},
  {Hardcastle}, {Duncan}, {R{\"o}ttgering}, {Best}, {Br{\"u}ggen}, {Chy{\.z}y},
  {Conselice}, {de Gasperin}, {Engels}, {G{\"u}rkan}, {Intema}, {Jarvis},
  {Mahony}, {Miley}, {Morabito}, {Prandoni}, {Sabater}, {Smith}, {Tasse}, {van
  der Werf}, \& {White}}]{2017MNRAS.469.3468C}
{Calistro Rivera}, G., {Williams}, W.~L., {Hardcastle}, M.~J., {et~al.} 2017,
  \mnras, 469, 3468

\bibitem[{{Capak} {et~al.}(2011){Capak}, {Riechers}, {Scoville}, {Carilli},
  {Cox}, {Neri}, {Robertson}, {Salvato}, {Schinnerer}, {Yan}, {Wilson}, {Yun},
  {Civano}, {Elvis}, {Karim}, {Mobasher}, \& {Staguhn}}]{2011Natur.470..233C}
{Capak}, P.~L., {Riechers}, D., {Scoville}, N.~Z., {et~al.} 2011, \nat, 470,
  233

\bibitem[{{Cappellari} {et~al.}(2011){Cappellari}, {Emsellem}, {Krajnovi{\'c}},
  {McDermid}, {Serra}, {Alatalo}, {Blitz}, {Bois}, {Bournaud}, {Bureau},
  {Davies}, {Davis}, {de Zeeuw}, {Khochfar}, {Kuntschner}, {Lablanche},
  {Morganti}, {Naab}, {Oosterloo}, {Sarzi}, {Scott}, {Weijmans}, \&
  {Young}}]{2011MNRAS.416.1680C}
{Cappellari}, M., {Emsellem}, E., {Krajnovi{\'c}}, D., {et~al.} 2011, \mnras,
  416, 1680

\bibitem[{{Carlstrom} {et~al.}(2011){Carlstrom}, {Ade}, {Aird}, {Benson},
  {Bleem}, {Busetti}, {Chang}, {Chauvin}, {Cho}, {Crawford}, {Crites}, {Dobbs},
  {Halverson}, {Heimsath}, {Holzapfel}, {Hrubes}, {Joy}, {Keisler}, {Lanting},
  {Lee}, {Leitch}, {Leong}, {Lu}, {Lueker}, {Luong-Van}, {McMahon}, {Mehl},
  {Meyer}, {Mohr}, {Montroy}, {Padin}, {Plagge}, {Pryke}, {Ruhl}, {Schaffer},
  {Schwan}, {Shirokoff}, {Spieler}, {Staniszewski}, {Stark}, {Tucker},
  {Vanderlinde}, {Vieira}, \& {Williamson}}]{2011PASP..123..568C}
{Carlstrom}, J.~E., {Ade}, P.~A.~R., {Aird}, K.~A., {et~al.} 2011, \pasp, 123,
  568

\bibitem[{{Casey} {et~al.}(2014){Casey}, {Narayanan}, \&
  {Cooray}}]{2014PhR...541...45C}
{Casey}, C.~M., {Narayanan}, D., \& {Cooray}, A. 2014, \physrep, 541, 45

\bibitem[{{Chabrier}(2003)}]{2003PASP..115..763C}
{Chabrier}, G. 2003, \pasp, 115, 763

\bibitem[{{Chapin} {et~al.}(2011){Chapin}, {Chapman}, {Coppin}, {Devlin},
  {Dunlop}, {Greve}, {Halpern}, {Hasselfield}, {Hughes}, {Ivison}, {Marsden},
  {Moncelsi}, {Netterfield}, {Pascale}, {Scott}, {Smail}, {Viero}, {Walter},
  {Weiss}, \& {van der Werf}}]{2011MNRAS.411..505C}
{Chapin}, E.~L., {Chapman}, S.~C., {Coppin}, K.~E., {et~al.} 2011, \mnras, 411,
  505

\bibitem[{{Chapman} {et~al.}(2020){Chapman}, {Aravena}, {Ashby}, {Canning}, {De
  Breuck}, {Endsley}, {Hayward}, {Hill}, {Marrone}, {Phadke}, {Scott},
  {Spilker}, {Stark}, {Vieira}, \& {Weiss}}]{2020hst..prop16237C}
{Chapman}, S.~C., {Aravena}, M.~A., {Ashby}, M. L.~N., {et~al.} 2020, {A
  massive protocluster at z=7 selected by the South Pole Telescope}, HST
  Proposal

\bibitem[{{Chapman} {et~al.}(2015){Chapman}, {Bertoldi}, {Smail}, {Steidel},
  {Blain}, {Geach}, {Gurwell}, {Ivison}, {Petitpas}, \&
  {Reddy}}]{2015MNRAS.453..951C}
{Chapman}, S.~C., {Bertoldi}, F., {Smail}, I., {et~al.} 2015, \mnras, 453, 951

\bibitem[{{Chapman} {et~al.}(2005){Chapman}, {Blain}, {Smail}, \&
  {Ivison}}]{2005ApJ...622..772C}
{Chapman}, S.~C., {Blain}, A.~W., {Smail}, I., \& {Ivison}, R.~J. 2005, \apj,
  622, 772

\bibitem[{{Charlot} \& {Fall}(2000)}]{2000ApJ...539..718C}
{Charlot}, S. \& {Fall}, S.~M. 2000, \apj, 539, 718

\bibitem[{{Chen} {et~al.}(2016){Chen}, {Smail}, {Ivison}, {Arumugam},
  {Almaini}, {Conselice}, {Geach}, {Hartley}, {Ma}, {Mortlock}, {Simpson},
  {Simpson}, {Swinbank}, {Aretxaga}, {Blain}, {Chapman}, {Dunlop}, {Farrah},
  {Halpern}, {Micha{\l}owski}, {van der Werf}, {Wilkinson}, \&
  {Zavala}}]{2016ApJ...820...82C}
{Chen}, C.-C., {Smail}, I., {Ivison}, R.~J., {et~al.} 2016, \apj, 820, 82

\bibitem[{{Cheng} {et~al.}(2019){Cheng}, {Clements}, {Greenslade}, {Cairns},
  {Andreani}, {Bremer}, {Conversi}, {Cooray}, {Dannerbauer}, {De Zotti},
  {Eales}, {Gonz{\'a}lez-Nuevo}, {Ibar}, {Leeuw}, {Ma}, {Micha{\l}owski},
  {Nayyeri}, {Riechers}, {Scott}, {Temi}, {Vaccari}, {Valtchanov}, {van
  Kampen}, \& {Wang}}]{2019MNRAS.490.3840C}
{Cheng}, T., {Clements}, D.~L., {Greenslade}, J., {et~al.} 2019, \mnras, 490,
  3840

\bibitem[{{Cheng} {et~al.}(2020){Cheng}, {Clements}, {Greenslade}, {Cairns},
  {Andreani}, {Bremer}, {Conversi}, {Cooray}, {Dannerbauer}, {De Zotti},
  {Eales}, {Gonz{\'a}lez-Nuevo}, {Ibar}, {Leeuw}, {Ma}, {Micha{\l}owski},
  {Nayyeri}, {Riechers}, {Scott}, {Temi}, {Vaccari}, {Valtchanov}, {van
  Kampen}, \& {Wang}}]{2020MNRAS.494.5985C}
{Cheng}, T., {Clements}, D.~L., {Greenslade}, J., {et~al.} 2020, \mnras, 494,
  5985

\bibitem[{{Chiang} {et~al.}(2014){Chiang}, {Overzier}, \&
  {Gebhardt}}]{2014ApJ...782L...3C}
{Chiang}, Y.-K., {Overzier}, R., \& {Gebhardt}, K. 2014, \apjl, 782, L3

\bibitem[{{Cimatti} {et~al.}(2008){Cimatti}, {Cassata}, {Pozzetti}, {Kurk},
  {Mignoli}, {Renzini}, {Daddi}, {Bolzonella}, {Brusa}, {Rodighiero},
  {Dickinson}, {Franceschini}, {Zamorani}, {Berta}, {Rosati}, \&
  {Halliday}}]{2008A&A...482...21C}
{Cimatti}, A., {Cassata}, P., {Pozzetti}, L., {et~al.} 2008, \aap, 482, 21

\bibitem[{{Clements} {et~al.}(2014){Clements}, {Braglia}, {Hyde},
  {P{\'e}rez-Fournon}, {Bock}, {Cava}, {Chapman}, {Conley}, {Cooray}, {Farrah},
  {Gonz{\'a}lez Solares}, {Marchetti}, {Marsden}, {Oliver}, {Roseboom},
  {Schulz}, {Smith}, {Vaccari}, {Vieira}, {Viero}, {Wang}, {Wardlow}, {Zemcov},
  \& {de Zotti}}]{2014MNRAS.439.1193C}
{Clements}, D.~L., {Braglia}, F.~G., {Hyde}, A.~K., {et~al.} 2014, \mnras, 439,
  1193

\bibitem[{{Cole} {et~al.}(2001){Cole}, {Norberg}, {Baugh}, {Frenk},
  {Bland-Hawthorn}, {Bridges}, {Cannon}, {Colless}, {Collins}, {Couch},
  {Cross}, {Dalton}, {De Propris}, {Driver}, {Efstathiou}, {Ellis},
  {Glazebrook}, {Jackson}, {Lahav}, {Lewis}, {Lumsden}, {Maddox}, {Madgwick},
  {Peacock}, {Peterson}, {Sutherland}, \& {Taylor}}]{2001MNRAS.326..255C}
{Cole}, S., {Norberg}, P., {Baugh}, C.~M., {et~al.} 2001, \mnras, 326, 255

\bibitem[{{Cowie} {et~al.}(1996){Cowie}, {Songaila}, {Hu}, \&
  {Cohen}}]{1996AJ....112..839C}
{Cowie}, L.~L., {Songaila}, A., {Hu}, E.~M., \& {Cohen}, J.~G. 1996, \aj, 112,
  839

\bibitem[{{Croom} {et~al.}(2005){Croom}, {Boyle}, {Shanks}, {Smith}, {Miller},
  {Outram}, {Loaring}, {Hoyle}, \& {da {\^A}ngela}}]{2005MNRAS.356..415C}
{Croom}, S.~M., {Boyle}, B.~J., {Shanks}, T., {et~al.} 2005, \mnras, 356, 415

\bibitem[{{Croom} {et~al.}(2004){Croom}, {Smith}, {Boyle}, {Shanks}, {Miller},
  {Outram}, \& {Loaring}}]{2004MNRAS.349.1397C}
{Croom}, S.~M., {Smith}, R.~J., {Boyle}, B.~J., {et~al.} 2004, \mnras, 349,
  1397

\bibitem[{{da Cunha} {et~al.}(2015){da Cunha}, {Walter}, {Smail}, {Swinbank},
  {Simpson}, {Decarli}, {Hodge}, {Weiss}, {van der Werf}, {Bertoldi},
  {Chapman}, {Cox}, {Danielson}, {Dannerbauer}, {Greve}, {Ivison}, {Karim}, \&
  {Thomson}}]{2015ApJ...806..110D}
{da Cunha}, E., {Walter}, F., {Smail}, I.~R., {et~al.} 2015, \apj, 806, 110

\bibitem[{{Daddi} {et~al.}(2009){Daddi}, {Dannerbauer}, {Stern}, {Dickinson},
  {Morrison}, {Elbaz}, {Giavalisco}, {Mancini}, {Pope}, \&
  {Spinrad}}]{2009ApJ...694.1517D}
{Daddi}, E., {Dannerbauer}, H., {Stern}, D., {et~al.} 2009, \apj, 694, 1517

\bibitem[{{De Lucia} {et~al.}(2006){De Lucia}, {Springel}, {White}, {Croton},
  \& {Kauffmann}}]{2006MNRAS.366..499D}
{De Lucia}, G., {Springel}, V., {White}, S. D.~M., {Croton}, D., \&
  {Kauffmann}, G. 2006, \mnras, 366, 499

\bibitem[{{Draine} {et~al.}(2014){Draine}, {Aniano}, {Krause}, {Groves},
  {Sandstrom}, {Braun}, {Leroy}, {Klaas}, {Linz}, {Rix}, {Schinnerer},
  {Schmiedeke}, \& {Walter}}]{2014ApJ...780..172D}
{Draine}, B.~T., {Aniano}, G., {Krause}, O., {et~al.} 2014, \apj, 780, 172

\bibitem[{{Dressler}(1980)}]{1980ApJ...236..351D}
{Dressler}, A. 1980, \apj, 236, 351

\bibitem[{{Duncan} {et~al.}(2021){Duncan}, {Kondapally}, {Brown}, {Bonato},
  {Best}, {R{\"o}ttgering}, {Bondi}, {Bowler}, {Cochrane}, {G{\"u}rkan},
  {Hardcastle}, {Jarvis}, {Kunert-Bajraszewska}, {Leslie}, {Ma{\l}ek},
  {Morabito}, {O'Sullivan}, {Prandoni}, {Sabater}, {Shimwell}, {Smith}, {Wang},
  {Wo{\l}owska}, \& {Tasse}}]{2021A&A...648A...4D}
{Duncan}, K.~J., {Kondapally}, R., {Brown}, M.~J.~I., {et~al.} 2021, \aap, 648,
  A4

\bibitem[{{Efstathiou} {et~al.}(2013){Efstathiou}, {Christopher}, {Verma}, \&
  {Siebenmorgen}}]{2013MNRAS.436.1873E}
{Efstathiou}, A., {Christopher}, N., {Verma}, A., \& {Siebenmorgen}, R. 2013,
  \mnras, 436, 1873

\bibitem[{{Efstathiou} \& {Rowan-Robinson}(1995)}]{1995MNRAS.273..649E}
{Efstathiou}, A. \& {Rowan-Robinson}, M. 1995, \mnras, 273, 649

\bibitem[{{Efstathiou} \& {Rowan-Robinson}(2003)}]{2003MNRAS.343..322E}
{Efstathiou}, A. \& {Rowan-Robinson}, M. 2003, \mnras, 343, 322

\bibitem[{{Efstathiou} {et~al.}(2000){Efstathiou}, {Rowan-Robinson}, \&
  {Siebenmorgen}}]{2000MNRAS.313..734E}
{Efstathiou}, A., {Rowan-Robinson}, M., \& {Siebenmorgen}, R. 2000, \mnras,
  313, 734

\bibitem[{{Eftekharzadeh} {et~al.}(2015){Eftekharzadeh}, {Myers}, {White},
  {Weinberg}, {Schneider}, {Shen}, {Font-Ribera}, {Ross}, {Paris}, \&
  {Streblyanska}}]{2015MNRAS.453.2779E}
{Eftekharzadeh}, S., {Myers}, A.~D., {White}, M., {et~al.} 2015, \mnras, 453,
  2779

\bibitem[{{Elbaz} {et~al.}(2007){Elbaz}, {Daddi}, {Le Borgne}, {Dickinson},
  {Alexander}, {Chary}, {Starck}, {Brandt}, {Kitzbichler}, {MacDonald},
  {Nonino}, {Popesso}, {Stern}, \& {Vanzella}}]{2007A&A...468...33E}
{Elbaz}, D., {Daddi}, E., {Le Borgne}, D., {et~al.} 2007, \aap, 468, 33

\bibitem[{{Fanidakis} {et~al.}(2013){Fanidakis}, {Macci{\`o}}, {Baugh},
  {Lacey}, \& {Frenk}}]{2013MNRAS.436..315F}
{Fanidakis}, N., {Macci{\`o}}, A.~V., {Baugh}, C.~M., {Lacey}, C.~G., \&
  {Frenk}, C.~S. 2013, \mnras, 436, 315

\bibitem[{{Farrah} {et~al.}(2004){Farrah}, {Geach}, {Fox}, {Serjeant},
  {Oliver}, {Verma}, {Kaviani}, \& {Rowan-Robinson}}]{2004MNRAS.349..518F}
{Farrah}, D., {Geach}, J., {Fox}, M., {et~al.} 2004, \mnras, 349, 518

\bibitem[{{Flores-Cacho} {et~al.}(2016){Flores-Cacho}, {Pierini}, {Soucail},
  {Montier}, {Dole}, {Pointecouteau}, {Pell{\'o}}, {Le Floc'h}, {Nesvadba},
  {Lagache}, {Guery}, \& {Ca{\~n}ameras}}]{2016A&A...585A..54F}
{Flores-Cacho}, I., {Pierini}, D., {Soucail}, G., {et~al.} 2016, \aap, 585, A54

\bibitem[{{Franck} \& {McGaugh}(2016)}]{2016ApJ...833...15F}
{Franck}, J.~R. \& {McGaugh}, S.~S. 2016, \apj, 833, 15

\bibitem[{{Fritz} {et~al.}(2006){Fritz}, {Franceschini}, \&
  {Hatziminaoglou}}]{2006MNRAS.366..767F}
{Fritz}, J., {Franceschini}, A., \& {Hatziminaoglou}, E. 2006, \mnras, 366, 767

\bibitem[{{Gao} {et~al.}(2021){Gao}, {Wang}, {Efstathiou}, {Ma{\l}ek}, {Best},
  {Bonato}, {Farrah}, {Kondapally}, {McCheyne}, \&
  {R{\"o}ttgering}}]{2021A&A...654A.117G}
{Gao}, F., {Wang}, L., {Efstathiou}, A., {et~al.} 2021, \aap, 654, A117

\bibitem[{{Garc{\'\i}a-Vergara} {et~al.}(2021){Garc{\'\i}a-Vergara}, {Rybak},
  {Hodge}, {Hennawi}, {Decarli}, {Gonz{\'a}lez-L{\'o}pez}, {Arrigoni-Battaia},
  {Aravena}, \& {Farina}}]{2021arXiv210909754G}
{Garc{\'\i}a-Vergara}, C., {Rybak}, M., {Hodge}, J., {et~al.} 2021, arXiv
  e-prints, arXiv:2109.09754

\bibitem[{{Geach} {et~al.}(2017){Geach}, {Dunlop}, {Halpern}, {Smail}, {van der
  Werf}, {Alexander}, {Almaini}, {Aretxaga}, {Arumugam}, {Asboth}, {Banerji},
  {Beanlands}, {Best}, {Blain}, {Birkinshaw}, {Chapin}, {Chapman}, {Chen},
  {Chrysostomou}, {Clarke}, {Clements}, {Conselice}, {Coppin}, {Cowley},
  {Danielson}, {Eales}, {Edge}, {Farrah}, {Gibb}, {Harrison}, {Hine}, {Hughes},
  {Ivison}, {Jarvis}, {Jenness}, {Jones}, {Karim}, {Koprowski}, {Knudsen},
  {Lacey}, {Mackenzie}, {Marsden}, {McAlpine}, {McMahon}, {Meijerink},
  {Micha{\l}owski}, {Oliver}, {Page}, {Peacock}, {Rigopoulou}, {Robson},
  {Roseboom}, {Rotermund}, {Scott}, {Serjeant}, {Simpson}, {Simpson}, {Smith},
  {Spaans}, {Stanley}, {Stevens}, {Swinbank}, {Targett}, {Thomson}, {Valiante},
  {Wake}, {Webb}, {Willott}, {Zavala}, \& {Zemcov}}]{2017MNRAS.465.1789G}
{Geach}, J.~E., {Dunlop}, J.~S., {Halpern}, M., {et~al.} 2017, \mnras, 465,
  1789

\bibitem[{{G{\'o}mez} {et~al.}(2003){G{\'o}mez}, {Nichol}, {Miller}, {Balogh},
  {Goto}, {Zabludoff}, {Romer}, {Bernardi}, {Sheth}, {Hopkins}, {Castander},
  {Connolly}, {Schneider}, {Brinkmann}, {Lamb}, {SubbaRao}, \&
  {York}}]{2003ApJ...584..210G}
{G{\'o}mez}, P.~L., {Nichol}, R.~C., {Miller}, C.~J., {et~al.} 2003, \apj, 584,
  210

\bibitem[{{Greenslade} {et~al.}(2018){Greenslade}, {Clements}, {Cheng}, {De
  Zotti}, {Scott}, {Valiante}, {Eales}, {Bremer}, {Dannerbauer}, {Birkinshaw},
  {Farrah}, {Harrison}, {Micha{\l}owski}, {Valtchanov}, {Oteo}, {Baes},
  {Cooray}, {Negrello}, {Wang}, {van der Werf}, {Dunne}, \&
  {Dye}}]{2018MNRAS.476.3336G}
{Greenslade}, J., {Clements}, D.~L., {Cheng}, T., {et~al.} 2018, \mnras, 476,
  3336

\bibitem[{{Guo} {et~al.}(2010){Guo}, {White}, {Li}, \&
  {Boylan-Kolchin}}]{2010MNRAS.404.1111G}
{Guo}, Q., {White}, S., {Li}, C., \& {Boylan-Kolchin}, M. 2010, \mnras, 404,
  1111

\bibitem[{{Hayashi} {et~al.}(2012){Hayashi}, {Kodama}, {Tadaki}, {Koyama}, \&
  {Tanaka}}]{2012ApJ...757...15H}
{Hayashi}, M., {Kodama}, T., {Tadaki}, K.-i., {Koyama}, Y., \& {Tanaka}, I.
  2012, \apj, 757, 15

\bibitem[{{Hickox} {et~al.}(2012){Hickox}, {Wardlow}, {Smail}, {Myers},
  {Alexander}, {Swinbank}, {Danielson}, {Stott}, {Chapman}, {Coppin}, {Dunlop},
  {Gawiser}, {Lutz}, {van der Werf}, \& {Wei{\ss}}}]{2012MNRAS.421..284H}
{Hickox}, R.~C., {Wardlow}, J.~L., {Smail}, I., {et~al.} 2012, \mnras, 421, 284

\bibitem[{{Hill} {et~al.}(2020){Hill}, {Chapman}, {Scott}, {Apostolovski},
  {Aravena}, {B{\'e}thermin}, {Bradford}, {Canning}, {De Breuck}, {Dong},
  {Gonzalez}, {Greve}, {Hayward}, {Hezaveh}, {Litke}, {Malkan}, {Marrone},
  {Phadke}, {Reuter}, {Rotermund}, {Spilker}, {Vieira}, \&
  {Wei{\ss}}}]{2020MNRAS.495.3124H}
{Hill}, R., {Chapman}, S., {Scott}, D., {et~al.} 2020, \mnras, 495, 3124

\bibitem[{{Hinshaw} {et~al.}(2013){Hinshaw}, {Larson}, {Komatsu}, {Spergel},
  {Bennett}, {Dunkley}, {Nolta}, {Halpern}, {Hill}, {Odegard}, {Page}, {Smith},
  {Weiland}, {Gold}, {Jarosik}, {Kogut}, {Limon}, {Meyer}, {Tucker}, {Wollack},
  \& {Wright}}]{2013ApJS..208...19H}
{Hinshaw}, G., {Larson}, D., {Komatsu}, E., {et~al.} 2013, \apjs, 208, 19

\bibitem[{{Huang} {et~al.}(2021){Huang}, {Matsuhara}, {Goto}, {Santos}, {Ho},
  {Kim}, {Hashimoto}, {Ikeda}, {Oi}, {Malkan}, {Pearson}, {Pollo}, {Serjeant},
  {Shim}, {Miyaji}, {Hwang}, {Durkalec}, {Poliszczuk}, {Greve}, {Pearson},
  {Toba}, {Lee}, {Kim}, {Toft}, {Jeong}, \& {Enokidani}}]{2021MNRAS.506.6063H}
{Huang}, T.-C., {Matsuhara}, H., {Goto}, T., {et~al.} 2021, \mnras, 506, 6063

\bibitem[{{Hurley} {et~al.}(2017){Hurley}, {Oliver}, {Betancourt}, {Clarke},
  {Cowley}, {Duivenvoorden}, {Farrah}, {Griffin}, {Lacey}, {Le Floc'h},
  {Papadopoulos}, {Sargent}, {Scudder}, {Vaccari}, {Valtchanov}, \&
  {Wang}}]{2017MNRAS.464..885H}
{Hurley}, P.~D., {Oliver}, S., {Betancourt}, M., {et~al.} 2017, \mnras, 464,
  885

\bibitem[{{Ilbert} {et~al.}(2010){Ilbert}, {Salvato}, {Le Floc'h}, {Aussel},
  {Capak}, {McCracken}, {Mobasher}, {Kartaltepe}, {Scoville}, {Sanders},
  {Arnouts}, {Bundy}, {Cassata}, {Kneib}, {Koekemoer}, {Le F{\`e}vre}, {Lilly},
  {Surace}, {Taniguchi}, {Tasca}, {Thompson}, {Tresse}, {Zamojski}, {Zamorani},
  \& {Zucca}}]{2010ApJ...709..644I}
{Ilbert}, O., {Salvato}, M., {Le Floc'h}, E., {et~al.} 2010, \apj, 709, 644

\bibitem[{{Jones} {et~al.}(2014){Jones}, {Blain}, {Stern}, {Assef}, {Bridge},
  {Eisenhardt}, {Petty}, {Wu}, {Tsai}, {Cutri}, {Wright}, \&
  {Yan}}]{2014MNRAS.443..146J}
{Jones}, S.~F., {Blain}, A.~W., {Stern}, D., {et~al.} 2014, \mnras, 443, 146

\bibitem[{{Kauffmann} \& {Charlot}(1998)}]{1998MNRAS.297L..23K}
{Kauffmann}, G. \& {Charlot}, S. 1998, \mnras, 297, L23

\bibitem[{{Kauffmann} {et~al.}(2004){Kauffmann}, {White}, {Heckman},
  {M{\'e}nard}, {Brinchmann}, {Charlot}, {Tremonti}, \&
  {Brinkmann}}]{2004MNRAS.353..713K}
{Kauffmann}, G., {White}, S. D.~M., {Heckman}, T.~M., {et~al.} 2004, \mnras,
  353, 713

\bibitem[{{Klypin} {et~al.}(2011){Klypin}, {Trujillo-Gomez}, \&
  {Primack}}]{2011ApJ...740..102K}
{Klypin}, A.~A., {Trujillo-Gomez}, S., \& {Primack}, J. 2011, \apj, 740, 102

\bibitem[{{Kondapally} {et~al.}(2021){Kondapally}, {Best}, {Hardcastle},
  {Nisbet}, {Bonato}, {Sabater}, {Duncan}, {McCheyne}, {Cochrane}, {Bowler},
  {Williams}, {Shimwell}, {Tasse}, {Croston}, {Goyal}, {Jamrozy}, {Jarvis},
  {Mahatma}, {R{\"o}ttgering}, {Smith}, {Wo{\l}owska}, {Bondi}, {Brienza},
  {Brown}, {Br{\"u}ggen}, {Chambers}, {Garrett}, {G{\"u}rkan}, {Huber},
  {Kunert-Bajraszewska}, {Magnier}, {Mingo}, {Mostert},
  {Nikiel-Wroczy{\'n}ski}, {O'Sullivan}, {Paladino}, {Ploeckinger}, {Prandoni},
  {Rosenthal}, {Schwarz}, {Shulevski}, {Wagenveld}, \&
  {Wang}}]{2021A&A...648A...3K}
{Kondapally}, R., {Best}, P.~N., {Hardcastle}, M.~J., {et~al.} 2021, \aap, 648,
  A3

\bibitem[{{Kubo} {et~al.}(2019){Kubo}, {Toshikawa}, {Kashikawa}, {Chiang},
  {Overzier}, {Uchiyama}, {Clements}, {Alexander}, {Matsuda}, {Kodama}, {Ono},
  {Goto}, {Cheng}, \& {Ito}}]{2019ApJ...887..214K}
{Kubo}, M., {Toshikawa}, J., {Kashikawa}, N., {et~al.} 2019, \apj, 887, 214

\bibitem[{{Kurk} {et~al.}(2001){Kurk}, {Pentericci}, {R{\"o}ttgering}, \&
  {Miley}}]{2001ApSSS.277..543K}
{Kurk}, J.~D., {Pentericci}, L., {R{\"o}ttgering}, H.~J.~A., \& {Miley}, G.~K.
  2001, Astrophysics and Space Science Supplement, 277, 543

\bibitem[{{Laigle} {et~al.}(2016){Laigle}, {McCracken}, {Ilbert}, {Hsieh},
  {Davidzon}, {Capak}, {Hasinger}, {Silverman}, {Pichon}, {Coupon}, {Aussel},
  {Le Borgne}, {Caputi}, {Cassata}, {Chang}, {Civano}, {Dunlop}, {Fynbo},
  {Kartaltepe}, {Koekemoer}, {Le F{\`e}vre}, {Le Floc'h}, {Leauthaud}, {Lilly},
  {Lin}, {Marchesi}, {Milvang-Jensen}, {Salvato}, {Sanders}, {Scoville},
  {Smolcic}, {Stockmann}, {Taniguchi}, {Tasca}, {Toft}, {Vaccari}, \&
  {Zabl}}]{2016ApJS..224...24L}
{Laigle}, C., {McCracken}, H.~J., {Ilbert}, O., {et~al.} 2016, \apjs, 224, 24

\bibitem[{{Laporte} {et~al.}(2015){Laporte}, {P{\'e}rez-Fournon}, {Calanog},
  {Cooray}, {Wardlow}, {Bock}, {Bridge}, {Burgarella}, {Bussmann},
  {Cabrera-Lavers}, {Casey}, {Clements}, {Conley}, {Dannerbauer}, {Farrah},
  {Fu}, {Gavazzi}, {Gonz{\'a}lez-Solares}, {Ivison}, {Lo Faro}, {Ma}, {Magdis},
  {Marques-Chaves}, {Mart{\'\i}nez-Navajas}, {Oliver}, {Osage}, {Riechers},
  {Rigopoulou}, {Scott}, {Streblyanska}, \& {Vieira}}]{2015ApJ...810..130L}
{Laporte}, N., {P{\'e}rez-Fournon}, I., {Calanog}, J.~A., {et~al.} 2015, \apj,
  810, 130

\bibitem[{{Lee} {et~al.}(2014){Lee}, {Dey}, {Hong}, {Reddy}, {Wilson},
  {Jannuzi}, {Inami}, \& {Gonzalez}}]{2014ApJ...796..126L}
{Lee}, K.-S., {Dey}, A., {Hong}, S., {et~al.} 2014, \apj, 796, 126

\bibitem[{{Lemaux} {et~al.}(2020){Lemaux}, {Cucciati}, {Le F{\`e}vre},
  {Zamorani}, {Lubin}, {Hathi}, {Ilbert}, {Pelliccia}, {Amor{\'\i}n},
  {Bardelli}, {Cassata}, {Gal}, {Garilli}, {Guaita}, {Giavalisco}, {Hung},
  {Koekemoer}, {Maccagni}, {Pentericci}, {Ribeiro}, {Schaerer}, {Shen},
  {Talia}, {Tomczak}, {Vanzella}, {Vergani}, \& {Zucca}}]{2020arXiv200903324L}
{Lemaux}, B.~C., {Cucciati}, O., {Le F{\`e}vre}, O., {et~al.} 2020, arXiv
  e-prints, arXiv:2009.03324

\bibitem[{{Lemaux} {et~al.}(2018){Lemaux}, {Le F{\`e}vre}, {Cucciati},
  {Ribeiro}, {Tasca}, {Zamorani}, {Ilbert}, {Thomas}, {Bardelli}, {Cassata},
  {Hathi}, {Pforr}, {Smol{\v{c}}i{\'c}}, {Delvecchio}, {Novak}, {Berta},
  {McCracken}, {Koekemoer}, {Amor{\'\i}n}, {Garilli}, {Maccagni}, {Schaerer},
  \& {Zucca}}]{2018A&A...615A..77L}
{Lemaux}, B.~C., {Le F{\`e}vre}, O., {Cucciati}, O., {et~al.} 2018, \aap, 615,
  A77

\bibitem[{{Lewis} {et~al.}(2002){Lewis}, {Balogh}, {De Propris}, {Couch},
  {Bower}, {Offer}, {Bland-Hawthorn}, {Baldry}, {Baugh}, {Bridges}, {Cannon},
  {Cole}, {Colless}, {Collins}, {Cross}, {Dalton}, {Driver}, {Efstathiou},
  {Ellis}, {Frenk}, {Glazebrook}, {Hawkins}, {Jackson}, {Lahav}, {Lumsden},
  {Maddox}, {Madgwick}, {Norberg}, {Peacock}, {Percival}, {Peterson},
  {Sutherland}, \& {Taylor}}]{2002MNRAS.334..673L}
{Lewis}, I., {Balogh}, M., {De Propris}, R., {et~al.} 2002, \mnras, 334, 673

\bibitem[{{Li} {et~al.}(2021){Li}, {Wang}, {Dannerbauer}, {Cai}, {Emonts},
  {Prochaska}, {Battaia}, {Neri}, {Zhang}, {Fan}, {Jin}, {Yoon}, \&
  {Bechtel}}]{2021ApJ...922..236L}
{Li}, Q., {Wang}, R., {Dannerbauer}, H., {et~al.} 2021, \apj, 922, 236

\bibitem[{{Longair} \& {Seldner}(1979)}]{1979MNRAS.189..433L}
{Longair}, M.~S. \& {Seldner}, M. 1979, \mnras, 189, 433

\bibitem[{{Lyke} {et~al.}(2020){Lyke}, {Higley}, {McLane}, {Schurhammer},
  {Myers}, {Ross}, {Dawson}, {Chabanier}, {Martini}, {Busca}, {Mas des
  Bourboux}, {Salvato}, {Streblyanska}, {Zarrouk}, {Burtin}, {Anderson},
  {Bautista}, {Bizyaev}, {Brandt}, {Brinkmann}, {Brownstein}, {Comparat},
  {Green}, {de la Macorra}, {Mu{\~n}oz Guti{\'e}rrez}, {Hou}, {Newman},
  {Palanque-Delabrouille}, {P{\^a}ris}, {Percival}, {Petitjean}, {Rich},
  {Rossi}, {Schneider}, {Smith}, {Vivek}, \& {Weaver}}]{2020ApJS..250....8L}
{Lyke}, B.~W., {Higley}, A.~N., {McLane}, J.~N., {et~al.} 2020, \apjs, 250, 8

\bibitem[{{Lynden-Bell}(1969)}]{1969Natur.223..690L}
{Lynden-Bell}, D. 1969, \nat, 223, 690

\bibitem[{{MacKenzie} {et~al.}(2017){MacKenzie}, {Scott}, {Bianconi},
  {Clements}, {Dole}, {Flores-Cacho}, {Guery}, {Kneissl}, {Lagache}, {Marleau},
  {Montier}, {Nesvadba}, {Pointecouteau}, \& {Soucail}}]{2017MNRAS.468.4006M}
{MacKenzie}, T.~P., {Scott}, D., {Bianconi}, M., {et~al.} 2017, \mnras, 468,
  4006

\bibitem[{{Magnelli} {et~al.}(2012){Magnelli}, {Lutz}, {Santini}, {Saintonge},
  {Berta}, {Albrecht}, {Altieri}, {Andreani}, {Aussel}, {Bertoldi},
  {B{\'e}thermin}, {Bongiovanni}, {Capak}, {Chapman}, {Cepa}, {Cimatti},
  {Cooray}, {Daddi}, {Danielson}, {Dannerbauer}, {Dunlop}, {Elbaz}, {Farrah},
  {F{\"o}rster Schreiber}, {Genzel}, {Hwang}, {Ibar}, {Ivison}, {Le Floc'h},
  {Magdis}, {Maiolino}, {Nordon}, {Oliver}, {P{\'e}rez Garc{\'\i}a},
  {Poglitsch}, {Popesso}, {Pozzi}, {Riguccini}, {Rodighiero}, {Rosario},
  {Roseboom}, {Salvato}, {Sanchez-Portal}, {Scott}, {Smail}, {Sturm},
  {Swinbank}, {Tacconi}, {Valtchanov}, {Wang}, \&
  {Wuyts}}]{2012A&A...539A.155M}
{Magnelli}, B., {Lutz}, D., {Santini}, P., {et~al.} 2012, \aap, 539, A155

\bibitem[{{Martinache} {et~al.}(2018){Martinache}, {Rettura}, {Dole},
  {Lehnert}, {Frye}, {Altieri}, {Beelen}, {B{\'e}thermin}, {Le Floc'h},
  {Giard}, {Hurier}, {Lagache}, {Montier}, {Omont}, {Pointecouteau},
  {Polletta}, {Puget}, {Scott}, {Soucail}, \& {Welikala}}]{2018A&A...620A.198M}
{Martinache}, C., {Rettura}, A., {Dole}, H., {et~al.} 2018, \aap, 620, A198

\bibitem[{{Miller} {et~al.}(2018){Miller}, {Chapman}, {Aravena}, {Ashby},
  {Hayward}, {Vieira}, {Wei{\ss}}, {Babul}, {B{\'e}thermin}, {Bradford},
  {Brodwin}, {Carlstrom}, {Chen}, {Cunningham}, {De Breuck}, {Gonzalez},
  {Greve}, {Harnett}, {Hezaveh}, {Lacaille}, {Litke}, {Ma}, {Malkan},
  {Marrone}, {Morningstar}, {Murphy}, {Narayanan}, {Pass}, {Perry}, {Phadke},
  {Rennehan}, {Rotermund}, {Simpson}, {Spilker}, {Sreevani}, {Stark},
  {Strandet}, \& {Strom}}]{2018Natur.556..469M}
{Miller}, T.~B., {Chapman}, S.~C., {Aravena}, M., {et~al.} 2018, \nat, 556, 469

\bibitem[{{Miller} {et~al.}(2015){Miller}, {Hayward}, {Chapman}, \&
  {Behroozi}}]{2015MNRAS.452..878M}
{Miller}, T.~B., {Hayward}, C.~C., {Chapman}, S.~C., \& {Behroozi}, P.~S. 2015,
  \mnras, 452, 878

\bibitem[{{Moster} {et~al.}(2010){Moster}, {Somerville}, {Maulbetsch}, {van den
  Bosch}, {Macci{\`o}}, {Naab}, \& {Oser}}]{2010ApJ...710..903M}
{Moster}, B.~P., {Somerville}, R.~S., {Maulbetsch}, C., {et~al.} 2010, \apj,
  710, 903

\bibitem[{{Myers} {et~al.}(2007){Myers}, {Brunner}, {Nichol}, {Richards},
  {Schneider}, \& {Bahcall}}]{2007ApJ...658...85M}
{Myers}, A.~D., {Brunner}, R.~J., {Nichol}, R.~C., {et~al.} 2007, \apj, 658, 85

\bibitem[{{Myers} {et~al.}(2006){Myers}, {Brunner}, {Richards}, {Nichol},
  {Schneider}, {Vanden Berk}, {Scranton}, {Gray}, \&
  {Brinkmann}}]{2006ApJ...638..622M}
{Myers}, A.~D., {Brunner}, R.~J., {Richards}, G.~T., {et~al.} 2006, \apj, 638,
  622

\bibitem[{{Negrello} {et~al.}(2005){Negrello}, {Gonz{\'a}lez-Nuevo},
  {Magliocchetti}, {Moscardini}, {De Zotti}, {Toffolatti}, \&
  {Danese}}]{2005MNRAS.358..869N}
{Negrello}, M., {Gonz{\'a}lez-Nuevo}, J., {Magliocchetti}, M., {et~al.} 2005,
  \mnras, 358, 869

\bibitem[{{Noll} {et~al.}(2009){Noll}, {Burgarella}, {Giovannoli}, {Buat},
  {Marcillac}, \& {Mu{\~n}oz-Mateos}}]{2009A&A...507.1793N}
{Noll}, S., {Burgarella}, D., {Giovannoli}, E., {et~al.} 2009, \aap, 507, 1793

\bibitem[{{Nowotka} {et~al.}(2022){Nowotka}, {Chen}, {Battaia}, {Fumagalli},
  {Cai}, {Lusso}, {Prochaska}, \& {Yang}}]{2022A&A...658A..77N}
{Nowotka}, M., {Chen}, C.-C., {Battaia}, F.~A., {et~al.} 2022, \aap, 658, A77

\bibitem[{{Overzier}(2016)}]{2016A&ARv..24...14O}
{Overzier}, R.~A. 2016, \aapr, 24, 14

\bibitem[{{Papovich}(2008)}]{2008ApJ...676..206P}
{Papovich}, C. 2008, \apj, 676, 206

\bibitem[{{Peng} {et~al.}(2010){Peng}, {Lilly}, {Kova{\v{c}}}, {Bolzonella},
  {Pozzetti}, {Renzini}, {Zamorani}, {Ilbert}, {Knobel}, {Iovino}, {Maier},
  {Cucciati}, {Tasca}, {Carollo}, {Silverman}, {Kampczyk}, {de Ravel},
  {Sanders}, {Scoville}, {Contini}, {Mainieri}, {Scodeggio}, {Kneib}, {Le
  F{\`e}vre}, {Bardelli}, {Bongiorno}, {Caputi}, {Coppa}, {de la Torre},
  {Franzetti}, {Garilli}, {Lamareille}, {Le Borgne}, {Le Brun}, {Mignoli},
  {Perez Montero}, {Pello}, {Ricciardelli}, {Tanaka}, {Tresse}, {Vergani},
  {Welikala}, {Zucca}, {Oesch}, {Abbas}, {Barnes}, {Bordoloi}, {Bottini},
  {Cappi}, {Cassata}, {Cimatti}, {Fumana}, {Hasinger}, {Koekemoer},
  {Leauthaud}, {Maccagni}, {Marinoni}, {McCracken}, {Memeo}, {Meneux}, {Nair},
  {Porciani}, {Presotto}, \& {Scaramella}}]{2010ApJ...721..193P}
{Peng}, Y.-j., {Lilly}, S.~J., {Kova{\v{c}}}, K., {et~al.} 2010, \apj, 721, 193

\bibitem[{{Pentericci} {et~al.}(2000){Pentericci}, {Kurk}, {R{\"o}ttgering},
  {Miley}, {van Breugel}, {Carilli}, {Ford}, {Heckman}, {McCarthy}, \&
  {Moorwood}}]{2000A&A...361L..25P}
{Pentericci}, L., {Kurk}, J.~D., {R{\"o}ttgering}, H.~J.~A., {et~al.} 2000,
  \aap, 361, L25

\bibitem[{{P{\'e}rez-Gonz{\'a}lez} {et~al.}(2008){P{\'e}rez-Gonz{\'a}lez},
  {Rieke}, {Villar}, {Barro}, {Blaylock}, {Egami}, {Gallego}, {Gil de Paz},
  {Pascual}, {Zamorano}, \& {Donley}}]{2008ApJ...675..234P}
{P{\'e}rez-Gonz{\'a}lez}, P.~G., {Rieke}, G.~H., {Villar}, V., {et~al.} 2008,
  \apj, 675, 234

\bibitem[{{Planck Collaboration} {et~al.}(2016){Planck Collaboration}, {Ade},
  {Aghanim}, {Arnaud}, {Aumont}, {Baccigalupi}, {Banday}, {Barreiro},
  {Bartolo}, {Battaner}, {Benabed}, {Benoit-L{\'e}vy}, {Bernard}, {Bersanelli},
  {Bielewicz}, {Bonaldi}, {Bonavera}, {Bond}, {Borrill}, {Bouchet},
  {Boulanger}, {Burigana}, {Butler}, {Calabrese}, {Catalano}, {Chiang},
  {Christensen}, {Clements}, {Colombo}, {Couchot}, {Coulais}, {Crill}, {Curto},
  {Cuttaia}, {Danese}, {Davies}, {Davis}, {de Bernardis}, {de Rosa}, {de
  Zotti}, {Delabrouille}, {Dickinson}, {Diego}, {Dole}, {Dor{\'e}}, {Douspis},
  {Ducout}, {Dupac}, {Elsner}, {En{\ss}lin}, {Eriksen}, {Falgarone}, {Finelli},
  {Flores-Cacho}, {Frailis}, {Fraisse}, {Franceschi}, {Galeotta}, {Galli},
  {Ganga}, {Giard}, {Giraud-H{\'e}raud}, {Gjerl{\o}w}, {Gonz{\'a}lez-Nuevo},
  {G{\'o}rski}, {Gregorio}, {Gruppuso}, {Gudmundsson}, {Hansen}, {Harrison},
  {Helou}, {Hern{\'a}ndez-Monteagudo}, {Herranz}, {Hildebrandt}, {Hivon},
  {Hobson}, {Hornstrup}, {Hovest}, {Huffenberger}, {Hurier}, {Jaffe}, {Jaffe},
  {Keih{\"a}nen}, {Keskitalo}, {Kisner}, {Kneissl}, {Knoche}, {Kunz},
  {Kurki-Suonio}, {Lagache}, {Lamarre}, {Lasenby}, {Lattanzi}, {Lawrence},
  {Leonardi}, {Levrier}, {Liguori}, {Lilje}, {Linden-V{\o}rnle},
  {L{\'o}pez-Caniego}, {Lubin}, {Mac{\'\i}as-P{\'e}rez}, {Maffei}, {Maggio},
  {Maino}, {Mandolesi}, {Mangilli}, {Maris}, {Martin},
  {Mart{\'\i}nez-Gonz{\'a}lez}, {Masi}, {Matarrese}, {Melchiorri}, {Mennella},
  {Migliaccio}, {Mitra}, {Miville-Desch{\^e}nes}, {Moneti}, {Montier},
  {Morgante}, {Mortlock}, {Munshi}, {Murphy}, {Nati}, {Natoli}, {Nesvadba},
  {Noviello}, {Novikov}, {Novikov}, {Oxborrow}, {Pagano}, {Pajot}, {Paoletti},
  {Partridge}, {Pasian}, {Pearson}, {Perdereau}, {Perotto}, {Pettorino},
  {Piacentini}, {Piat}, {Plaszczynski}, {Pointecouteau}, {Polenta}, {Pratt},
  {Prunet}, {Puget}, {Rachen}, {Reinecke}, {Remazeilles}, {Renault}, {Renzi},
  {Ristorcelli}, {Rocha}, {Rosset}, {Rossetti}, {Roudier},
  {Rubi{\~n}o-Mart{\'\i}n}, {Rusholme}, {Sandri}, {Santos}, {Savelainen},
  {Savini}, {Scott}, {Spencer}, {Stolyarov}, {Stompor}, {Sudiwala}, {Sunyaev},
  {Suur-Uski}, {Sygnet}, {Tauber}, {Terenzi}, {Toffolatti}, {Tomasi},
  {Tristram}, {Tucci}, {T{\"u}rler}, {Umana}, {Valenziano}, {Valiviita}, {Van
  Tent}, {Vielva}, {Villa}, {Wade}, {Wandelt}, {Wehus}, {Welikala}, {Yvon},
  {Zacchei}, \& {Zonca}}]{2016A&A...596A.100P}
{Planck Collaboration}, {Ade}, P.~A.~R., {Aghanim}, N., {et~al.} 2016, \aap,
  596, A100

\bibitem[{{Porciani} {et~al.}(2004){Porciani}, {Magliocchetti}, \&
  {Norberg}}]{2004MNRAS.355.1010P}
{Porciani}, C., {Magliocchetti}, M., \& {Norberg}, P. 2004, \mnras, 355, 1010

\bibitem[{{Pozzetti} {et~al.}(2010){Pozzetti}, {Bolzonella}, {Zucca},
  {Zamorani}, {Lilly}, {Renzini}, {Moresco}, {Mignoli}, {Cassata}, {Tasca},
  {Lamareille}, {Maier}, {Meneux}, {Halliday}, {Oesch}, {Vergani}, {Caputi},
  {Kova{\v{c}}}, {Cimatti}, {Cucciati}, {Iovino}, {Peng}, {Carollo}, {Contini},
  {Kneib}, {Le F{\'e}vre}, {Mainieri}, {Scodeggio}, {Bardelli}, {Bongiorno},
  {Coppa}, {de la Torre}, {de Ravel}, {Franzetti}, {Garilli}, {Kampczyk},
  {Knobel}, {Le Borgne}, {Le Brun}, {Pell{\`o}}, {Perez Montero},
  {Ricciardelli}, {Silverman}, {Tanaka}, {Tresse}, {Abbas}, {Bottini}, {Cappi},
  {Guzzo}, {Koekemoer}, {Leauthaud}, {Maccagni}, {Marinoni}, {McCracken},
  {Memeo}, {Porciani}, {Scaramella}, {Scarlata}, \&
  {Scoville}}]{2010A&A...523A..13P}
{Pozzetti}, L., {Bolzonella}, M., {Zucca}, E., {et~al.} 2010, \aap, 523, A13

\bibitem[{{Riebe} {et~al.}(2013){Riebe}, {Partl}, {Enke}, {Forero-Romero},
  {Gottl{\"o}ber}, {Klypin}, {Lemson}, {Prada}, {Primack}, {Steinmetz}, \&
  {Turchaninov}}]{2013AN....334..691R}
{Riebe}, K., {Partl}, A.~M., {Enke}, H., {et~al.} 2013, Astronomische
  Nachrichten, 334, 691

\bibitem[{{Riechers} {et~al.}(2014){Riechers}, {Carilli}, {Capak}, {Scoville},
  {Smol{\v{c}}i{\'c}}, {Schinnerer}, {Yun}, {Cox}, {Bertoldi}, {Karim}, \&
  {Yan}}]{2014ApJ...796...84R}
{Riechers}, D.~A., {Carilli}, C.~L., {Capak}, P.~L., {et~al.} 2014, \apj, 796,
  84

\bibitem[{{Rodr{\'\i}guez-Puebla} {et~al.}(2015){Rodr{\'\i}guez-Puebla},
  {Avila-Reese}, {Yang}, {Foucaud}, {Drory}, \& {Jing}}]{2015ApJ...799..130R}
{Rodr{\'\i}guez-Puebla}, A., {Avila-Reese}, V., {Yang}, X., {et~al.} 2015,
  \apj, 799, 130

\bibitem[{{Rodr{\'\i}guez-Puebla} {et~al.}(2016){Rodr{\'\i}guez-Puebla},
  {Behroozi}, {Primack}, {Klypin}, {Lee}, \& {Hellinger}}]{2016MNRAS.462..893R}
{Rodr{\'\i}guez-Puebla}, A., {Behroozi}, P., {Primack}, J., {et~al.} 2016,
  \mnras, 462, 893

\bibitem[{{Ross} {et~al.}(2009){Ross}, {Shen}, {Strauss}, {Vanden Berk},
  {Connolly}, {Richards}, {Schneider}, {Weinberg}, {Hall}, {Bahcall}, \&
  {Brunner}}]{2009ApJ...697.1634R}
{Ross}, N.~P., {Shen}, Y., {Strauss}, M.~A., {et~al.} 2009, \apj, 697, 1634

\bibitem[{{Rotermund} {et~al.}(2021){Rotermund}, {Chapman}, {Phadke}, {Hill},
  {Pass}, {Aravena}, {Ashby}, {Babul}, {B{\'e}thermin}, {Canning}, {de Breuck},
  {Dong}, {Gonzalez}, {Hayward}, {Jarugula}, {Marrone}, {Narayanan}, {Reuter},
  {Scott}, {Spilker}, {Vieira}, {Wang}, \& {Weiss}}]{2021MNRAS.502.1797R}
{Rotermund}, K.~M., {Chapman}, S.~C., {Phadke}, K.~A., {et~al.} 2021, \mnras,
  502, 1797

\bibitem[{{Rowan-Robinson}(2000)}]{2000MNRAS.316..885R}
{Rowan-Robinson}, M. 2000, \mnras, 316, 885

\bibitem[{{Sabater} {et~al.}(2021){Sabater}, {Best}, {Tasse}, {Hardcastle},
  {Shimwell}, {Nisbet}, {Jelic}, {Callingham}, {R{\"o}ttgering}, {Bonato},
  {Bondi}, {Ciardi}, {Cochrane}, {Jarvis}, {Kondapally}, {Koopmans},
  {O'Sullivan}, {Prandoni}, {Schwarz}, {Smith}, {Wang}, {Williams}, \&
  {Zaroubi}}]{2021A&A...648A...2S}
{Sabater}, J., {Best}, P.~N., {Tasse}, C., {et~al.} 2021, \aap, 648, A2

\bibitem[{{Salpeter}(1955)}]{1955ApJ...121..161S}
{Salpeter}, E.~E. 1955, \apj, 121, 161

\bibitem[{{Salpeter}(1964)}]{1964ApJ...140..796S}
{Salpeter}, E.~E. 1964, \apj, 140, 796

\bibitem[{{Santos} {et~al.}(2021){Santos}, {Goto}, {Kim}, {Wang}, {Ho},
  {Hashimoto}, {Huang}, {Lu}, {On}, {Wong}, {Hsiao}, {Pollo}, {Malkan},
  {Miyaji}, {Toba}, {Kilerci-Eser}, {Ma{\l}ek}, {Hwang}, {Jeong}, {Shim},
  {Pearson}, {Poliszczuk}, \& {Chen}}]{2021MNRAS.507.3070S}
{Santos}, D. J.~D., {Goto}, T., {Kim}, S.~J., {et~al.} 2021, \mnras, 507, 3070

\bibitem[{{Scoville} {et~al.}(2013){Scoville}, {Arnouts}, {Aussel}, {Benson},
  {Bongiorno}, {Bundy}, {Calvo}, {Capak}, {Carollo}, {Civano}, {Dunlop},
  {Elvis}, {Faisst}, {Finoguenov}, {Fu}, {Giavalisco}, {Guo}, {Ilbert},
  {Iovino}, {Kajisawa}, {Kartaltepe}, {Leauthaud}, {Le F{\`e}vre}, {LeFloch},
  {Lilly}, {Liu}, {Manohar}, {Massey}, {Masters}, {McCracken}, {Mobasher},
  {Peng}, {Renzini}, {Rhodes}, {Salvato}, {Sanders}, {Sarvestani}, {Scarlata},
  {Schinnerer}, {Sheth}, {Shopbell}, {Smol{\v{c}}i{\'c}}, {Taniguchi},
  {Taylor}, {White}, \& {Yan}}]{2013ApJS..206....3S}
{Scoville}, N., {Arnouts}, S., {Aussel}, H., {et~al.} 2013, \apjs, 206, 3

\bibitem[{{Shen} {et~al.}(2007){Shen}, {Strauss}, {Oguri}, {Hennawi}, {Fan},
  {Richards}, {Hall}, {Gunn}, {Schneider}, {Szalay}, {Thakar}, {Vanden Berk},
  {Anderson}, {Bahcall}, {Connolly}, \& {Knapp}}]{2007AJ....133.2222S}
{Shen}, Y., {Strauss}, M.~A., {Oguri}, M., {et~al.} 2007, \aj, 133, 2222

\bibitem[{{Shimakawa} {et~al.}(2014){Shimakawa}, {Kodama}, {Tadaki}, {Tanaka},
  {Hayashi}, \& {Koyama}}]{2014MNRAS.441L...1S}
{Shimakawa}, R., {Kodama}, T., {Tadaki}, K.~I., {et~al.} 2014, \mnras, 441, L1

\bibitem[{{Shirley} {et~al.}(2019){Shirley}, {Roehlly}, {Hurley}, {Buat},
  {Campos Varillas}, {Duivenvoorden}, {Duncan}, {Efstathiou}, {Farrah},
  {Gonz{\'a}lez Solares}, {Malek}, {Marchetti}, {McCheyne}, {Papadopoulos},
  {Pons}, {Scipioni}, {Vaccari}, \& {Oliver}}]{2019MNRAS.490..634S}
{Shirley}, R., {Roehlly}, Y., {Hurley}, P.~D., {et~al.} 2019, \mnras, 490, 634

\bibitem[{{Smail} {et~al.}(2003){Smail}, {Ivison}, {Gilbank}, {Dunlop}, {Keel},
  {Motohara}, \& {Stevens}}]{2003ApJ...583..551S}
{Smail}, I., {Ivison}, R.~J., {Gilbank}, D.~G., {et~al.} 2003, \apj, 583, 551

\bibitem[{{Stach} {et~al.}(2021){Stach}, {Smail}, {Amvrosiadis}, {Swinbank},
  {Dudzevi{\v{c}}i{\={u}}t{\.{e}}}, {Geach}, {Almaini}, {Birkin}, {Chen},
  {Conselice}, {Cooke}, {Coppin}, {Dunlop}, {Farrah}, {Ikarashi}, {Ivison}, \&
  {Wardlow}}]{2021MNRAS.504..172S}
{Stach}, S.~M., {Smail}, I., {Amvrosiadis}, A., {et~al.} 2021, \mnras, 504, 172

\bibitem[{{Stalevski} {et~al.}(2012){Stalevski}, {Fritz}, {Baes}, {Nakos}, \&
  {Popovi{\'c}}}]{2012MNRAS.420.2756S}
{Stalevski}, M., {Fritz}, J., {Baes}, M., {Nakos}, T., \& {Popovi{\'c}},
  L.~{\v{C}}. 2012, \mnras, 420, 2756

\bibitem[{{Steidel} {et~al.}(2005){Steidel}, {Adelberger}, {Shapley}, {Erb},
  {Reddy}, \& {Pettini}}]{2005ApJ...626...44S}
{Steidel}, C.~C., {Adelberger}, K.~L., {Shapley}, A.~E., {et~al.} 2005, \apj,
  626, 44

\bibitem[{{Swinbank} {et~al.}(2014){Swinbank}, {Simpson}, {Smail}, {Harrison},
  {Hodge}, {Karim}, {Walter}, {Alexander}, {Brandt}, {de Breuck}, {da Cunha},
  {Chapman}, {Coppin}, {Danielson}, {Dannerbauer}, {Decarli}, {Greve},
  {Ivison}, {Knudsen}, {Lagos}, {Schinnerer}, {Thomson}, {Wardlow}, {Wei{\ss}},
  \& {van der Werf}}]{2014MNRAS.438.1267S}
{Swinbank}, A.~M., {Simpson}, J.~M., {Smail}, I., {et~al.} 2014, \mnras, 438,
  1267

\bibitem[{{Tasse} {et~al.}(2021){Tasse}, {Shimwell}, {Hardcastle},
  {O'Sullivan}, {van Weeren}, {Best}, {Bester}, {Hugo}, {Smirnov}, {Sabater},
  {Calistro-Rivera}, {de Gasperin}, {Morabito}, {R{\"o}ttgering}, {Williams},
  {Bonato}, {Bondi}, {Botteon}, {Br{\"u}ggen}, {Brunetti}, {Chy{\.z}y},
  {Garrett}, {G{\"u}rkan}, {Jarvis}, {Kondapally}, {Mandal}, {Prandoni},
  {Repetti}, {Retana-Montenegro}, {Schwarz}, {Shulevski}, \&
  {Wiaux}}]{2021A&A...648A...1T}
{Tasse}, C., {Shimwell}, T., {Hardcastle}, M.~J., {et~al.} 2021, \aap, 648, A1

\bibitem[{{Thomas} {et~al.}(2005){Thomas}, {Maraston}, {Bender}, \& {Mendes de
  Oliveira}}]{2005ApJ...621..673T}
{Thomas}, D., {Maraston}, C., {Bender}, R., \& {Mendes de Oliveira}, C. 2005,
  \apj, 621, 673

\bibitem[{{Toshikawa} {et~al.}(2018){Toshikawa}, {Uchiyama}, {Kashikawa},
  {Ouchi}, {Overzier}, {Ono}, {Harikane}, {Ishikawa}, {Kodama}, {Matsuda},
  {Lin}, {Onoue}, {Tanaka}, {Nagao}, {Akiyama}, {Komiyama}, {Goto}, \&
  {Lee}}]{2018PASJ...70S..12T}
{Toshikawa}, J., {Uchiyama}, H., {Kashikawa}, N., {et~al.} 2018, \pasj, 70, S12

\bibitem[{{Trainor} \& {Steidel}(2012)}]{2012ApJ...752...39T}
{Trainor}, R.~F. \& {Steidel}, C.~C. 2012, \apj, 752, 39

\bibitem[{{Tran} {et~al.}(2010){Tran}, {Papovich}, {Saintonge}, {Brodwin},
  {Dunlop}, {Farrah}, {Finkelstein}, {Finkelstein}, {Lotz}, {McLure},
  {Momcheva}, \& {Willmer}}]{2010ApJ...719L.126T}
{Tran}, K.-V.~H., {Papovich}, C., {Saintonge}, A., {et~al.} 2010, \apjl, 719,
  L126

\bibitem[{{Uchiyama} {et~al.}(2018){Uchiyama}, {Toshikawa}, {Kashikawa},
  {Overzier}, {Chiang}, {Marinello}, {Tanaka}, {Niino}, {Ishikawa}, {Onoue},
  {Ichikawa}, {Akiyama}, {Coupon}, {Harikane}, {Imanishi}, {Kodama},
  {Komiyama}, {Lee}, {Lin}, {Miyazaki}, {Nagao}, {Nishizawa}, {Ono}, {Ouchi},
  \& {Wang}}]{2018PASJ...70S..32U}
{Uchiyama}, H., {Toshikawa}, J., {Kashikawa}, N., {et~al.} 2018, \pasj, 70, S32

\bibitem[{{Wang} {et~al.}(2021{\natexlab{a}}){Wang}, {Hill}, {Chapman},
  {Wei{\ss}}, {Scott}, {Apostolovski}, {Aravena}, {Archipley}, {B{\'e}thermin},
  {Canning}, {De Breuck}, {Dong}, {Everett}, {Gonzalez}, {Greve}, {Hayward},
  {Hezaveh}, {Jarugula}, {Marrone}, {Phadke}, {Reuter}, {Rotermund}, {Spilker},
  \& {Vieira}}]{2021MNRAS.508.3754W}
{Wang}, G. C.~P., {Hill}, R., {Chapman}, S.~C., {et~al.} 2021{\natexlab{a}},
  \mnras, 508, 3754

\bibitem[{{Wang} {et~al.}(2013){Wang}, {Farrah}, {Oliver}, {Amblard},
  {B{\'e}thermin}, {Bock}, {Conley}, {Cooray}, {Halpern}, {Heinis}, {Ibar},
  {Ilbert}, {Ivison}, {Marsden}, {Roseboom}, {Rowan-Robinson}, {Schulz},
  {Smith}, {Viero}, \& {Zemcov}}]{2013MNRAS.431..648W}
{Wang}, L., {Farrah}, D., {Oliver}, S.~J., {et~al.} 2013, \mnras, 431, 648

\bibitem[{{Wang} {et~al.}(2021{\natexlab{b}}){Wang}, {Gao}, {Best}, {Duncan},
  {Hardcastle}, {Kondapally}, {Ma{\l}ek}, {McCheyne}, {Sabater}, {Shimwell},
  {Tasse}, {Bonato}, {Bondi}, {Cochrane}, {Farrah}, {G{\"u}rkan}, {Haskell},
  {Pearson}, {Prandoni}, {R{\"o}ttgering}, {Smith}, {Vaccari}, \&
  {Williams}}]{2021A&A...648A...8W}
{Wang}, L., {Gao}, F., {Best}, P.~N., {et~al.} 2021{\natexlab{b}}, \aap, 648,
  A8

\bibitem[{{Wang} {et~al.}(2019){Wang}, {Gao}, {Duncan}, {Williams},
  {Rowan-Robinson}, {Sabater}, {Shimwell}, {Bonato}, {Calistro-Rivera},
  {Chy{\.z}y}, {Farrah}, {G{\"u}rkan}, {Hardcastle}, {McCheyne}, {Prandoni},
  {Read}, {R{\"o}ttgering}, \& {Smith}}]{2019A&A...631A.109W}
{Wang}, L., {Gao}, F., {Duncan}, K.~J., {et~al.} 2019, \aap, 631, A109

\bibitem[{{Wang} {et~al.}(2015){Wang}, {Viero}, {Ross}, {Asboth},
  {B{\'e}thermin}, {Bock}, {Clements}, {Conley}, {Cooray}, {Farrah}, {Hajian},
  {Han}, {Lagache}, {Marsden}, {Myers}, {Norberg}, {Oliver}, {Page},
  {Symeonidis}, {Schulz}, {Wang}, \& {Zemcov}}]{2015MNRAS.449.4476W}
{Wang}, L., {Viero}, M., {Ross}, N.~P., {et~al.} 2015, \mnras, 449, 4476

\bibitem[{{White} {et~al.}(2012){White}, {Myers}, {Ross}, {Schlegel},
  {Hennawi}, {Shen}, {McGreer}, {Strauss}, {Bolton}, {Bovy}, {Fan},
  {Miralda-Escude}, {Palanque-Delabrouille}, {Paris}, {Petitjean}, {Schneider},
  {Viel}, {Weinberg}, {Yeche}, {Zehavi}, {Pan}, {Snedden}, {Bizyaev},
  {Brewington}, {Brinkmann}, {Malanushenko}, {Malanushenko}, {Oravetz},
  {Simmons}, {Sheldon}, \& {Weaver}}]{2012MNRAS.424..933W}
{White}, M., {Myers}, A.~D., {Ross}, N.~P., {et~al.} 2012, \mnras, 424, 933

\bibitem[{{Wilkinson} {et~al.}(2017){Wilkinson}, {Almaini}, {Chen}, {Smail},
  {Arumugam}, {Blain}, {Chapin}, {Chapman}, {Conselice}, {Cowley}, {Dunlop},
  {Farrah}, {Geach}, {Hartley}, {Ivison}, {Maltby}, {Micha{\l}owski},
  {Mortlock}, {Scott}, {Simpson}, {Simpson}, {van der Werf}, \&
  {Wild}}]{2017MNRAS.464.1380W}
{Wilkinson}, A., {Almaini}, O., {Chen}, C.-C., {et~al.} 2017, \mnras, 464, 1380

\bibitem[{{York} {et~al.}(2000){York}, {Adelman}, {Anderson}, {Anderson},
  {Annis}, {Bahcall}, {Bakken}, {Barkhouser}, {Bastian}, {Berman}, {Boroski},
  {Bracker}, {Briegel}, {Briggs}, {Brinkmann}, {Brunner}, {Burles}, {Carey},
  {Carr}, {Castander}, {Chen}, {Colestock}, {Connolly}, {Crocker}, {Csabai},
  {Czarapata}, {Davis}, {Doi}, {Dombeck}, {Eisenstein}, {Ellman}, {Elms},
  {Evans}, {Fan}, {Federwitz}, {Fiscelli}, {Friedman}, {Frieman}, {Fukugita},
  {Gillespie}, {Gunn}, {Gurbani}, {de Haas}, {Haldeman}, {Harris}, {Hayes},
  {Heckman}, {Hennessy}, {Hindsley}, {Holm}, {Holmgren}, {Huang}, {Hull},
  {Husby}, {Ichikawa}, {Ichikawa}, {Ivezi{\'c}}, {Kent}, {Kim}, {Kinney},
  {Klaene}, {Kleinman}, {Kleinman}, {Knapp}, {Korienek}, {Kron}, {Kunszt},
  {Lamb}, {Lee}, {Leger}, {Limmongkol}, {Lindenmeyer}, {Long}, {Loomis},
  {Loveday}, {Lucinio}, {Lupton}, {MacKinnon}, {Mannery}, {Mantsch}, {Margon},
  {McGehee}, {McKay}, {Meiksin}, {Merelli}, {Monet}, {Munn}, {Narayanan},
  {Nash}, {Neilsen}, {Neswold}, {Newberg}, {Nichol}, {Nicinski}, {Nonino},
  {Okada}, {Okamura}, {Ostriker}, {Owen}, {Pauls}, {Peoples}, {Peterson},
  {Petravick}, {Pier}, {Pope}, {Pordes}, {Prosapio}, {Rechenmacher}, {Quinn},
  {Richards}, {Richmond}, {Rivetta}, {Rockosi}, {Ruthmansdorfer}, {Sandford},
  {Schlegel}, {Schneider}, {Sekiguchi}, {Sergey}, {Shimasaku}, {Siegmund},
  {Smee}, {Smith}, {Snedden}, {Stone}, {Stoughton}, {Strauss}, {Stubbs},
  {SubbaRao}, {Szalay}, {Szapudi}, {Szokoly}, {Thakar}, {Tremonti}, {Tucker},
  {Uomoto}, {Vanden Berk}, {Vogeley}, {Waddell}, {Wang}, {Watanabe},
  {Weinberg}, {Yanny}, {Yasuda}, \& {SDSS Collaboration}}]{2000AJ....120.1579Y}
{York}, D.~G., {Adelman}, J., {Anderson}, John~E., J., {et~al.} 2000, \aj, 120,
  1579

\end{thebibliography}

\begin{appendix}
\section{Estimate stellar mass using Random Forest}\label{RF}

To explore if there is any dependence on the method by which stellar masses are estimated, we repeat the analysis on the distribution of stellar mass of neighbours with an alternative method to derive stellar masses for the IRAC-selected sources. We adopt the Random Forest (RF) method to derive stellar mass estimates in EN1 and trained a RF estimator with COSMOS2015 data \citep{2016ApJS..224...24L}. We use the \texttt{RandomForestRegressor} included in Python package \texttt{scikit-learn} to fit the decision trees in the Random Forest, which creates a regression model that can be applied to the EN1 field data. We first selected the COSMOS2015 galaxies to have $>3-\sigma$ detection in \textit{Spitzer} IRAC1 band at 3.6 $\mu$m, which is the same criterion we use to reduce spurious sources in Section \ref{data2}. Galaxies with stellar masses below $10^{7.5} \, \mathrm{M_{\sun}}$ are discarded as they may provide us with incorrect low values. We divided the COSMOS2015 data into three redshift-matched sub-samples: train, test and validation, containing 180 931, 54 240 and 488 281 galaxies respectively. We use seven overlapping bands between COSMOS2015 catalog and EN1 source catalog ($u$ band data from Canada-France Hawaii Telescope; $i, r, z$ and $y$ from SUBARU; 3.6 $\mu$m and 4.5 $\mu$m from \textit{Spitzer}) together with the photometric redshift information, as the features for the regression model.

We use different combinations of values for the parameters max\_depth (maximum depth of the three), max\_features (number of features to be consider for the best split) and n\_extimators (number of trees in the forest), inside \texttt{RandomForestRegressor} to select the best parameters in order to train the data. We start with a random search to narrow down the range and then we use the grid search to select the best parameters. We found that the best values for the \texttt{RandomForestRegressor} were max\_depth = 31, max\_features = log2(n\_features) and n\_extimators = 300. This led to an accuracy of $\sim$ 99.08\% and average error of $\sim$ 0.084 dex for the stellar mass in both the test and validation samples. With the trained regression model, we estimate the stellar masses for 275,672 galaxies in the EN1 field (see Section \ref{data2}). We compare them with our stellar mass estimates using CIGALE SED fitting code, finding a mean difference and standard deviation of $0.41\pm{0.27}$ dex. However, COSMOS2015 adopted the \citet{2003PASP..115..763C} IMF while we use the \citet{1955ApJ...121..161S} IMF in this paper. Considering this difference, the mean difference reduces to $\sim$0.18 dex. We also compare the RF stellar masses of HLIRG neighbours with that of random galaxy neighbours and only find weak excess as discussed in Section \ref{re2}. Thus, we conclude that the lack of significant excess of HLIRG neighbours compared to random galaxy neighbours is not due to the method used to derive stellar masses. 

\section{Spectroscopic follow-up observation of the HLIRG EN1\_19770}\label{ob}

\begin{figure*}
\resizebox{\hsize}{!}{\includegraphics[width=\linewidth]{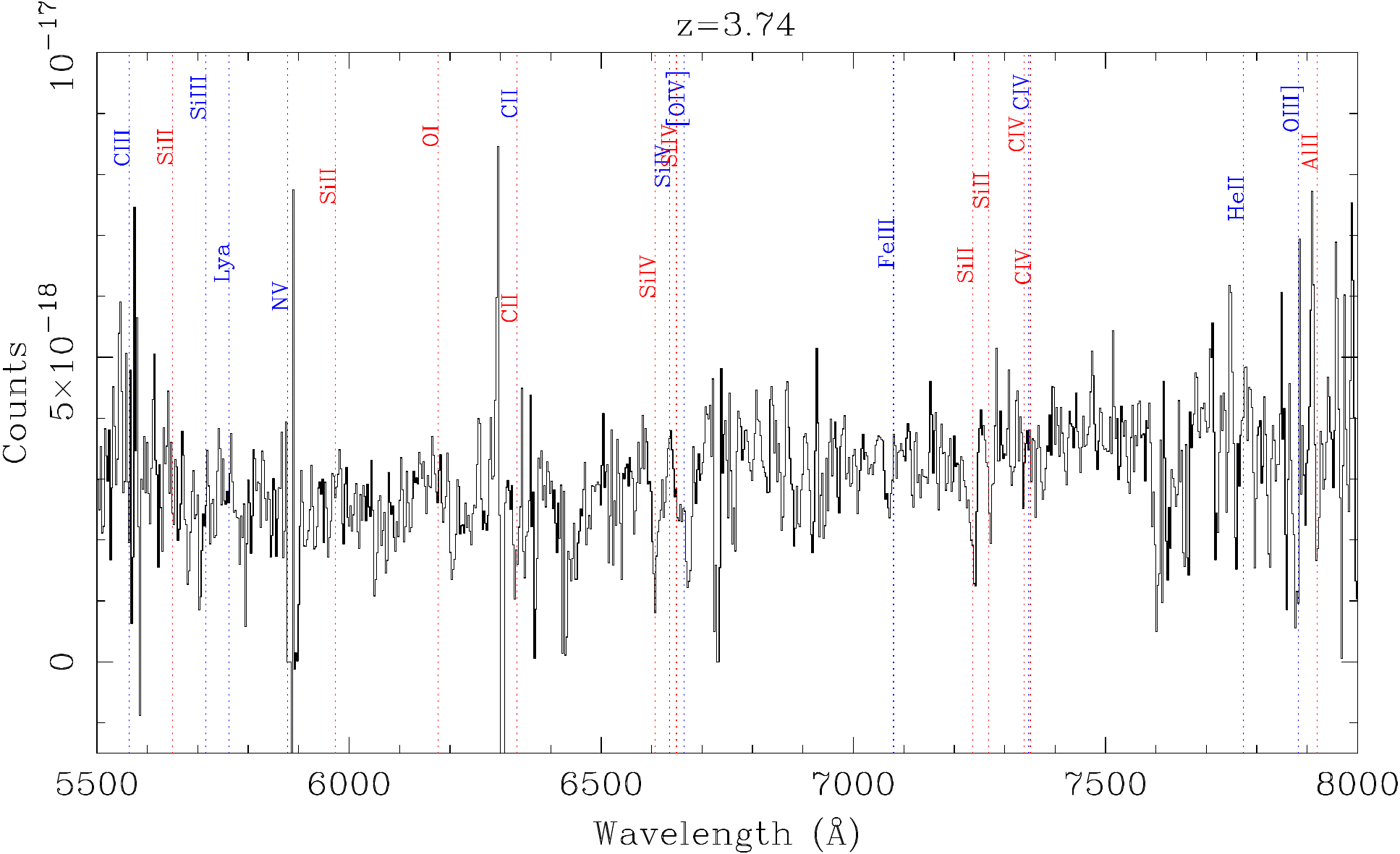}}
\caption{The optical spectrum for EN1\_19770. We retrieve a spectroscopic redshift of 3.74 based on the detection of Si IV and Si II absorption lines.}
\label{spectrum}
\end{figure*}

Follow-up observation of EN1\_19770 was carried out on June 20th of 2021 on the Subaru 8m Telescope using the Faint Object Camera and Spectrograph (FOCAS) instrument in clear sky conditions under long slit mode, the B300 grism and a 0.5" slit. Two 900 second integrations were taken along with associated acquisition and calibration observations. We reduce FOCAS observations for EN1\_19770, a galaxy near the field of view of galaxy LEDA 2514599. We follow the cookbook v1.0.3 of the instrument, where we perform the different steps from the raw images using IRAF version 2.16 and the FOCASRED package. First, we subtract the offset of the images (i.e. bias) from the read-out process of the charge-coupled device (CCD). Second, we correct the intrinsic performance of the CCDs by diving the images with the flat-field taken before the observations. For this step, we use the \texttt{flatnorm} task to normalise the spectrum using an specified region of the average detector image. Then, we correct the distortion pattern on the FOCAS data using the \texttt{distcalib} task. We calibrate the wavelength using the night sky lines for the object and the emission light from a ThAr lamp for the standard star (HZ44). We perform the sky-subtraction using the \texttt{background} task in IRAF by fitting the sky region close to the two galaxies located in the same slit (EN1\_19770 and LEDA 2514599). We calibrate the flux with HZ44 to them combine the different exposures to remove the cosmic rays. Finally, we correct by interstellar extinction and heliocentric velocity and extract the final 1D spectrum, as shown in Figure \ref{spectrum}. We derive a spectroscopic redshift of z = 3.74 based on the detection of Si IV and Si II absorption lines as shown in the figure. This source has a photometric redshift of 3.91 with a minimum value 3.82 in \citet{2021A&A...648A...4D}. Thus, the photometric redshift estimate agrees well with the derived spectroscopic redshift.

\section{Overdensity of the most promising protocluster candidates}

In the left panel of each subfigure in Figure \ref{fig:all}, we present the 250 $\mu$m flux density distribution of neighbours within 100$\arcsec$ of the most promising protocluster candidates. The median value and uncertainty are drawn from 100 realizations. In the right panel of each subfigure, we plot the overdensity parameter calculates as $\frac{N_{HLIRG}-N_{random}}{N_{random}}$ where $N_{HLIRG}$ is the number of neighbours aroud HLIRGs within difference radius and $N_{random}$ is the mean number of neighbours around random positions drawn from 100 random selections. Values above zero indicate an overdensity.

\newpage

\begin{figure*}
\begin{subfigure}{.5\textwidth}
  \centering
  \includegraphics[width=\linewidth]{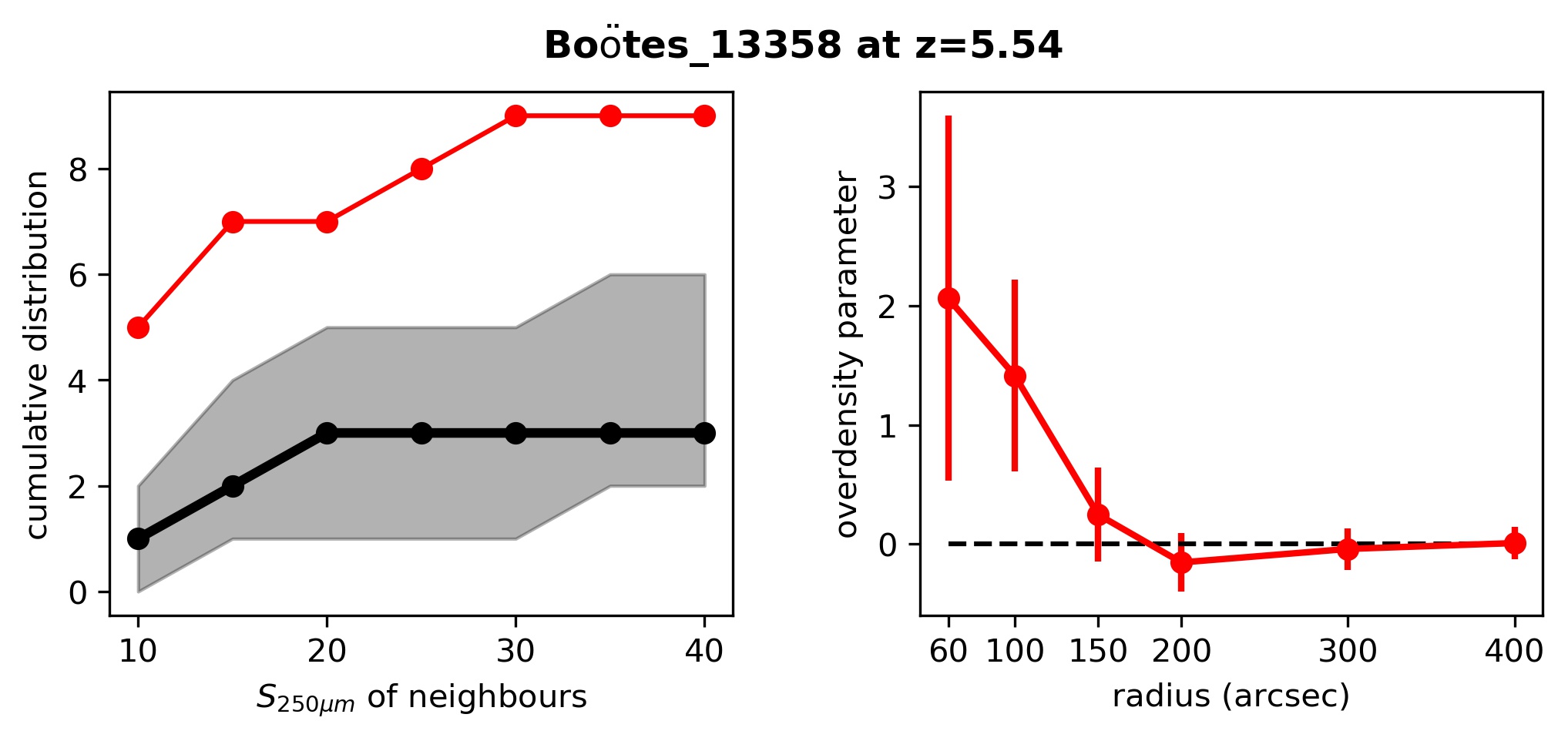}
\end{subfigure}
\begin{subfigure}{.5\textwidth}
  \centering
  \includegraphics[width=\linewidth]{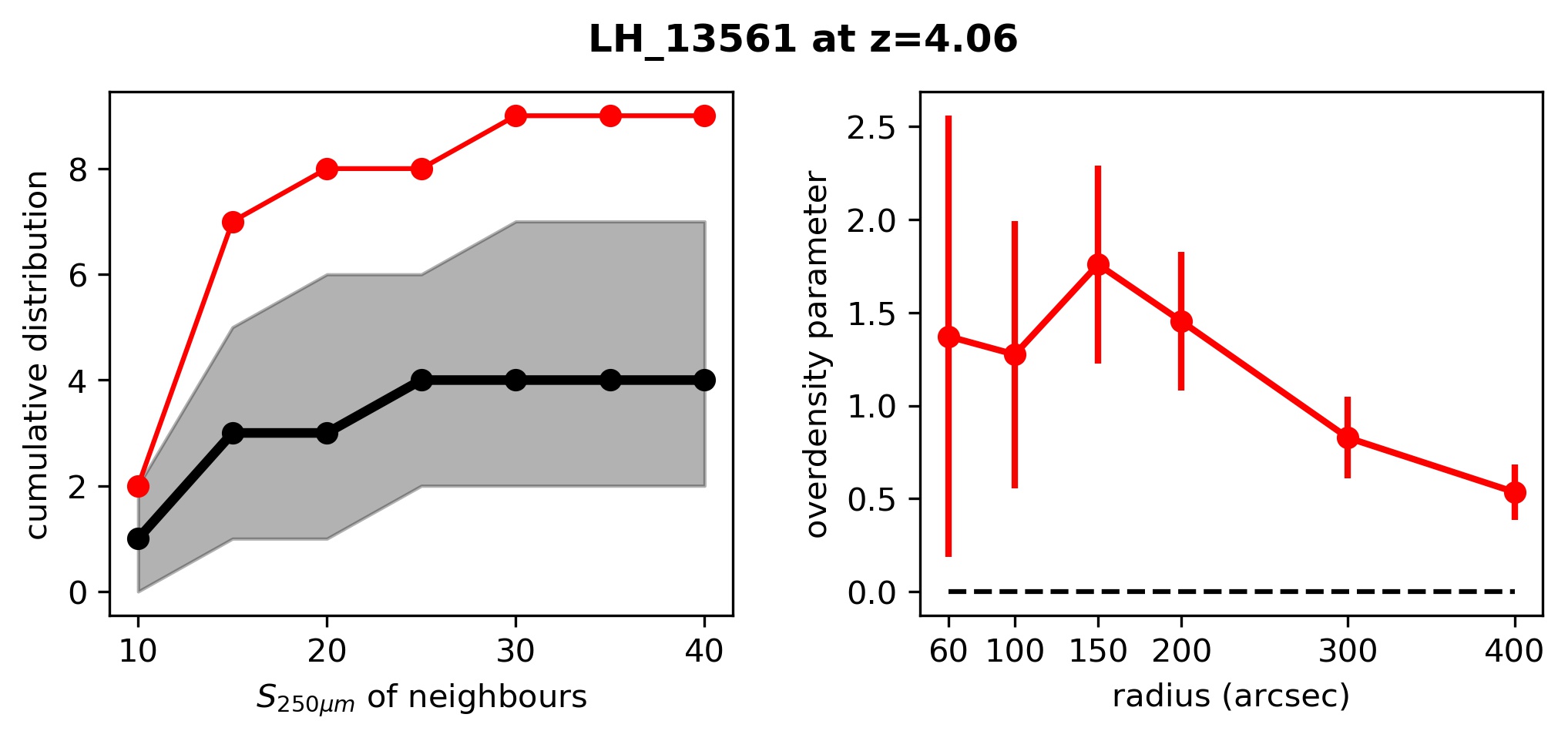}
\end{subfigure}

\begin{subfigure}{.5\textwidth}
  \centering
  \includegraphics[width=\linewidth]{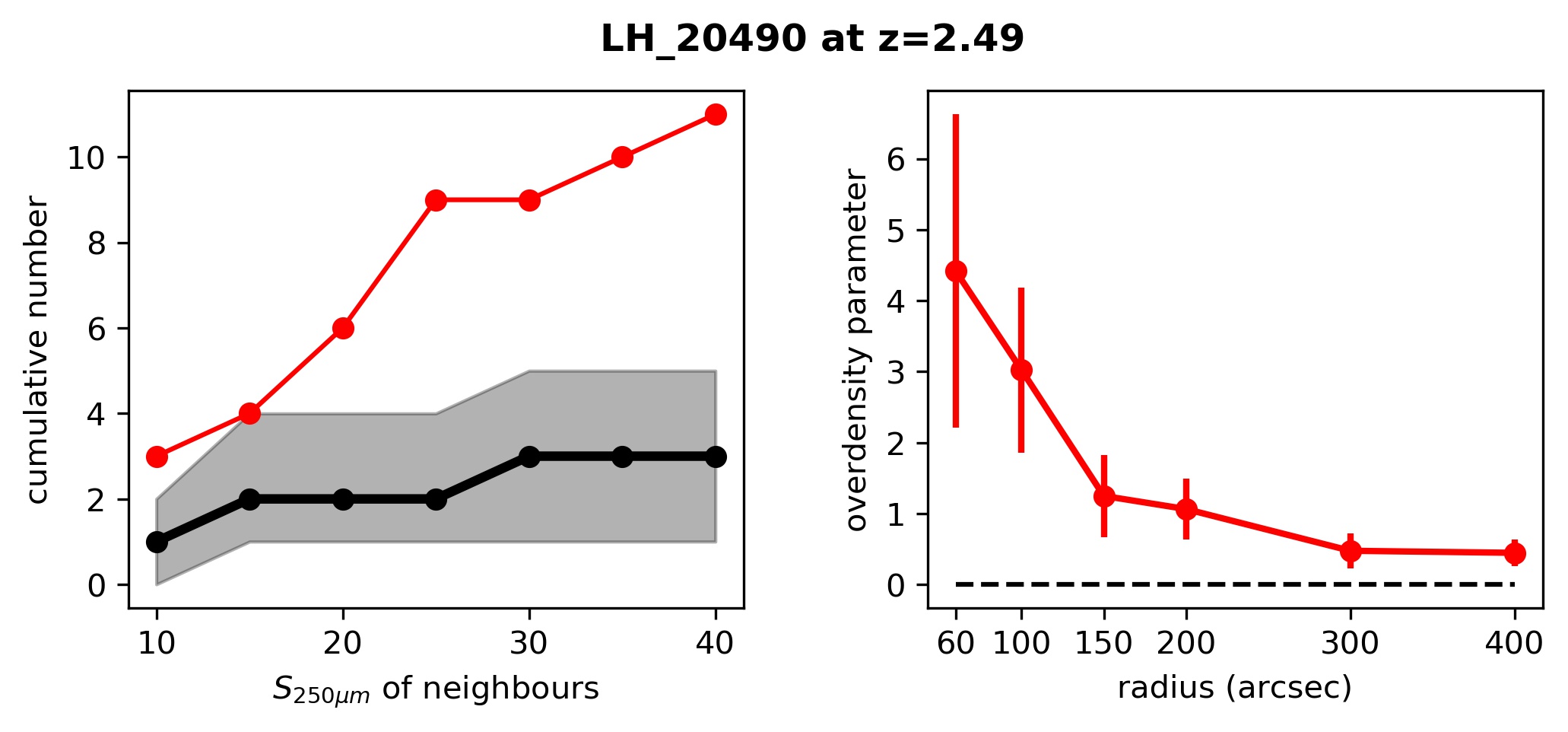}
\end{subfigure}
\begin{subfigure}{.5\textwidth}
  \centering
  \includegraphics[width=\linewidth]{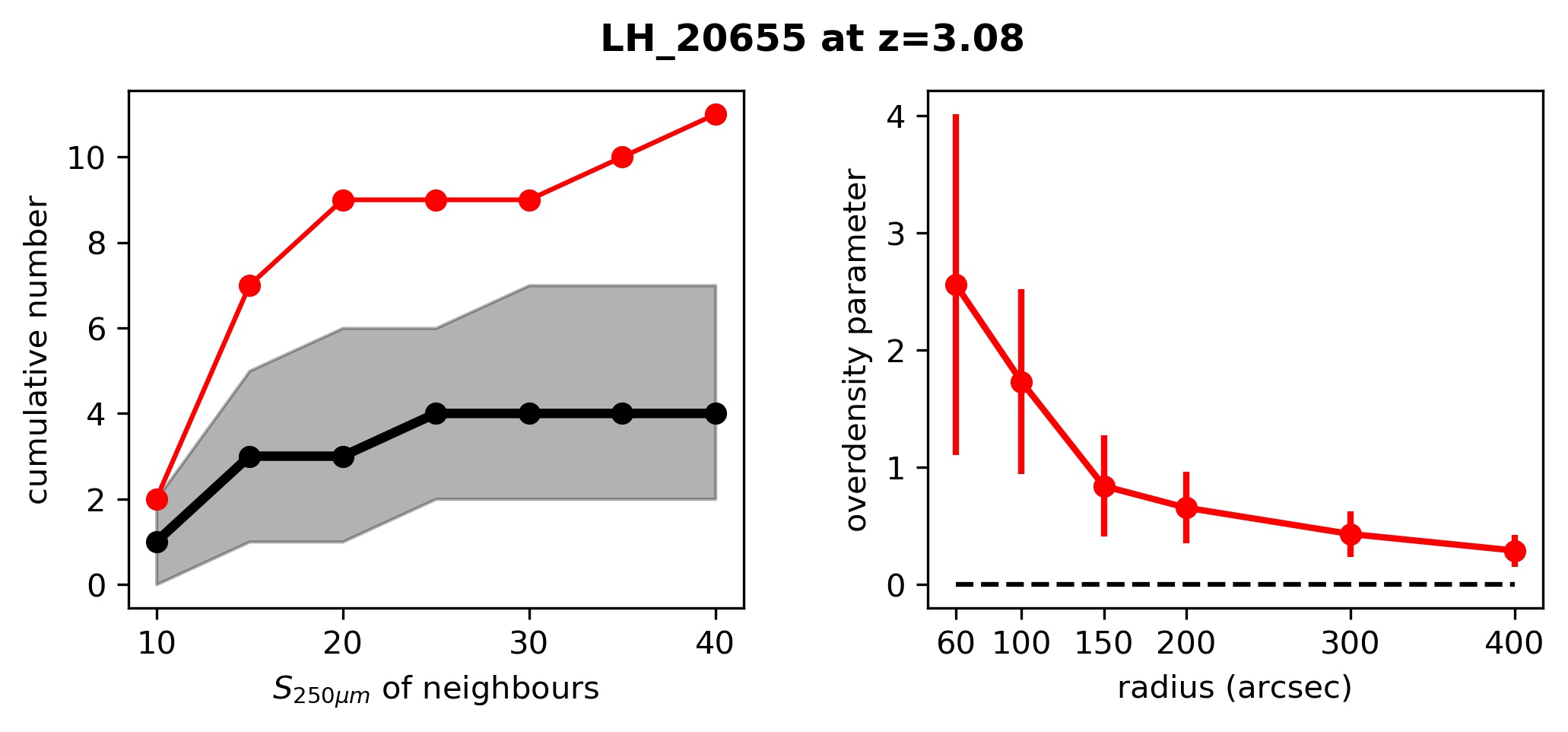}
\end{subfigure}

\begin{subfigure}{.5\textwidth}
  \centering
  \includegraphics[width=\linewidth]{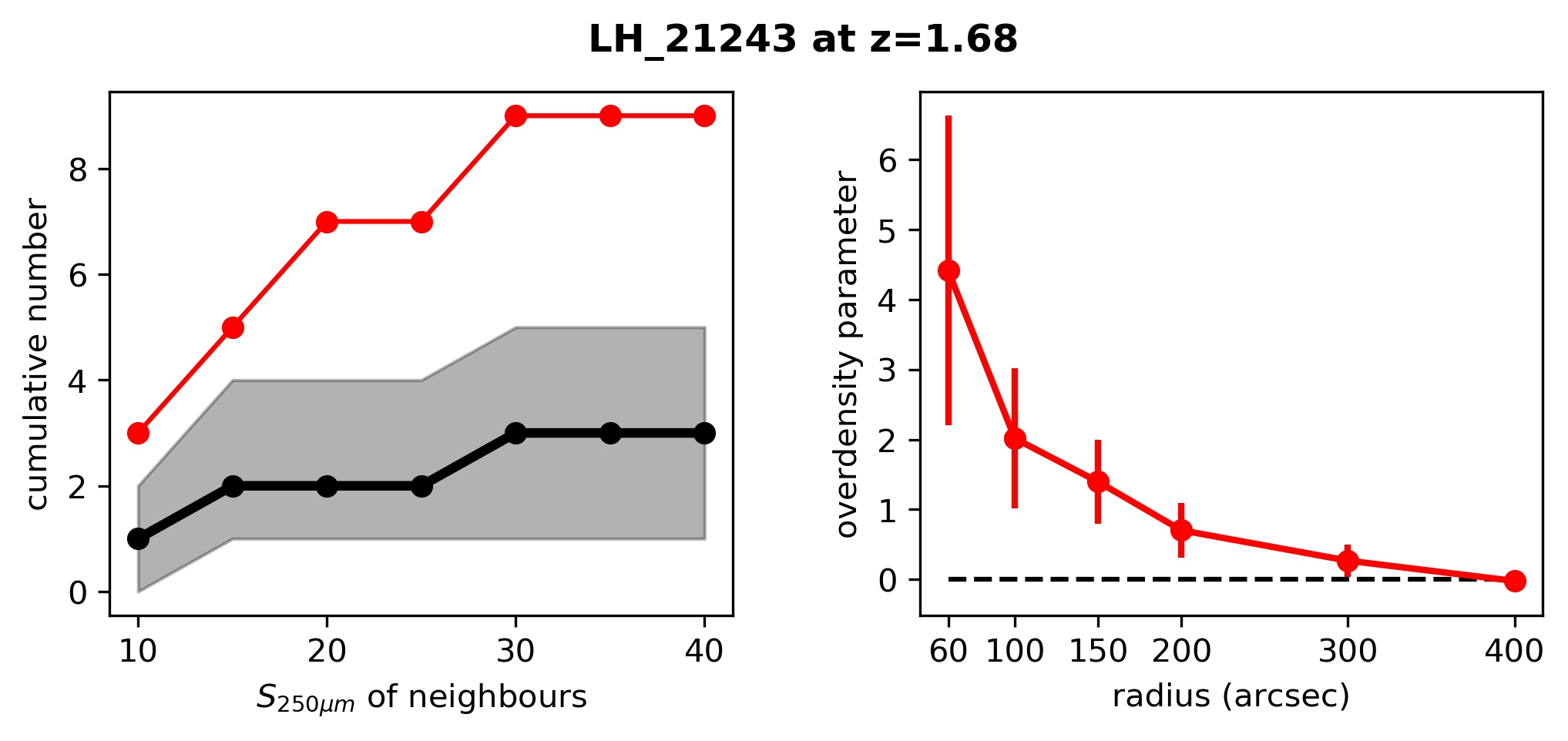}
\end{subfigure}
\begin{subfigure}{.5\textwidth}
  \centering
  \includegraphics[width=\linewidth]{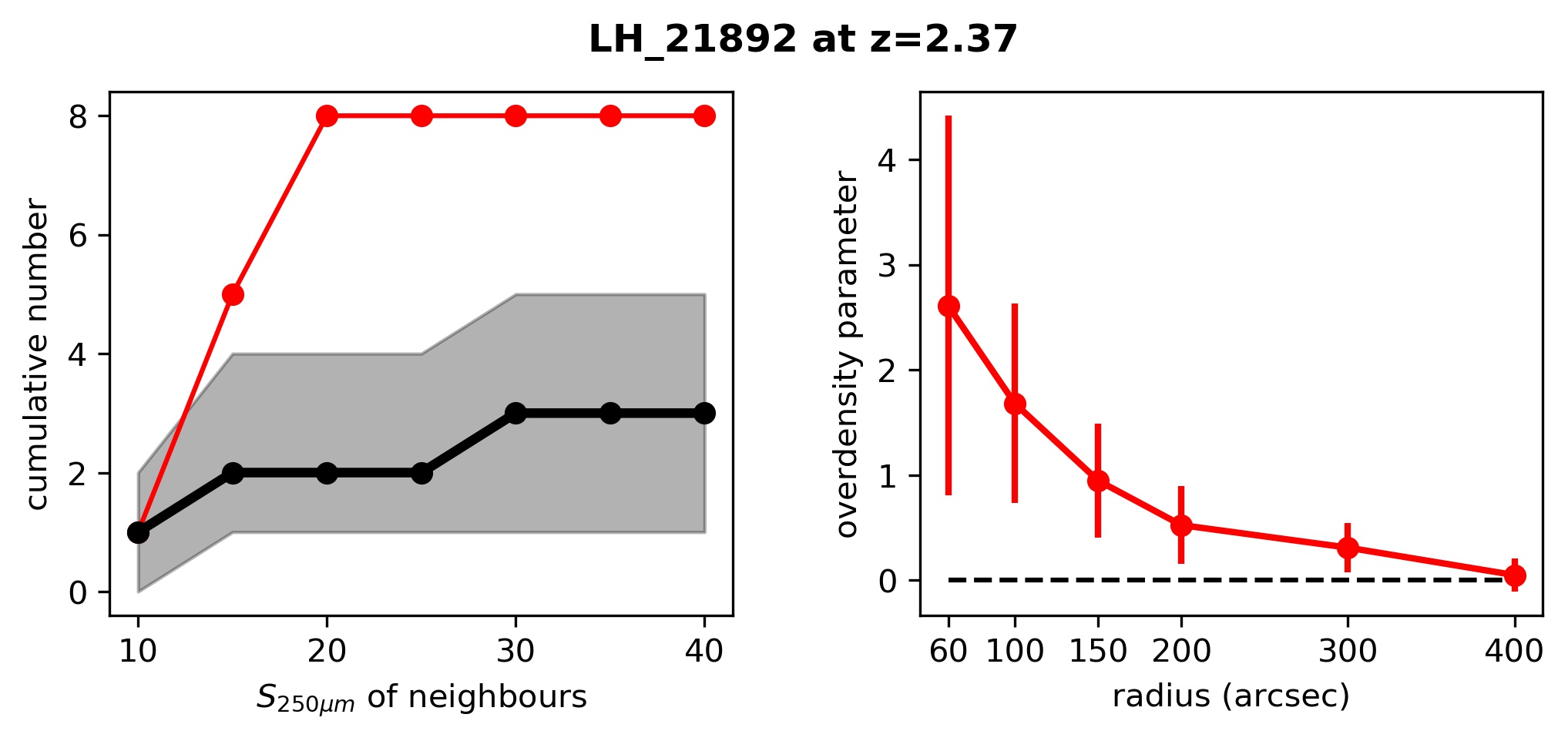}
\end{subfigure}

\begin{subfigure}{.5\textwidth}
  \centering
  \includegraphics[width=\linewidth]{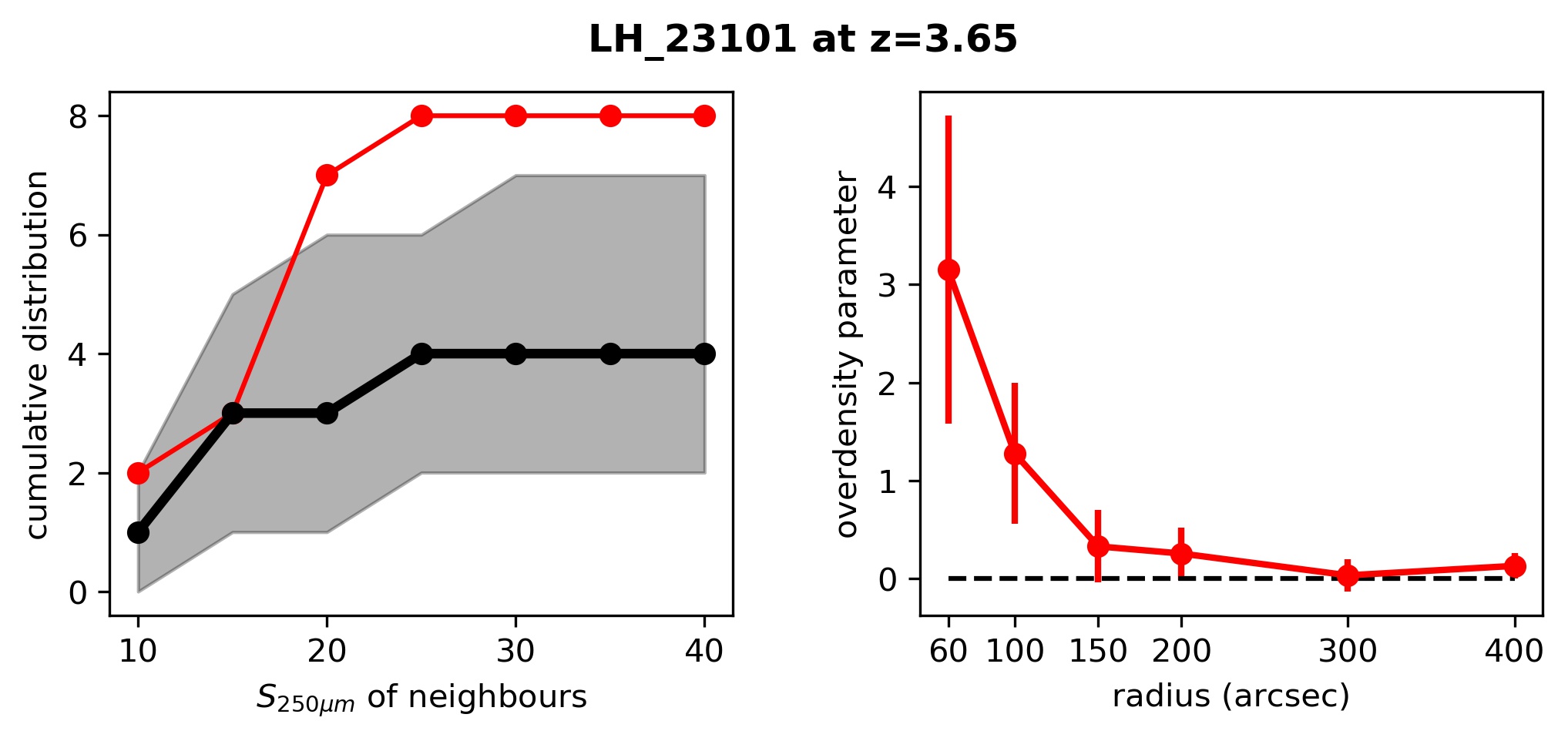}
\end{subfigure}
\begin{subfigure}{.5\textwidth}
  \centering
  \includegraphics[width=\linewidth]{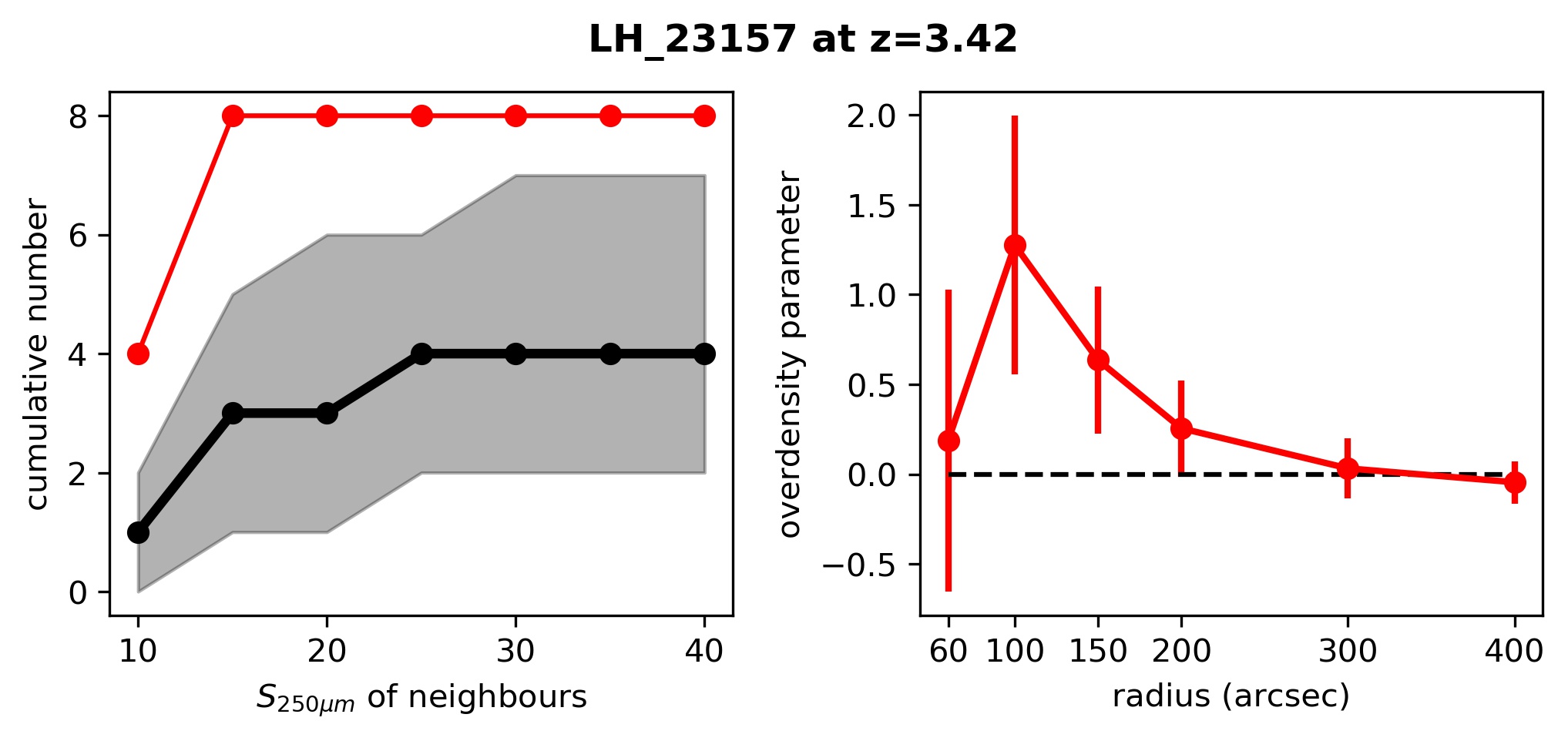}
\end{subfigure}

\begin{subfigure}{.5\textwidth}
  \centering
  \includegraphics[width=\linewidth]{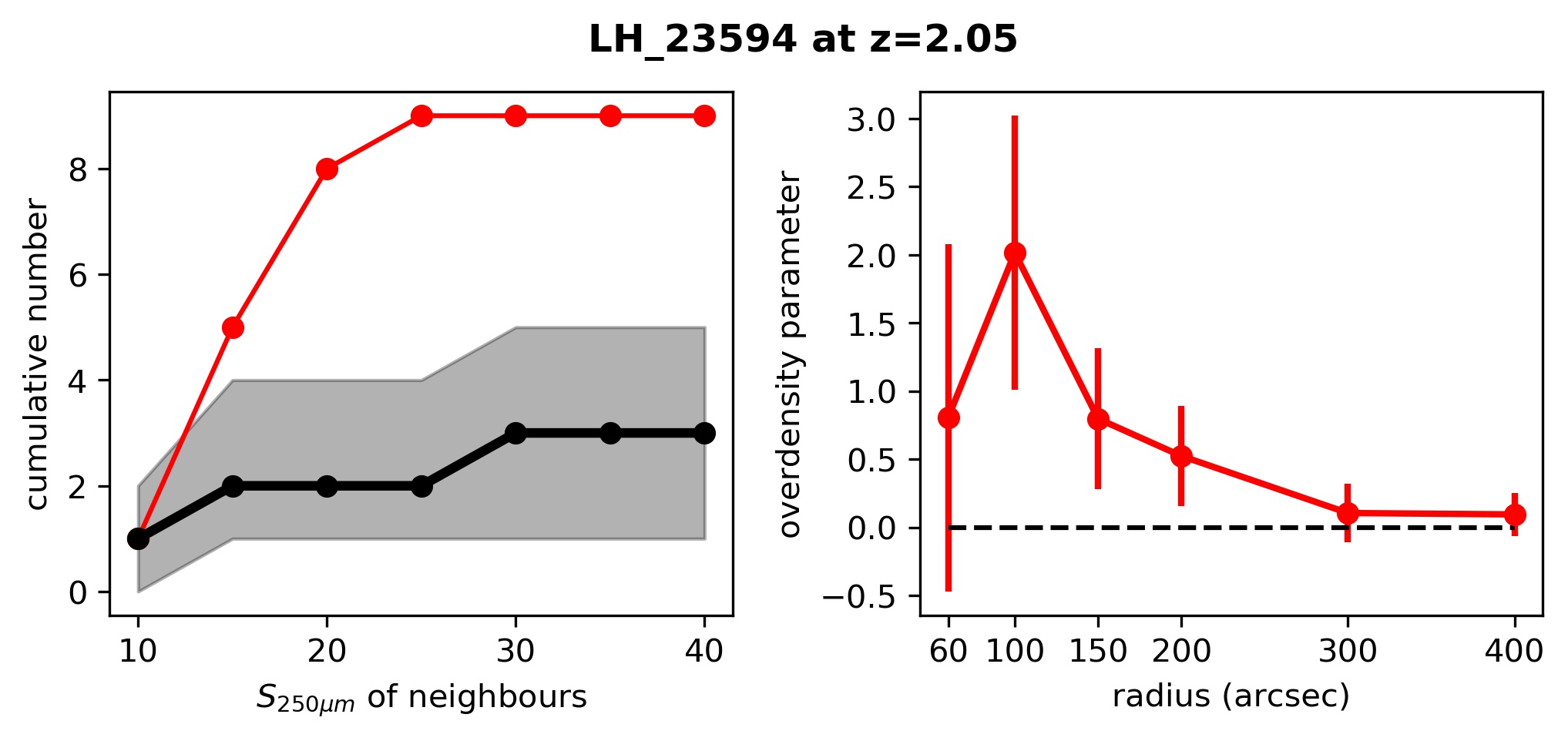}
\end{subfigure}
\begin{subfigure}{.5\textwidth}
  \centering
  \includegraphics[width=\linewidth]{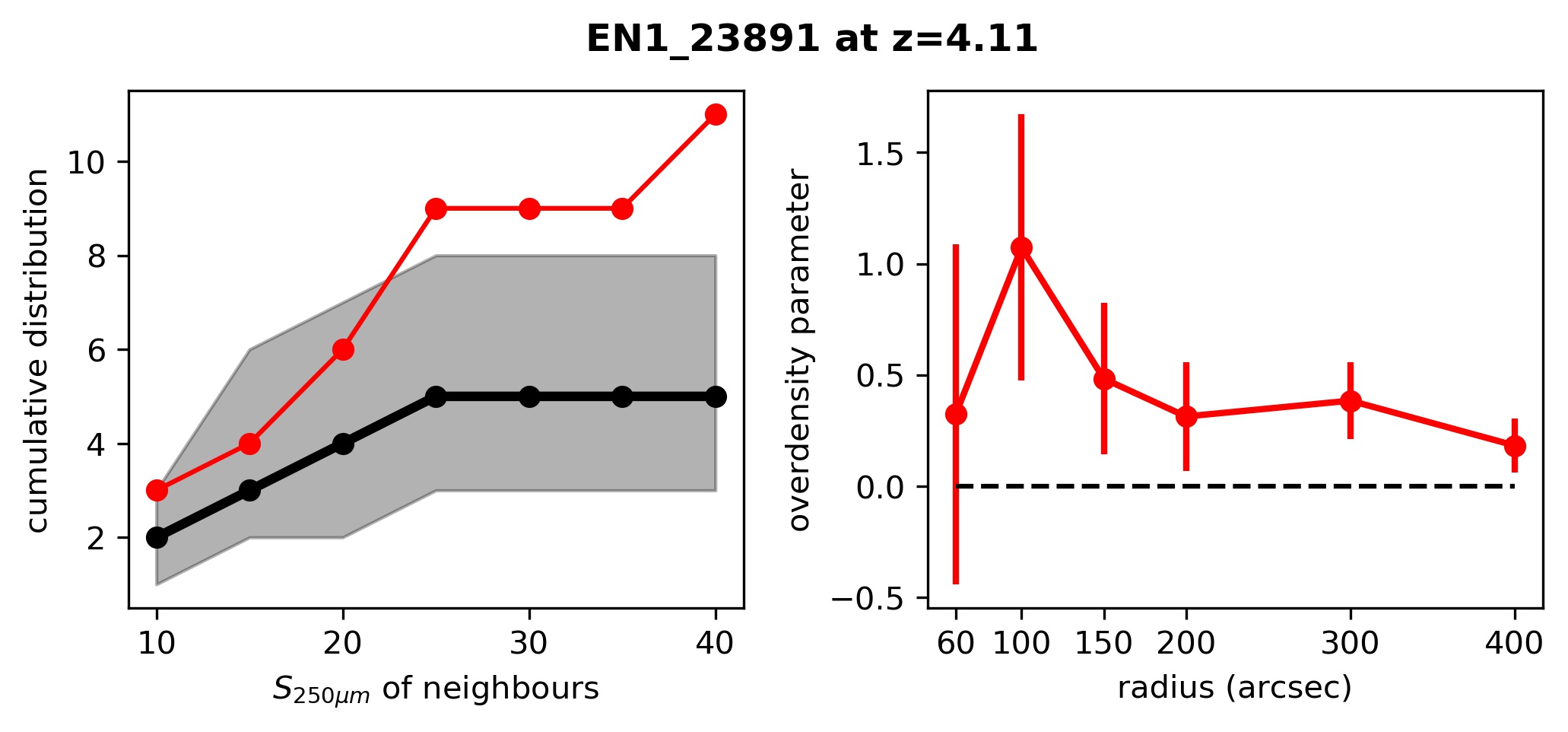}
\end{subfigure}
\caption{Left: Cumulative 250 $\mu$m flux density distribution of HLIRG neighbours within 100$\arcsec$ compared with random neighbours. Right: overdensity parameters measured within different radius.}
\label{fig:all}
\end{figure*}

\renewcommand{\thefigure}{}
\addtocounter{figure}{-1}

\begin{figure*}

\begin{subfigure}{.5\textwidth}
  \centering
  \includegraphics[width=\linewidth]{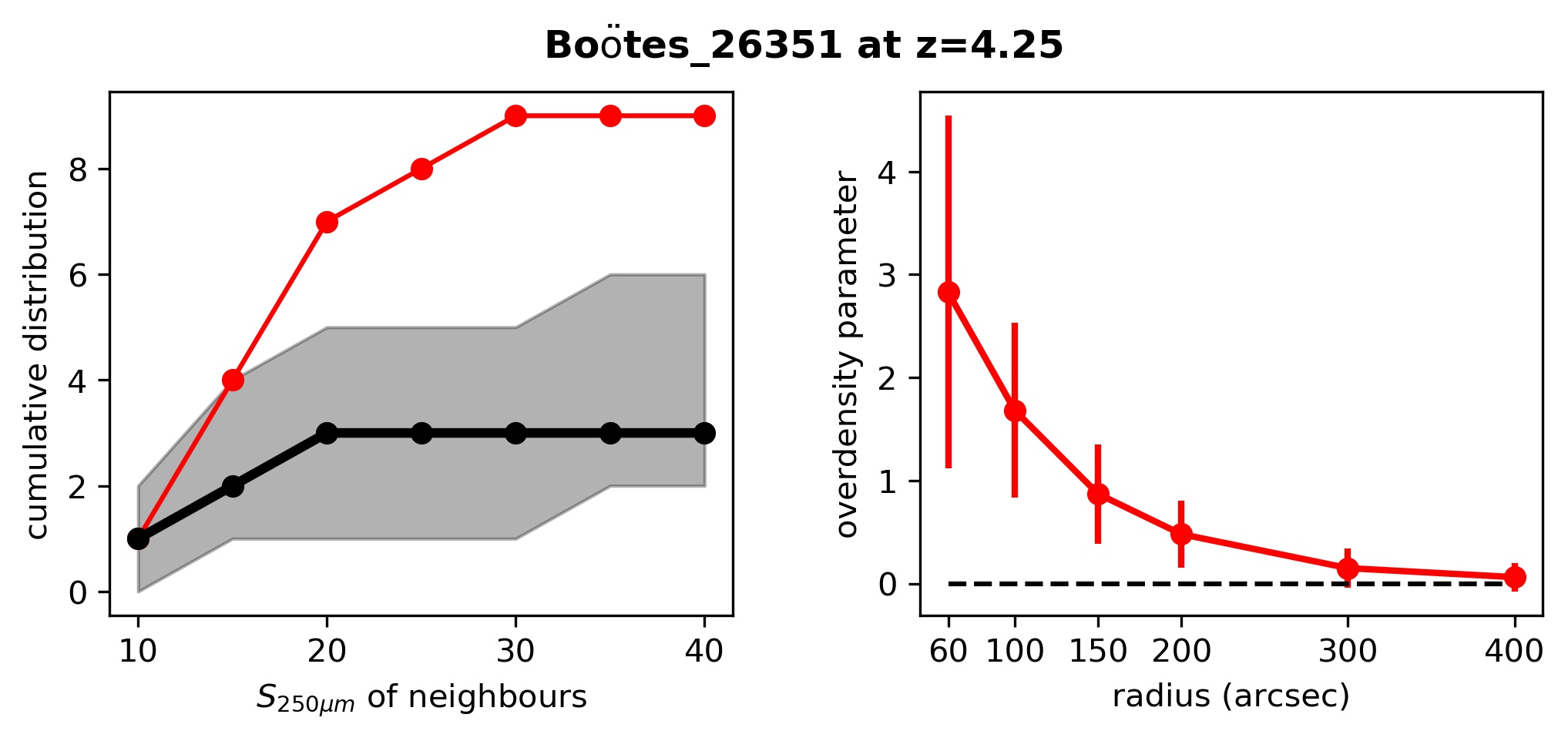}
\end{subfigure}
\begin{subfigure}{.5\textwidth}
  \centering
  \includegraphics[width=\linewidth]{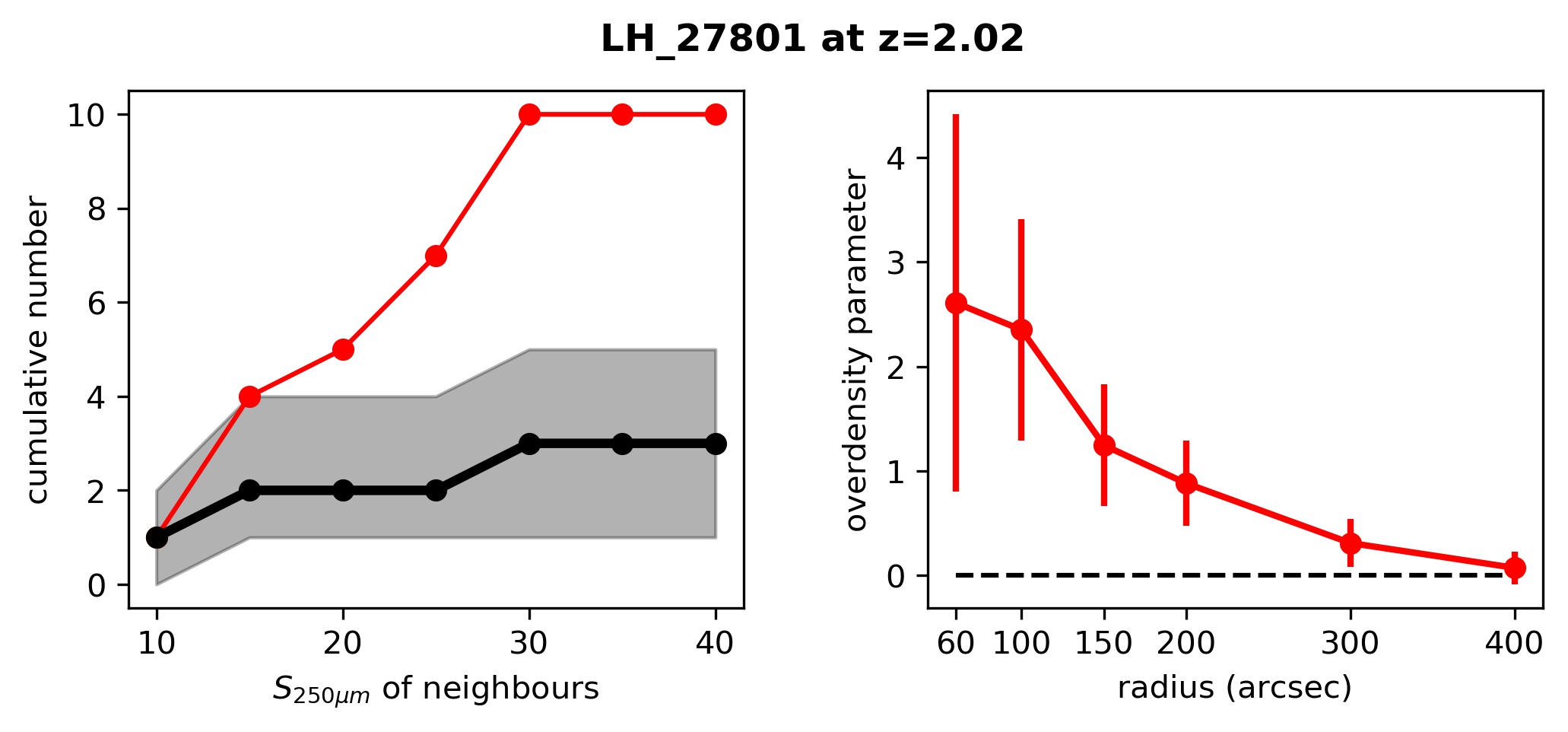}
\end{subfigure}

\begin{subfigure}{.5\textwidth}
  \centering
  \includegraphics[width=\linewidth]{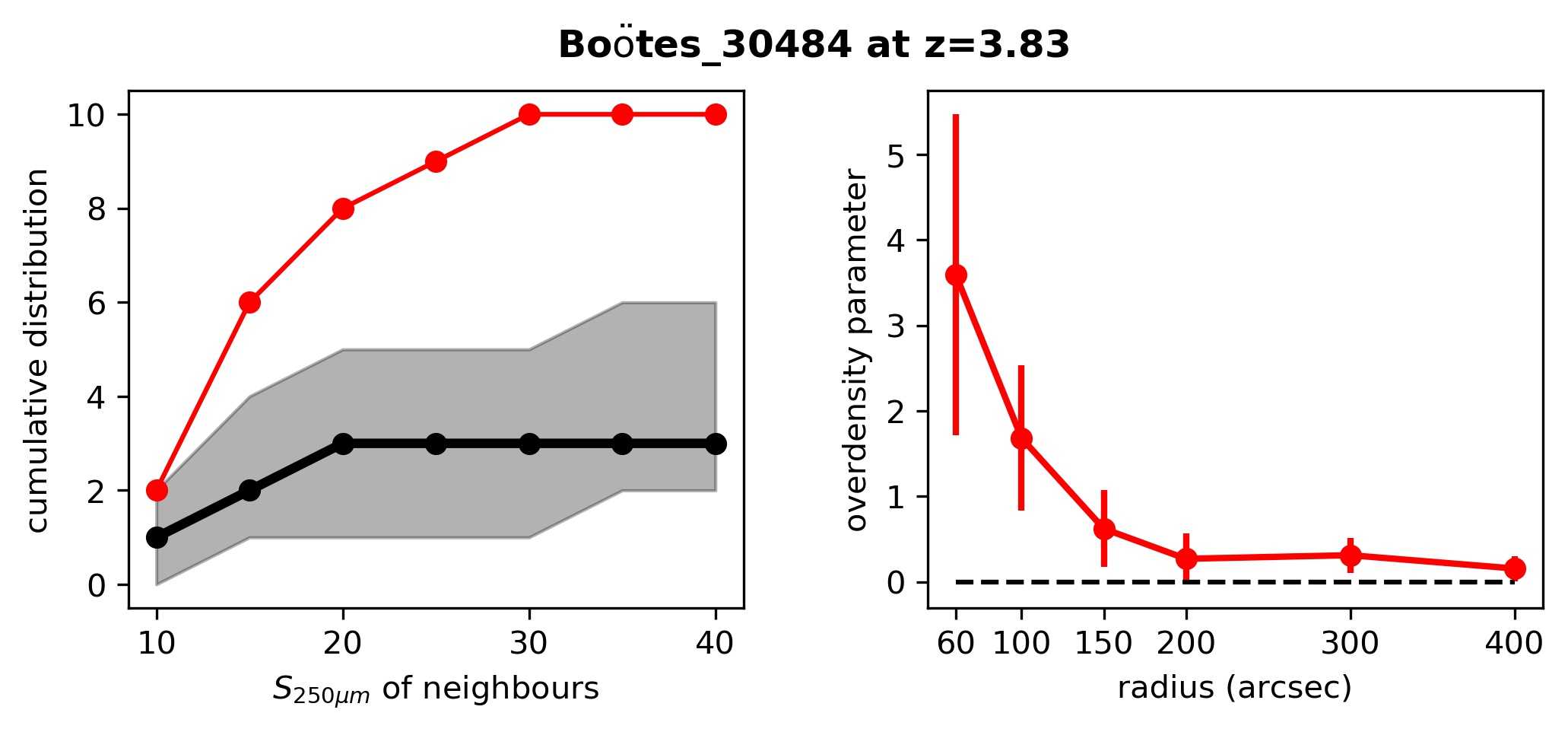}
\end{subfigure}
\begin{subfigure}{.5\textwidth}
  \centering
  \includegraphics[width=\linewidth]{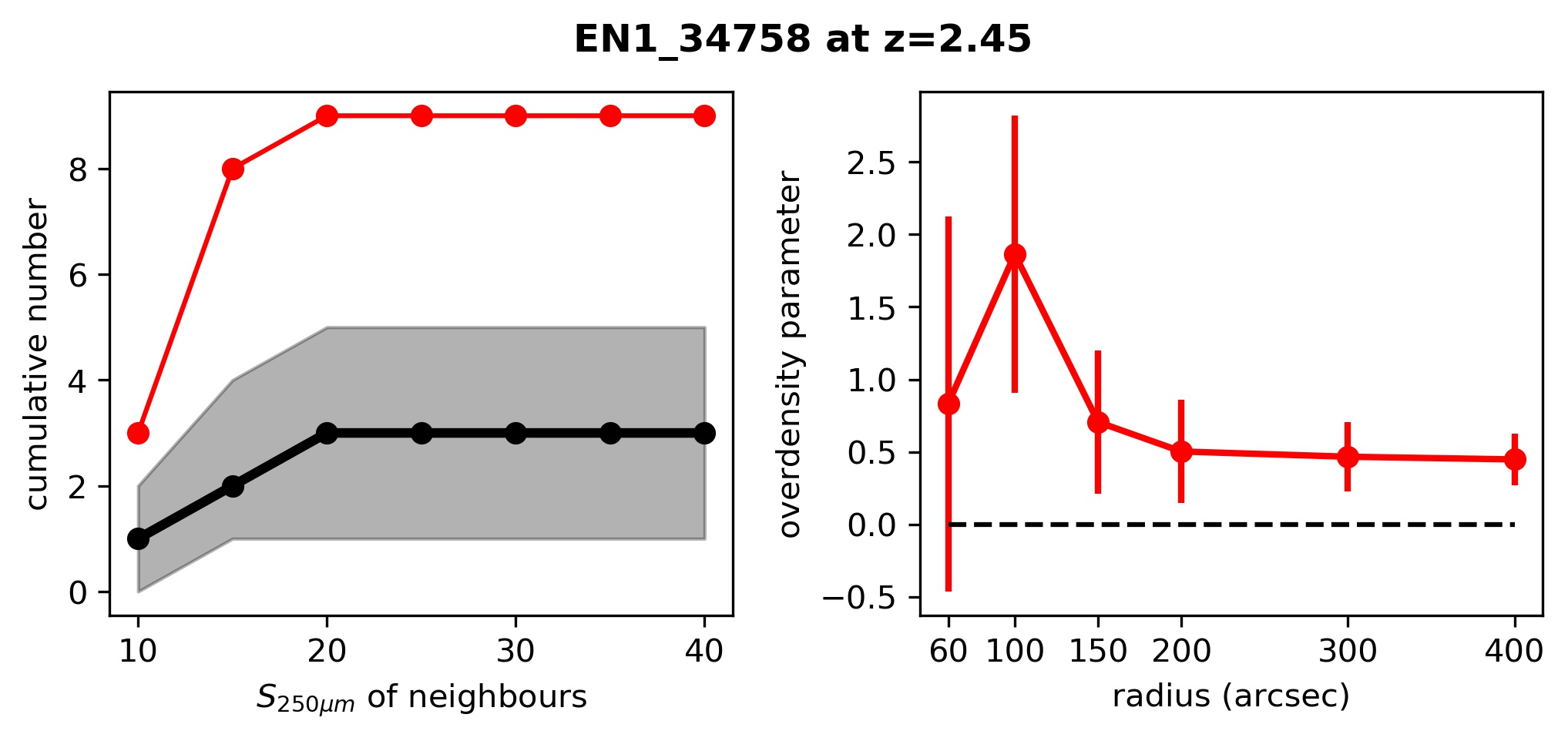}
\end{subfigure}

\begin{subfigure}{.5\textwidth}
  \centering
  \includegraphics[width=\linewidth]{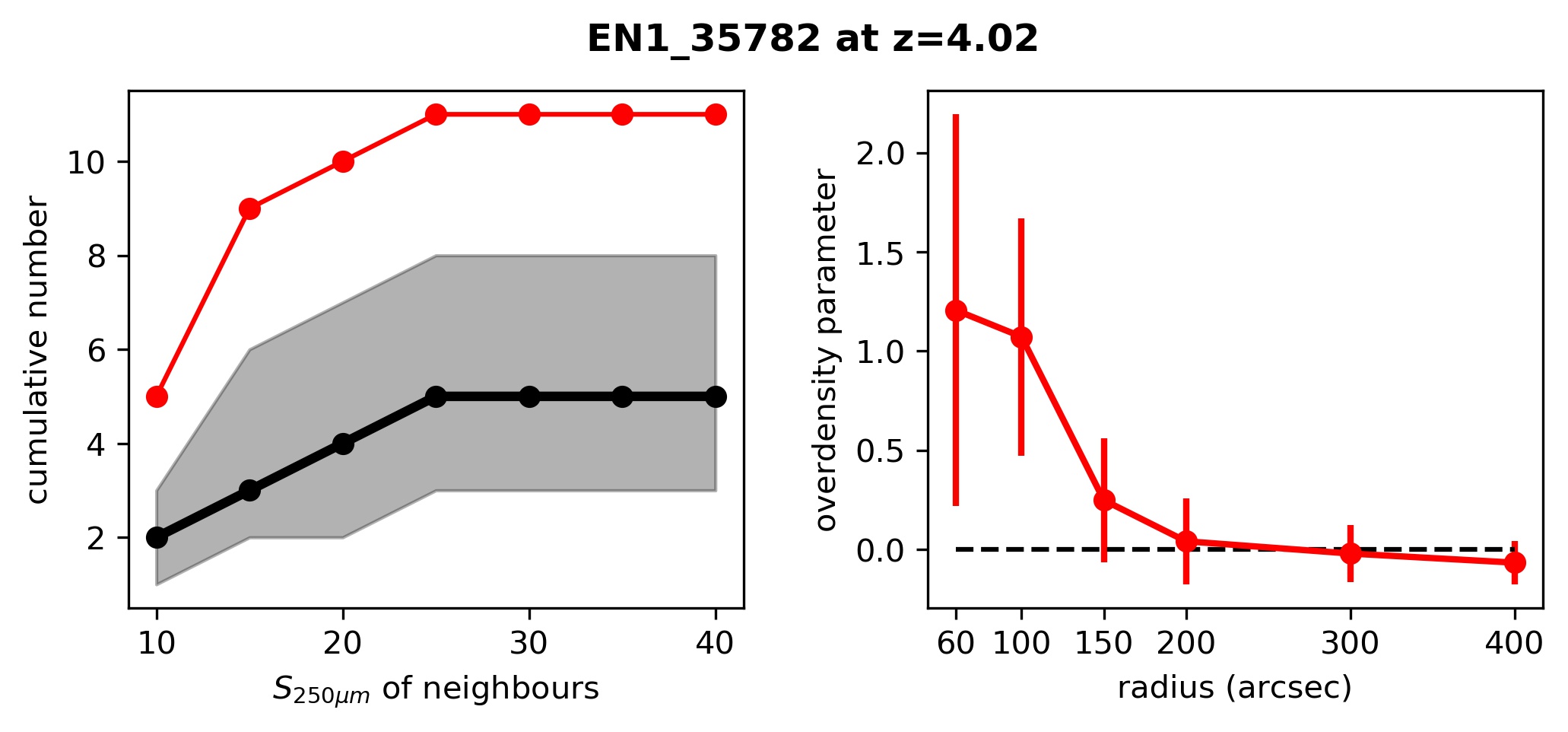}
\end{subfigure}
\begin{subfigure}{.5\textwidth}
  \centering
  \includegraphics[width=\linewidth]{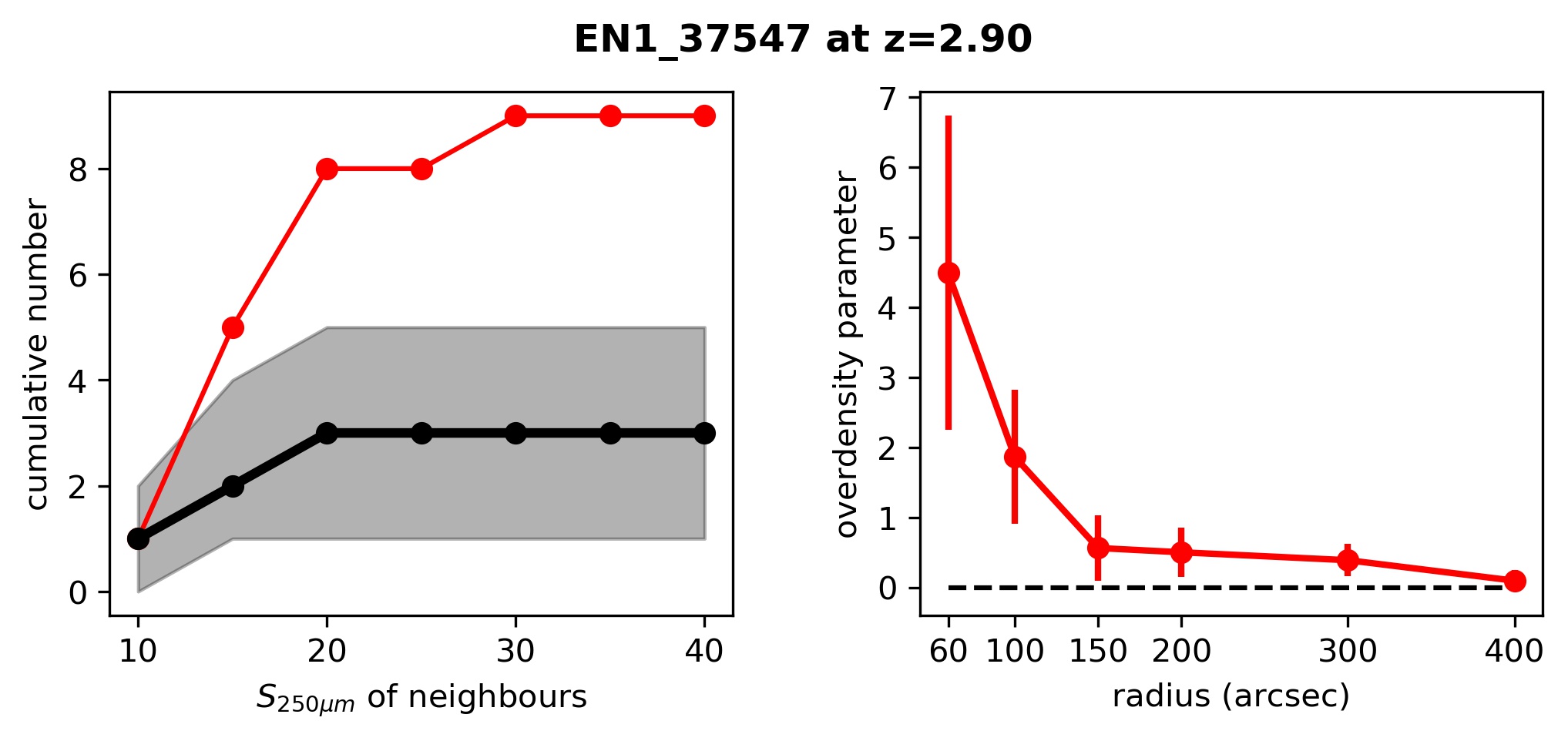}
\end{subfigure}

\begin{subfigure}{.5\textwidth}
  \centering
  \includegraphics[width=\linewidth]{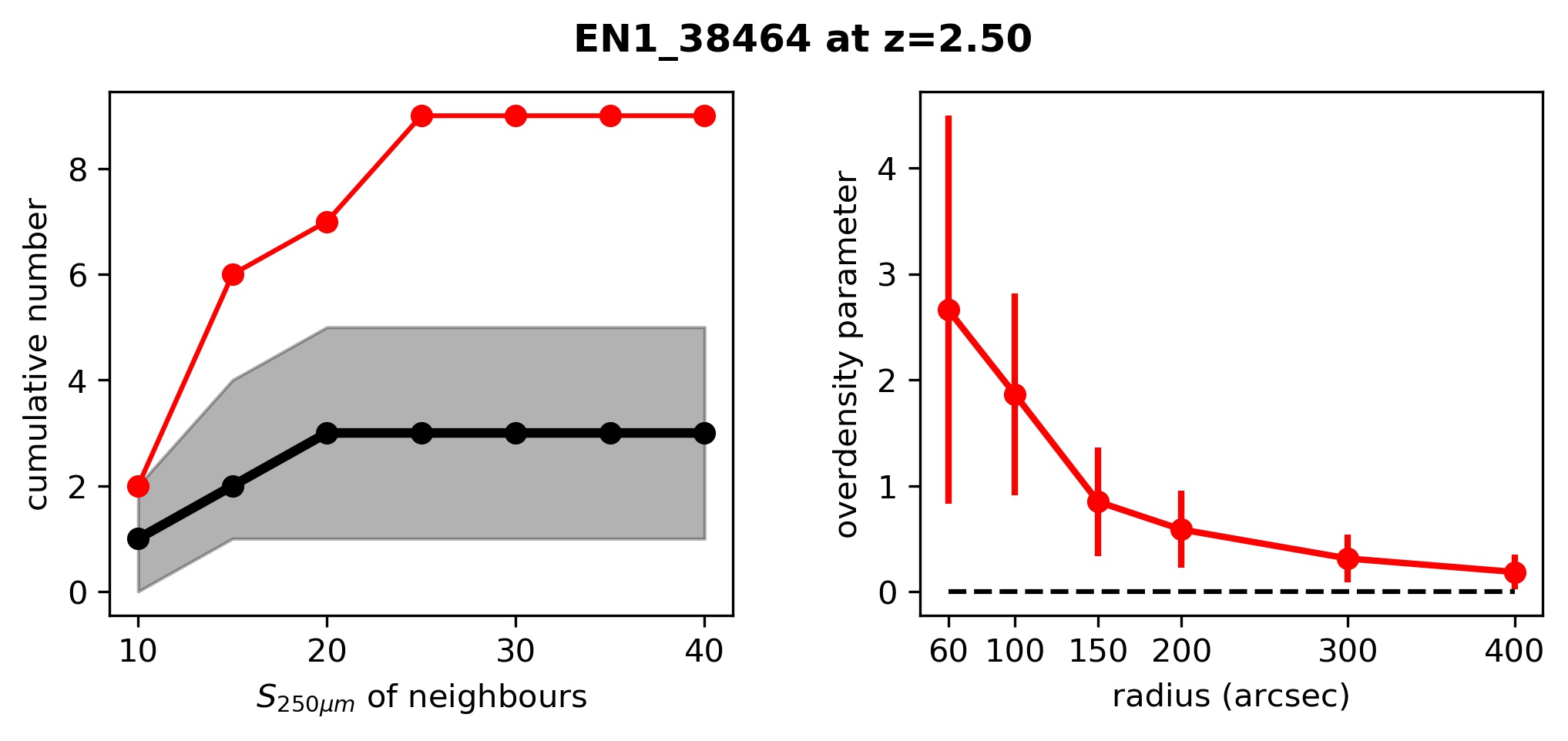}
\end{subfigure}
\begin{subfigure}{.5\textwidth}
  \centering
  \includegraphics[width=\linewidth]{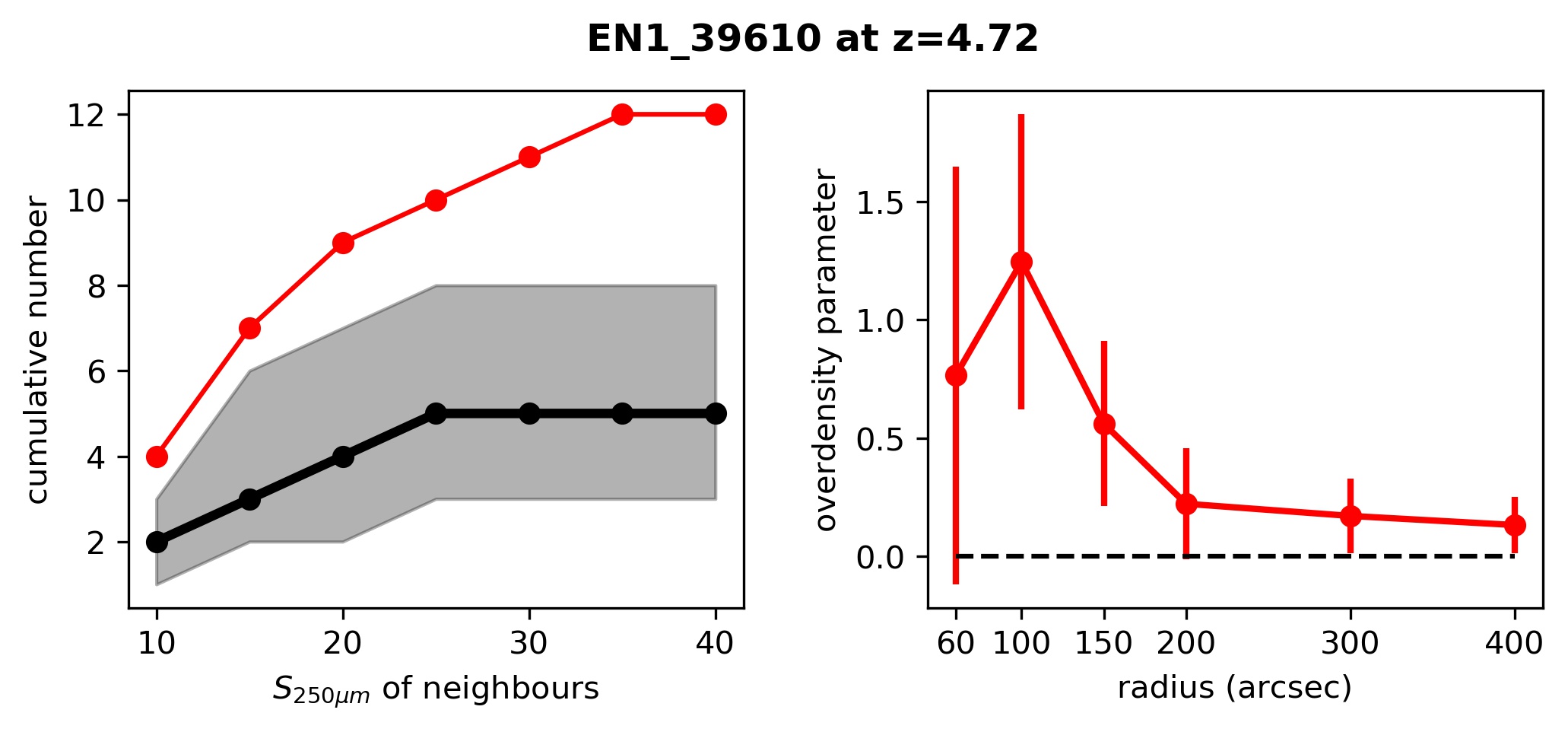}
\end{subfigure}

\begin{subfigure}{.5\textwidth}
  \centering
  \includegraphics[width=\linewidth]{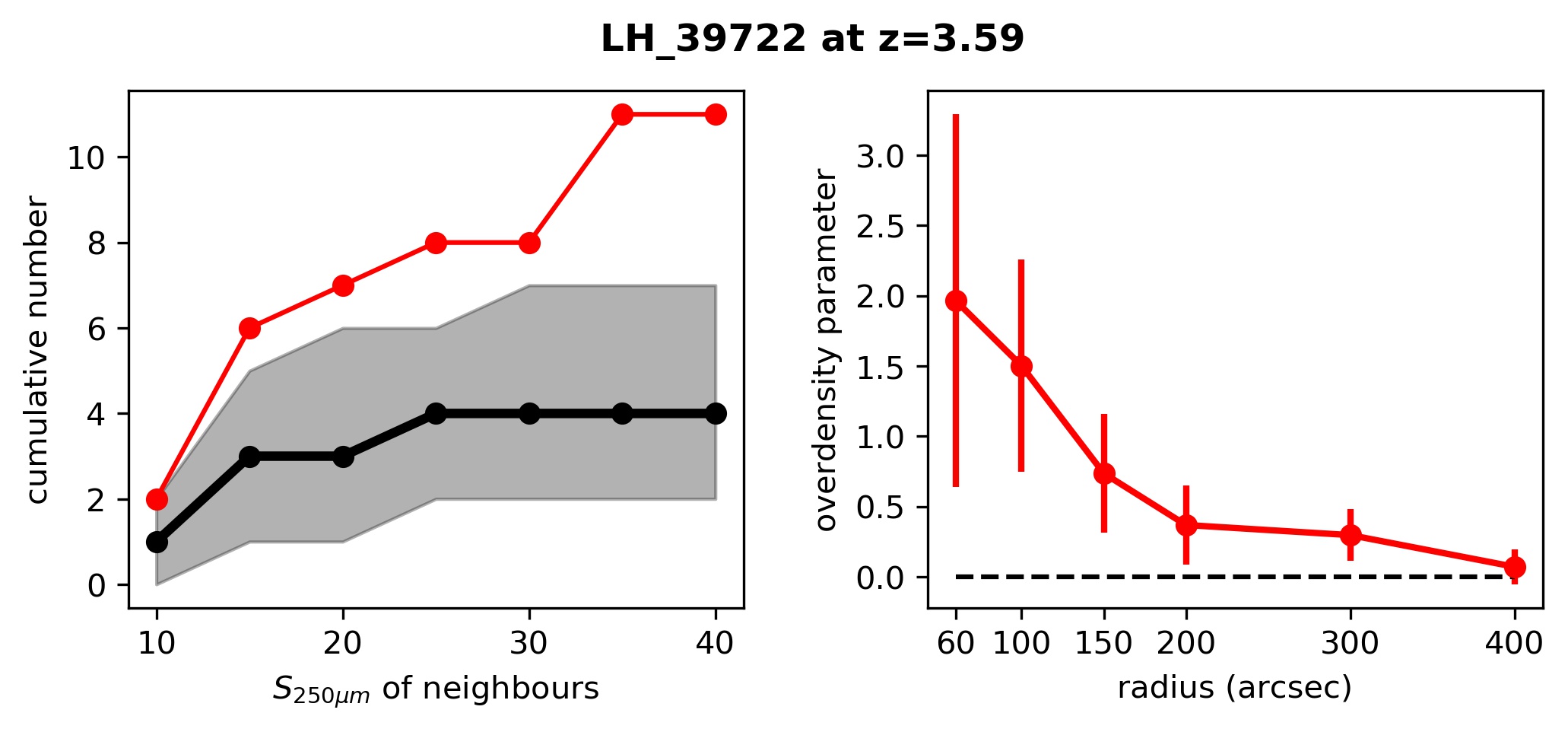}
\end{subfigure}
\begin{subfigure}{.5\textwidth}
  \centering
  \includegraphics[width=\linewidth]{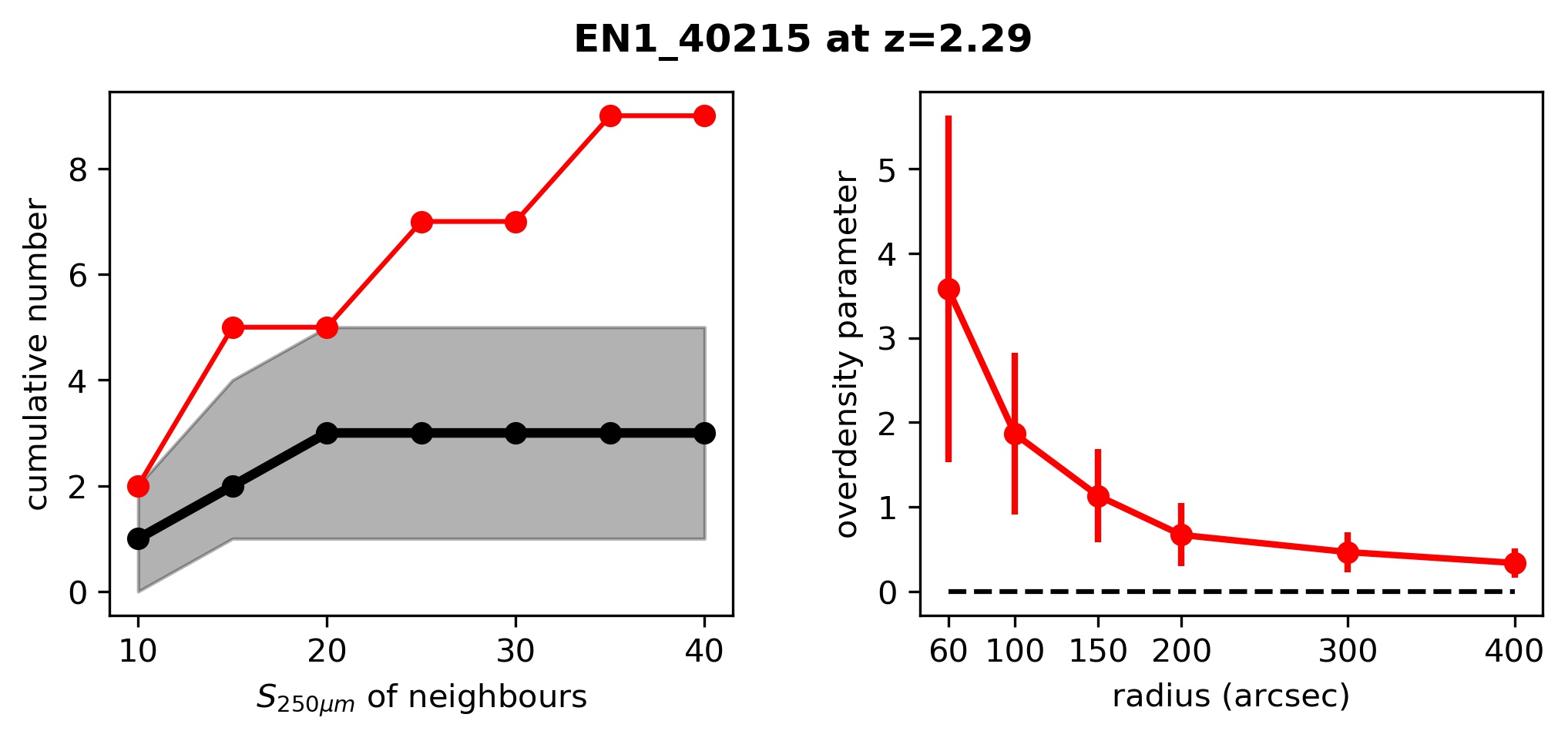}
\end{subfigure}

\caption{Continued}
\end{figure*}

\renewcommand{\thefigure}{}
\addtocounter{figure}{-1}

\begin{figure*}

\begin{subfigure}{.5\textwidth}
  \centering
  \includegraphics[width=\linewidth]{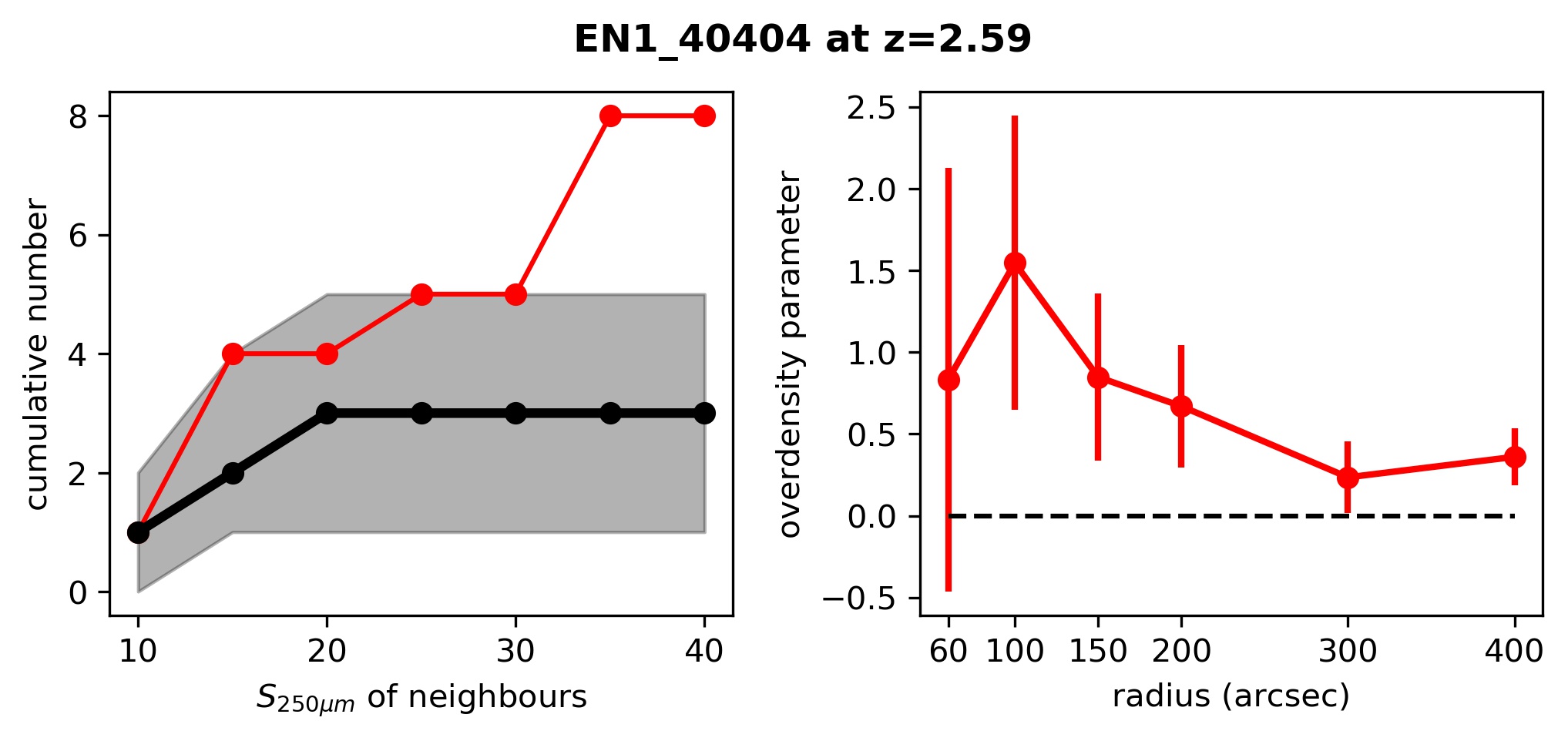}
\end{subfigure}
\begin{subfigure}{.5\textwidth}
  \centering
  \includegraphics[width=\linewidth]{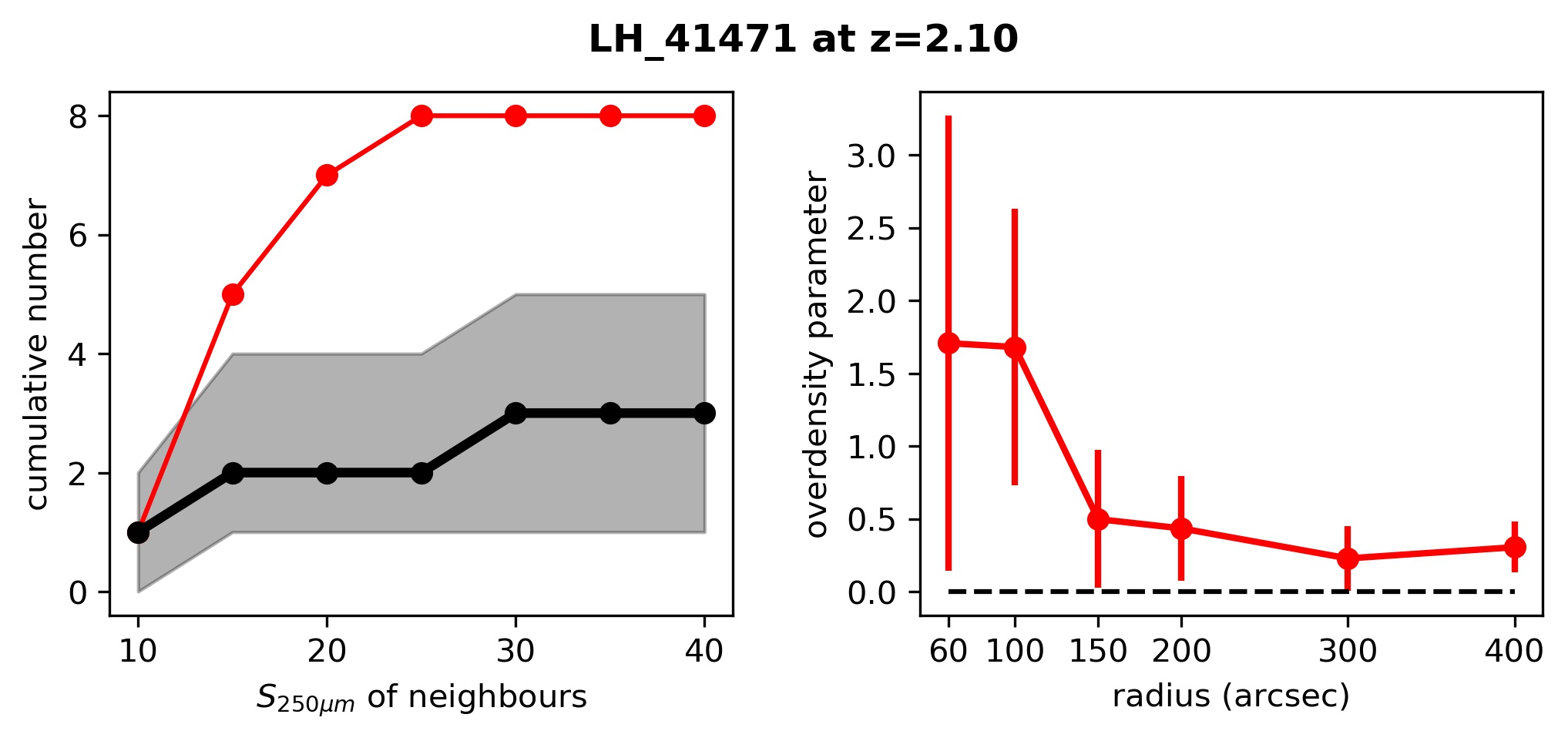}
\end{subfigure}

\begin{subfigure}{.5\textwidth}
  \centering
  \includegraphics[width=\linewidth]{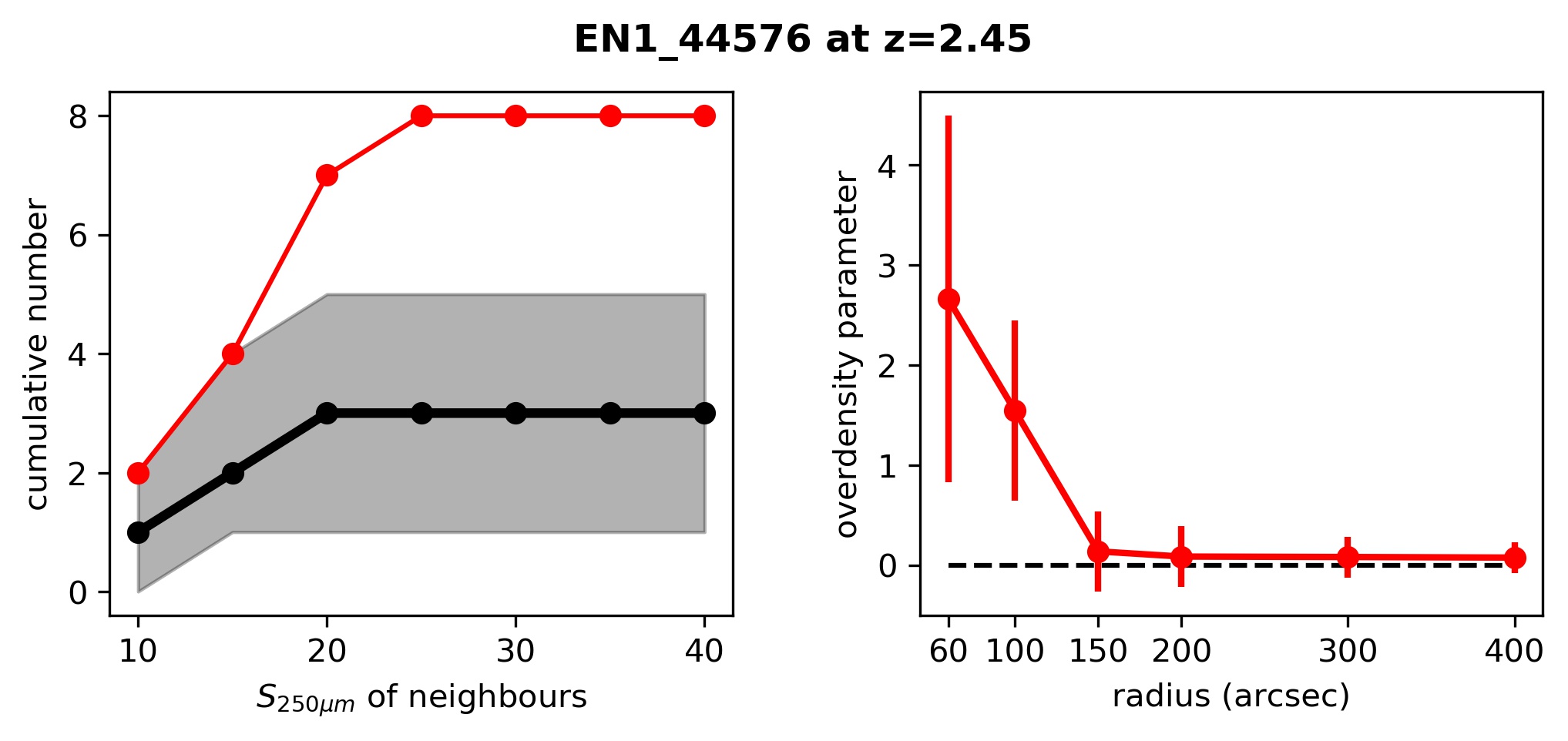}
\end{subfigure}
\begin{subfigure}{.5\textwidth}
  \centering
  \includegraphics[width=\linewidth]{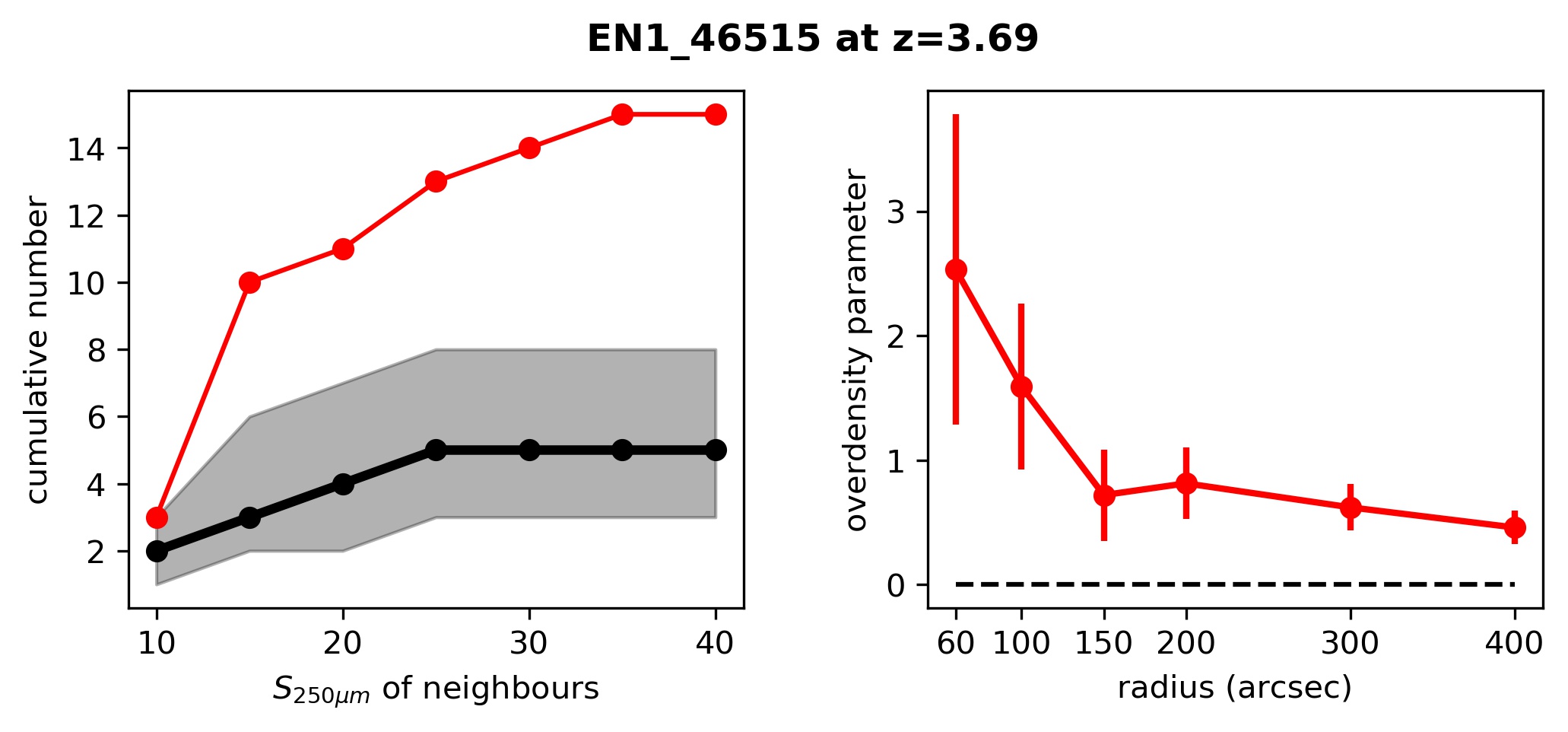}
\end{subfigure}

\begin{subfigure}{.5\textwidth}
  \centering
  \includegraphics[width=\linewidth]{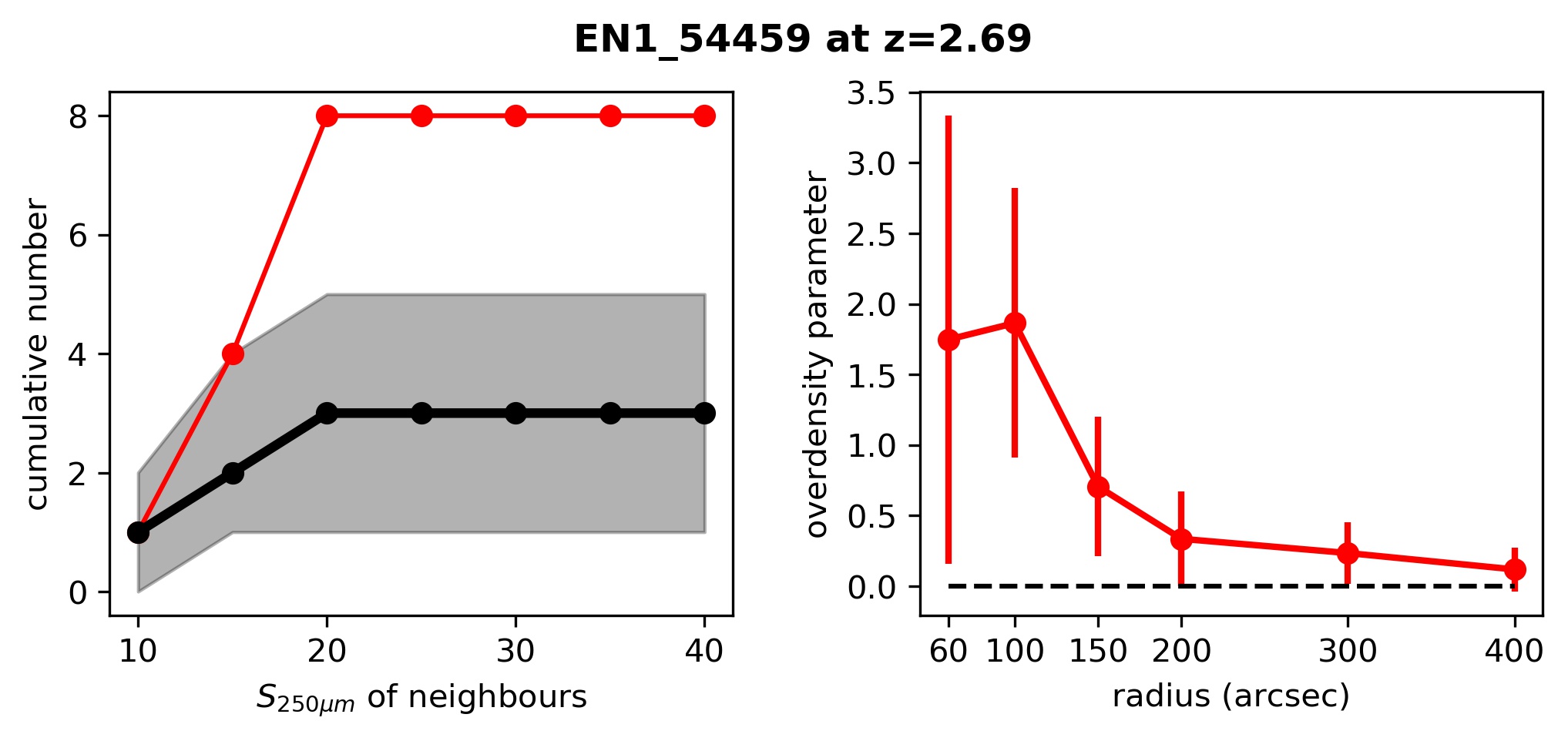}
\end{subfigure}
\begin{subfigure}{.5\textwidth}
  \centering
  \includegraphics[width=\linewidth]{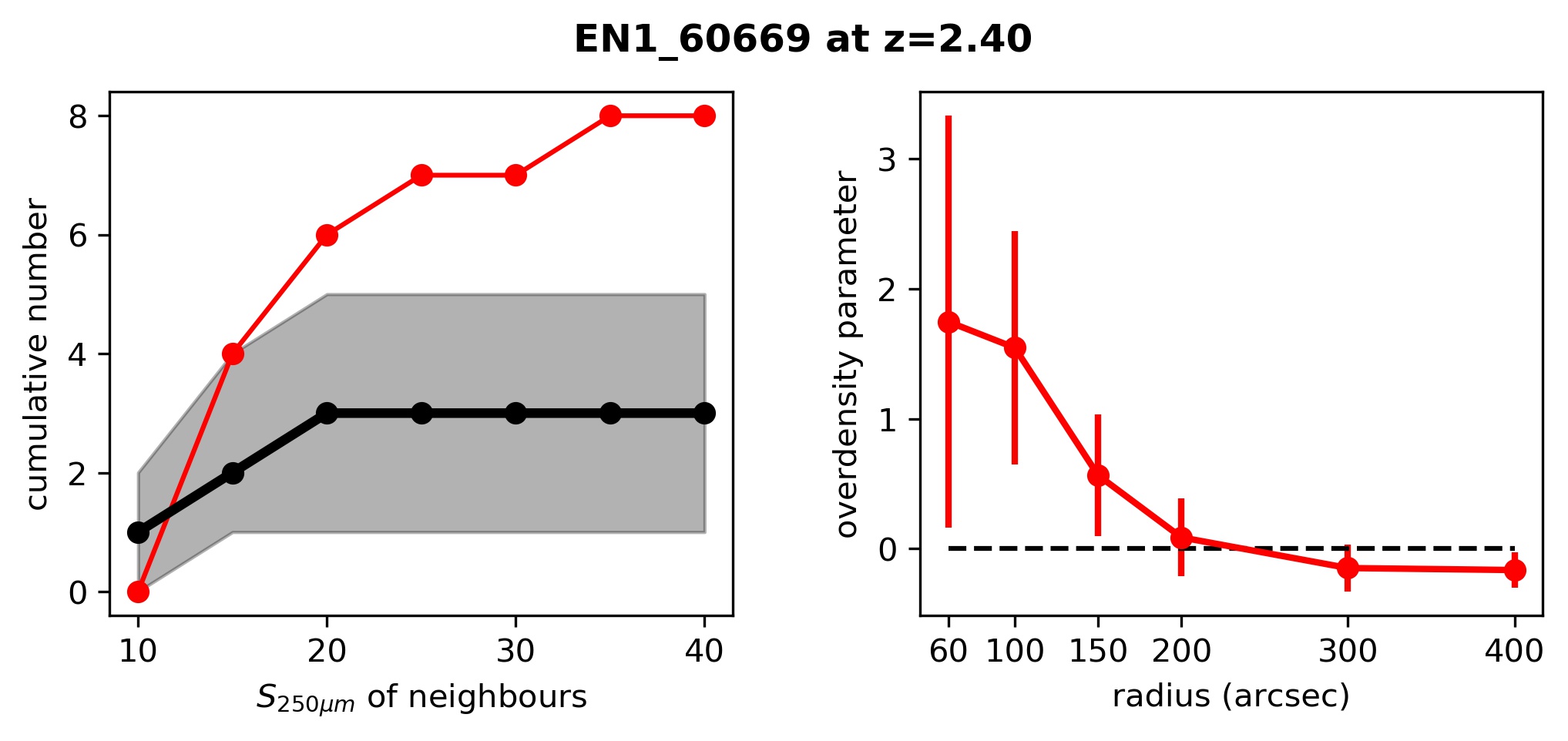}
\end{subfigure}

\caption{Continued}
\end{figure*}

\renewcommand{\thefigure}{\arabic{figure}}

\end{appendix}

\end{document}